\documentclass[LATO1COL]{WileyNJD-v2}
\usepackage[numbers,super,sort&compress]{NJDnatbib}

\makeatletter
\renewcommand{\@biblabel}[1]{\quad#1.}
\makeatother

\usepackage{hyperref}
\renewcommand{\doi}[1]{doi: \href{https://doi.org/#1}{#1}}

\usepackage[utf8]{inputenc}

\usepackage{graphicx}
\usepackage{caption}
\usepackage{subcaption}
\usepackage{sidecap}

\usepackage{amsmath}
\usepackage{color}
\usepackage{ifpdf}

\usepackage{hyperref}

\usepackage{threeparttable}
\usepackage{booktabs}
\usepackage{float}
\usepackage{multirow}
\usepackage{rotating}
\usepackage{setspace}
\usepackage{makecell}
\usepackage{moresize}
\usepackage{tabularx}
\usepackage{algorithm}
\usepackage{algpseudocode}
\usepackage{listings}
\usepackage{paralist}

\usepackage{xspace}
\usepackage{keyval}
\usepackage{nicefrac}

\usepackage{textcomp}

\usepackage{breakurl}
\usepackage{url}
\usepackage{adjustbox}

\usepackage{textcomp}

\let\carriagereturn\undefined
\usepackage{dingbat}

\definecolor{mauve}{rgb}{0.58,0,0.82}
\definecolor{gray}{rgb}{0.5,0.5,0.5}
\definecolor{listinggray}{gray}{0.95}
\definecolor{darkgray}{gray}{0.7}
\definecolor{commentgreen}{rgb}{0, 0.4, 0}
\definecolor{darkblue}{rgb}{0, 0, 0.4}
\definecolor{middleblue}{rgb}{0, 0, 0.7}
\definecolor{darkred}{rgb}{0.4, 0, 0}
\definecolor{brown}{rgb}{0.5, 0.5, 0}

\newcommand\pythonstyle{\lstset{
    language=Python,
    basicstyle=\scriptsize\ttfamily,
    otherkeywords={self},             
    keywordstyle=\bfseries\color{blue},
    emph={MyClass,__init__},          
    emphstyle=\ttfamily\bfseries\color{deepred},    
    stringstyle=\color{mauve},
    commentstyle=\color{gray}\textit,      
    frame=tb,                         
    showstringspaces=false,           %
    breaklines=true,
    prebreak=\carriagereturn,
    numberstyle=\tiny\color{gray},
    numbers=left,
    stepnumber=1
}}

\lstnewenvironment{python}[1][]
{
\pythonstyle
\lstset{#1}
}
{}

\newcommand{\package}[1]{\textsl{#1}}

\newcommand{\tcomp}{\ensuremath{{t}_{\text{comp}}}\xspace}
\newcommand{\tIO}{\ensuremath{{t}_{\text{I/O}}}\xspace}
\newcommand{\tcomm}{\ensuremath{t_{\text{comm}}}\xspace}

\newcommand{\RcompIO}{\ensuremath{R_{\text{comp/IO}}}\xspace}
\newcommand{\Rcompcomm}{\ensuremath{R_{\text{comp/comm}}}\xspace}

\floatstyle{ruled}

\ifpdf
\DeclareGraphicsExtensions{.pdf, .jpg, .tif}
\else
\DeclareGraphicsExtensions{.ps,  .eps, .jpg}
\fi

\articletype{Research Article}%

\received{\color{red}Added at production}
\revised{\color{red}Added at production}
\accepted{\color{red}Added at production}
        
\begin{document}

\title{Parallel Performance of Molecular Dynamics Trajectory Analysis}

\author[1]{Mahzad Khoshlessan}
\author[2]{Ioannis Paraskevakos}
\author[3]{Geoffrey C. Fox}
\author[2]{Shantenu Jha}
\author[1,4]{Oliver Beckstein*}

\authormark{KHOSHLESSAN \textsc{et al}}

\address[1]{\orgdiv{Department of Physics}, \orgname{Arizona State University},
  \orgaddress{Tempe, AZ 85281}, \country{USA}}
\address[2]{\orgdiv{Department of Electrical \& Computer Engineering},
  \orgname{Rutgers University}, \orgaddress{Piscataway, NJ 08854}, \country{USA}}
\address[3]{\orgdiv{Digital Science Center}, \orgname{Indiana University},
  \orgaddress{Bloomington, IN 47405}, \country{USA}}
\address[4]{\orgdiv{Center for Biological Physics}, \orgname{Arizona State University},
  \orgaddress{Tempe, AZ 85281}, \country{USA}}

\corres{*Oliver Beckstein, \orgdiv{Department of Physics}, \orgname{Arizona State University},
  \orgaddress{Tempe, AZ 85281}, \country{USA}. \email{oliver.beckstein@asu.edu}}
    


\abstract[Summary]{%
The performance of biomolecular molecular dynamics simulations has steadily increased on modern high performance computing resources but acceleration of the analysis of the output trajectories has lagged behind so that analyzing simulations is becoming a bottleneck.
To close this gap, we studied the performance of parallel trajectory analysis with MPI and the Python \package{MDAnalysis} library on three different XSEDE supercomputers where trajectories were read from a Lustre parallel file system. 
Strong scaling performance was impeded by stragglers, MPI processes that were slower than the typical process. 
Stragglers were less prevalent for compute-bound workloads, thus pointing to file reading as a bottleneck for scaling.
However, a more complicated picture emerged in which both the computation and the data ingestion exhibited close to ideal strong scaling behavior whereas stragglers were primarily caused by either large MPI communication costs or long times to open the single shared trajectory file.
We improved overall strong scaling performance by either subfiling (splitting the trajectory into separate files) or MPI-IO with Parallel HDF5 trajectory files.
The parallel HDF5 approach resulted in near ideal strong scaling on up to 384 cores (16 nodes), thus reducing trajectory analysis times by two orders of magnitude compared to the serial approach.
}


\maketitle


\section{Introduction}
\label{sec:introduction}

Molecular dynamics (MD) simulations are a powerful method to generate new insights into the function of biomolecules \cite{Borhani:2012mi, Dror:2012cr, Orozco:2014dq, Perilla:2015kx, Bottaro:2018aa}.
These simulations produce trajectories---time series of atomic coordinates---that now routinely include millions of time steps and can measure Terabytes in size.
These trajectories need to be analyzed using statistical mechanics approaches \cite{Tuckerman:2010cr, Mura:2014kx} but because of the increasing size of the data, trajectory analysis is becoming a bottleneck in typical biomolecular simulation scientific workflows~\cite{Cheatham:2015}.
Many data analysis tools and libraries have been developed to extract the desired information from the output trajectories from MD simulations ~\cite{nmoldyn, nmoldyn-2012, Hum96, Hinsen:2000kx, Grant:2006ud, himach-2008, Romo:2009zr, Romo:2014bh, Michaud-Agrawal:2011fu, Gowers:2016aa, cpptraj-2013, McGibbon:2015aa, pteros2015, Doerr:2016aa} but few can efficiently use modern High Performance Computing (HPC) resources to accelerate the analysis stage.
MD trajectory analysis primarily requires \emph{reading} of data from the file system; the processed output data are typically negligible in size compared to the input data and therefore we exclusively investigate the reading aspects of trajectory I/O (i.e., the ``I'').
We focus on the \package{MDAnalysis} package \cite{Gowers:2016aa,Michaud-Agrawal:2011fu}, which is an open-source object-oriented Python library for structural and temporal analysis of MD simulation trajectories and individual protein structures.
Although \package{MDAnalysis} accelerates selected algorithms with OpenMP, it is not clear how to best use it for scaling up analysis on multi-node supercomputers.
Here we discuss the challenges and lessons-learned for making parallel analysis on HPC resources feasible with \package{MDAnalysis}, which should also be broadly applicable to other general purpose trajectory analysis libraries.

Previously, we had used a parallel split-apply-combine  approach \cite{Wickham:2011aa} to study the performance of the commonly performed ``RMSD fitting'' analysis problem~\cite{Khoshlessan:2017ab, ICCP-2018, Fan:2019aa}, which calculates the minimal root mean squared distance (RMSD) of the positions of a subset of atoms to a reference conformation under optimization of rigid body translations and rotations \cite{Liu:2010kx, Lea96, Mura:2014kx}.
We had investigated two parallel implementations, one using \package{Dask}~\cite{Rocklin:2015aa} and one using the message passing interface (MPI) with \package{mpi4py}~\cite{Dalcin:2011aa, Dalcin:2005aa}. 
For both \package{Dask} and MPI, we had previously only been able to obtain good strong scaling performance within a single node.
Beyond a single node performance had dropped due to \emph{straggler} tasks, a subset of tasks that had performed abnormally slower than the typical task execution times; the total execution time had become dominated by stragglers and overall performance had decreased.
Stragglers are a well-known challenge to improving performance on HPC resources \cite{Garraghan2016} but there has been little discussion of their impact in the biomolecular simulation community.

In the present study, we analyzed the MPI case in more detail to better understand the origin of stragglers with the goal to find  parallelization approaches to speed up parallel post-processing of MD trajectories in the \package{MDAnalysis} library.
We especially wanted to make efficient use of the resources provided by current supercomputers such as multiple nodes with hundreds of CPU cores and a Lustre parallel file system.

As in our previous study\cite{Khoshlessan:2017ab} we selected the commonly used RMSD algorithm implemented in \package{MDAnalysis} as a typical use case.
We employed the single program multiple data (SPMD) paradigm to parallelize this algorithm on three different HPC resources (XSEDE's \emph{SDSC Comet}, \emph{LSU SuperMic}, and \emph{PSC Bridges} \cite{xsede}).
With SPMD, each process executes essentially the same operations on different parts of the data.
The three clusters differed in their architecture but all used Lustre as their parallel file system.
We used Python (\url{https://www.python.org/}), a machine-independent, byte-code interpreted, object-oriented programming language, which is well-established in the biomolecular simulation community with good support for parallel programming for HPC~\cite{Dalcin:2011aa, GAiN}. 
We found that communication and reading I/O were the two main scalability bottlenecks, with some indication that read I/O might have been interfering with the communications.
Stragglers and therefore poor scaling were a consequence of inefficient use of the parallel Lustre file system.
We therefore focused on two different approaches to better leverage Lustre and mitigate I/O bottlenecks: MPI parallel I/O (MPI-IO) with the HDF5 file format and subfiling (trajectory file splitting).
MPI-IO eliminated stragglers and improved the performance with near ideal scaling, $S(N) = N$, i.e., the speed-up $S$ scaled linearly with the number $N$ of CPU cores while exhibiting a slope of one.

The paper is organized as follows:
We first review stragglers and existing approaches to parallelizing MD trajectory analysis in section \ref{sec:background}.
We describe the software packages and algorithms in section \ref{sec:packages} and the benchmarking environment in section \ref{sec:system}.
Section \ref{sec:methods} explains how we measured performance.
The main results are presented in section \ref{sec:experiments}, with section \ref{sec:clusters} demonstrating reproducibility on different supercomputers.
We provide general guidelines and lessons-learned in section \ref{sec:guidelines} and finish with conclusions in section \ref{sec:conclusions}.


\section{Background and Related Work}
\label{sec:background}

In our previous work, we found that straightforward implementation of simple parallelization with a split-apply-combine algorithm in Python failed to scale beyond a single compute node~\cite{Khoshlessan:2017ab} because a few tasks (MPI-ranks or Dask~\cite{Rocklin:2015aa} processes) took much longer than the typical task and so limited the overall performance.
However, the cause for these \emph{straggler} tasks remained obscure.
Here, we studied the straggler problem in the context of an MPI-parallelized trajectory analysis algorithm in Python and investigated solutions to overcome it.
We briefly review stragglers in section~\ref{sec:stragglers} and summarize existing approaches to parallel trajectory analysis in section~\ref{sec:otherparallel}.

\subsection{Stragglers}
\label{sec:stragglers}

\emph{Stragglers} or \emph{outliers} were traditionally considered in the context of MapReduce jobs that consist of multiple tasks that all have to finish for the job to succeed: A straggler was a task that took an ``unusually long time to complete'' \cite{Dean2008} and therefore substantially impeded job completion.
In general, any component of a parallel workflow whose runtime exceeds a typical run time (for example, 1.5 times the median runtime) can be considered a straggler \cite{Ananthanarayanan:2010aa}.
Stragglers are a challenge for improving performance on HPC resources \cite{Garraghan2016}; they are a known problem in frameworks such as MapReduce~\cite{Dean2008, Ananthanarayanan:2010aa}, Spark~\cite{Kyong2017,Ousterhout2017,Gittens2016,Yang2016}, Hadoop~\cite{Dean2008}, cloud data centers~\cite{Kirpichov2016,Garraghan2016}, and have a high impact on performance and energy consumption of big data systems~\cite{Tien-2017}.
Both internal and external factors are known to contribute to stragglers. 
Internal factors include heterogeneous capacity of worker nodes and resource competition due to other tasks running on the same worker node.
External factors include resource competition due to co-hosted applications, input data skew, remote input or output source being too slow,  faulty hardware~\cite{Chen2014,Dean2008}, and node mis-configuration~\cite{Dean2008}.
Competition over scarce resources \cite{Ananthanarayanan:2010aa}, in particular the network bandwidth, was found to lead to stragglers in writing on Lustre file systems \cite{Xie:2012aa}.
Garbage collection~\cite{Kyong2017,Ousterhout2017}, Java virtual machine (JVM) positioning to cores~\cite{Kyong2017}, delays introduced while the tasks move from the scheduler to execution~\cite{Gittens2016}, disk I/O during shuffling, Java's just-in-time compilation~\cite{Ousterhout2017}, output skew~\cite{Ousterhout2017}, high CPU utilization, disk utilization, unhandled I/O access requests, and network package loss~\cite{Garraghan2016} were also among other external factors that might introduce stragglers.
A wide variety of approaches have been investigated for detecting and mitigating stragglers, including tuning resource allocation and parallelism such as breaking the workload into many small tasks that are dynamically scheduled at runtime~\cite{Rosen2012}, slow Node-Threshold~\cite{Dean2008}, speculative execution~\cite{Dean2008} and cause/resource-aware task management \cite{Ananthanarayanan:2010aa}, sampling or data distribution estimation techniques, SkewTune to avoid data imbalance~\cite{Kwon2012}, dynamic work rebalancing~\cite{Kirpichov2016}, blocked time analysis~\cite{Ousterhout2015}, and intelligent scheduling~\cite{AWE-WQ2014}. 

In the present study, we analyzed MD trajectories in parallel with MPI and Python and observed large variations in the completion time of individual MPI ranks.
These variations bore some similarity to the straggler tasks observed in MapReduce frameworks so we approached analyzing and eliminating them in a similar fashion by systematically looking at different components of the problem, including read I/O from the shared Lustre file system and MPI communication.
Even though we specifically worked in with the \package{MDAnalysis} package, all these principles and techniques are potentially applicable to MPI-parallelized data analysis in other Python-based libraries.

\subsection{Other Packages with Parallel Analysis Capabilities}
\label{sec:otherparallel}

Different approaches to parallelizing the analysis of MD trajectories have been proposed.
HiMach~\cite{himach-2008} introduces scalable and flexible parallel Python framework to deal with massive MD trajectories, by combining and extending Google's MapReduce and the VMD analysis tool~\cite{Hum96}. 
HiMach's runtime is responsible for parallelizing and distributing Map and Reduce classes to assigned cores.
HiMach uses parallel I/O for file access during map tasks and \texttt{MPI\_Allgather} in the reduction process. 
HiMach, however, does not discuss parallel analysis of analysis types that cannot be implemented via MapReduce.
Furthermore, HiMach is not available under an open source license, which makes it difficult to integrate community contributions and add new state-of-the-art methods.

Wu et. al.~\cite{Wu_et.al} present a scalable parallel framework for distributed-memory MD simulation data analysis.
This work consists of an interface that allows a user to write analysis programs sequentially, and the machinery that ensures these programs execute in parallel automatically. 
Parallelization is performed over domains in the simulation system via domain decomposition and the introduction of ghost atoms to include appropriate nearest neighbor interactions.
The HDF5 file format is used for parallel reading and writing.
This work focuses on applications in the materials science and does not consider parallelization over trajectory blocks.

Zazen~\cite{Zazen} is a novel task-assignment protocol to overcome the I/O bottleneck for many I/O bound tasks. This protocol caches a copy of simulation output files on the local disks of the compute nodes of a cluster, and uses co-located data access with computation. 
Zazen is implemented in a parallel disk cache system and avoids the overhead associated with querying metadata servers by reading data in parallel from local disks.
This approach has also been used to improve the performance of HiMach~\cite{himach-2008}.
It, however, advocates a specific architecture where a parallel supercomputer, which runs the simulations, immediately pushes the trajectory data to a local analysis cluster where trajectory fragments are cached on node-local disks.
In the absence of such a specific  workflow, one would need to stage the trajectory across nodes, and the time for data distribution is likely to reduce any gains from the parallel analysis.

VMD~\cite{Hum96, VMD2013} provides molecular visualization and analysis tool through algorithmic and memory efficiency improvements, vectorization of key CPU algorithms, GPU analysis and visualization algorithms, and good parallel I/O performance on supercomputers. It is one of the most advanced programs for the visualization and analysis of MD simulations. It is, however, a large monolithic program, that is primarily driven through its built-in Tcl interface (or less frequently, through its Python interface) and thus is less well suited as a library that allows the rapid development of new algorithms or integration into workflows.

CPPTRAJ~\cite{cpptraj-2013} offers multiple levels of parallelization (MPI and OpenMP) in a monolithic C++ implementation.
It can process single, multiple, and ensembles of trajectories in parallel without changes to input scripts \cite{Roe:2018aa}. 
A Python API exists in the form of the pytraj package (\url{https://github.com/Amber-MD/pytraj}), which has its own implementation of parallelization based on Python's multiprocessing or MPI (via \package{mpi4py} \cite{Dalcin:2011aa, Dalcin:2005aa}).

pyPcazip~\cite{pyPcazip} is a suite of software tools written in Python for compression and analysis of MD simulation data, in particular ensembles of trajectories. 
pyPcazip is MPI parallelized and is specific to PCA-based investigations of MD trajectories and supports a wide variety of trajectory file formats (based on the capabilities of the underlying MDTraj package~\cite{McGibbon:2015aa}).
pyPcazip can take one or many input MD trajectory files and convert them into a highly compressed, HDF5-based pcz format with insignificant loss of information.
However, the package does not support general purpose analysis.

\textit{In situ} analysis is an approach to execute analysis simultaneously with the running MD simulation so that I/O bottlenecks are  mitigated~\cite{Malakar-etal, Johnston:2017aa}.
\citet{Malakar-etal} studied the scalability challenges of time and space shared modes of analyzing large-scale MD simulations through a topology-aware mapping for simulation and analysis using the LAMMPS code.
Similarly, Taufer and colleagues \cite{Johnston:2017aa} presented their own framework for \textit{in situ} analysis, which is based on fast on-the-fly calculation of metadata that characterizes protein substructures via maximum eigenvalues of distance matrices.
These metadata are used to index trajectory frames and enable targeted analysis of trajectory subsets.
Both studies provide important ideas and approaches towards moving towards online-analysis in conjunction with a running simulation but are limited in generality.

All of the above frameworks provide tools for parallel analysis of MD trajectories. 
Although straggler tasks are a common challenge arising in parallel analysis and are well-known in the data analysis community (see Section \ref{sec:stragglers}), there is, to our knowledge, little discussion about this problem in the biomolecular simulation community.
Our own experience with a MapReduce approach in \package{MDAnalysis}~\cite{Khoshlessan:2017ab, Fan:2019aa} suggested that stragglers might be a somewhat under-appreciated problem.
Therefore, in the present work we wanted to better understand requirements for efficient parallel analysis of MD trajectories in \package{MDAnalysis}, but to also provide more general guidance that could benefit developments in other libraries.


\section{Algorithms and Software Packages}
\label{sec:packages}

For our investigation of parallel trajectory analysis we focus on using MPI as the standard approach to parallelization in HPC.
We employ the Python language, which is widely used in the scientific community because it facilitates rapid development of small scripts and code prototypes as well as development of large applications and highly portable and reusable modules and libraries.
We use the \package{MDAnalysis} library to calculate a ``RMSD time series'' (explained in section \ref{sec:mda}) as a representative use case.
Further details on the software packages are provided in sections \ref{sec:methods-mpi4py}--\ref{sec:methods-hdf5}.

\subsection{RMSD Calculation with MDAnalysis}
\label{sec:mda}

Simulation data exist in trajectories in the form of time series of atom positions and sometimes velocities.
Trajectories come in a plethora of different and idiosyncratic file formats. 
\package{MDAnalysis} \cite{Gowers:2016aa, Michaud-Agrawal:2011fu} is a widely used open source library to analyze trajectory files with an object oriented interface. 
The library is written in Python, with time critical code in C/C++/Cython. 
\package{MDAnalysis} supports most file formats of simulation packages including CHARMM \cite{Brooks:2009pt}, Gromacs \cite{Abraham:2015aa}, Amber \cite{Case:2005uq}, and NAMD \cite{Phillips:2005ek} and the Protein Data Bank \cite{Burley:2018aa} format.
At its core, it reads trajectory data in different formats and makes them available through a uniform API; specifically, coordinates are represented as standard NumPy arrays \cite{Van-Der-Walt:2011aa}.

\begin{figure}[!htb]
  \centering
  \includegraphics[width=7cm]{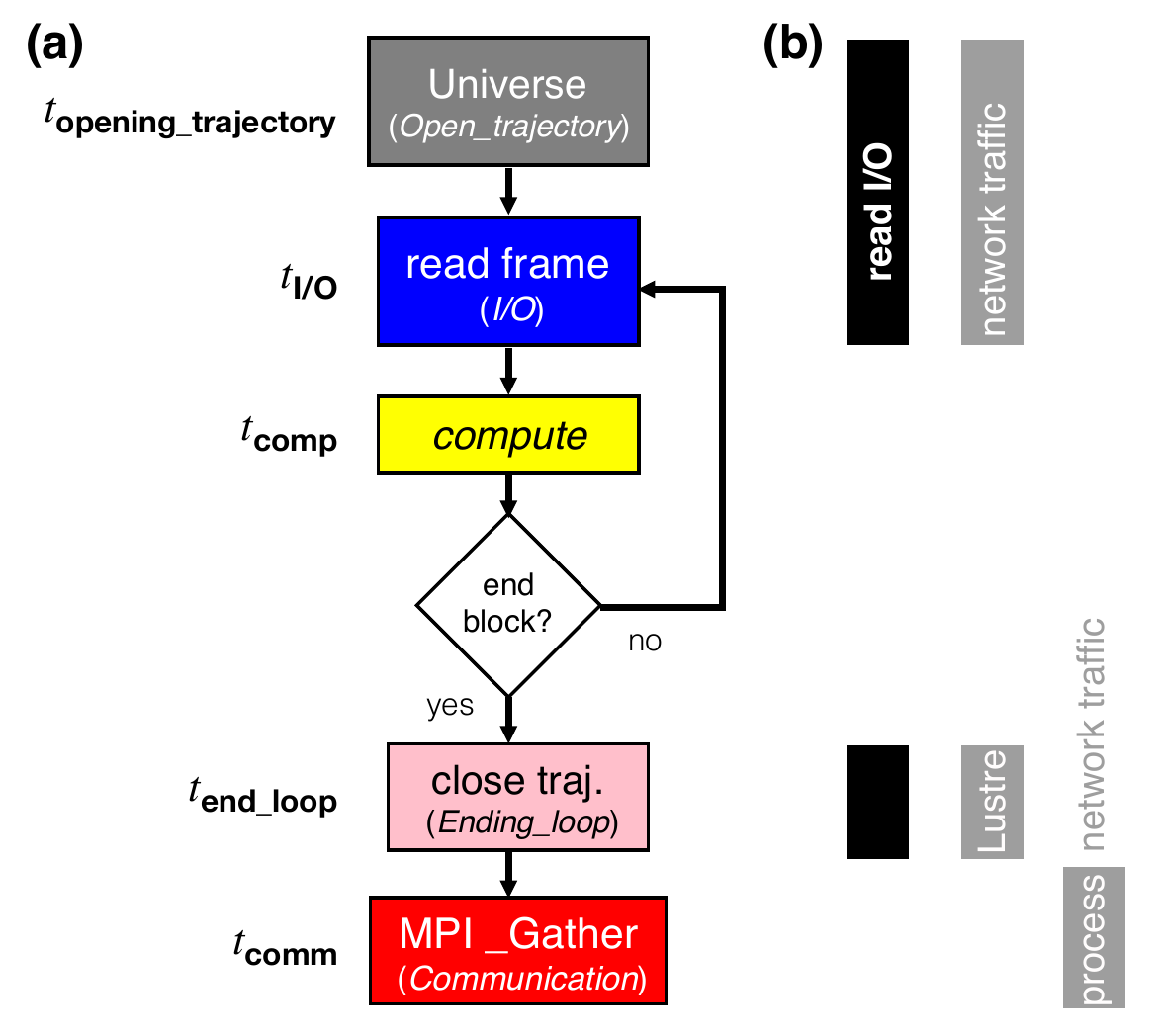}
  \caption{Flow chart of the MPI-parallelized RMSD algorithm, Algorithm~\ref{alg:RMSD}.
    \textbf{(a)} Each MPI process performs the same steps but reads trajectory frames from different blocks of the trajectory.
    The color scheme and labels in italics correspond to the colors and labels for measured timing quantities in the following graphs (e.g., Figs.~\protect\ref{fig:ScalingComputeIO} and \protect\ref{fig:MPIranks}).
    The names of the corresponding timing quantities from Table \protect\ref{tab:notation} are listed next to each step.
    \textbf{(b)} Steps that access the shared Lustre file system with read I/O are included in the black bars; steps that communicate via the shared InfiniBand network are included in the gray ``network traffic'' bars.
    The Lustre file system is accessed through the network and hence all I/O steps also use the network (gray ``Lustre network traffic'' bars).
    Processes only communicate over the network (gray ``process network traffic'' bar) when results are communicated back to process rank 0 in the \emph{Communication} step.
  }
  \label{fig:flowchart}
\end{figure}

As a test case that is representative of a common task in the analysis of biomolecular simulation trajectories we calculated the time series of the minimal structural root mean square distance  (\textbf{RMSD}) after rigid body superposition \cite{Lea96, Mura:2014kx}.
The RMSD is used to show the rigidity of protein domains and more generally characterizes structural changes.
It is calculated as a function of time $t$ as
\begin{equation}
  \label{eq:rmsd}
  \text{RMSD}(t) = \min_{\mathsf{R}, \mathbf{t}} %
  \sqrt{\frac{1}{N} \sum_{i=1}^{N} \left[ %
      (\mathsf{R}\cdot\mathbf{x}_{i}(t) + \mathbf{t}) - \mathbf{x}_{i}^{\text{ref}} \right]^{2}}
\end{equation}
where $\mathbf{x}_{i}(t)$ is the position of atom $i$ at time $t$, $\mathbf{x}_{i}^{\text{ref}}$ is its position in a reference structure and the distance between these two is minimized by finding the optimum $3\times3$ rotation matrix $\mathsf{R}$ and translation vector $\mathbf{t}$. 
The optimum rigid body superposition was calculated with the QCPROT algorithm~\cite{Liu:2010kx,Theobald:2005vn} (implemented in Cython and available through the \texttt{MDAnalysis.analysis.rms} module \cite{Gowers:2016aa}).

The RMSD trajectory analysis was parallelized with a simple \emph{split-apply-combine} approach \cite{Wickham:2011aa} as outlined in the flow chart in Figure~\ref{fig:flowchart}, with further details available in Algorithm~\ref{alg:RMSD}.
Each MPI process loads the core MDAnalysis data structure (called the \texttt{Universe}), which includes loading a shared ``topology'' file with the simulation system information and opening the shared trajectory file.
Each process operates on a different block of frames (\emph{split}) and iterates through them by reading the coordinates of a single frame into memory and performing the RMSD computation with them (\emph{apply}).
Once all frames in the block are processed, the trajectory file is closed and results are communicated to MPI rank 0 using \texttt{MPI\_Gather()} (\emph{combine}).

The RMSD was determined for a subset of protein atoms, the $N=146$  C$_{\alpha}$ atoms in the so-called ``core'' domain of our test system, the protein adenylate kinase \cite{Seyler:2014il} (see section \ref{sec:data} for further details).
The time complexity for the RMSD Algorithm~\ref{alg:RMSD} is $\mathcal{O}(T \times N^{2})$ where $T$ is the number of frames in the trajectory and $N$ the number of particles included in the RMSD calculation \cite{Liu:2010kx}.

\begin{algorithm}[ht]
	\scriptsize
	\caption{MPI-parallel Multi-frame RMSD Algorithm}
	\label{alg:RMSD}
	\hspace*{\algorithmicindent} \textbf{Input:} \emph{size}: total number of frames \\
	\hspace*{\algorithmicindent} \emph{ref}: mobile group in the initial frame which will be considered as reference \\
	\hspace*{\algorithmicindent} \emph{start \& stop}: starting and stopping frame index\\
	\hspace*{\algorithmicindent} \emph{topology \& trajectory}: files to read the data structure from \\
	\hspace*{\algorithmicindent} \textbf{Output:} calculated RMSD arrays
	\begin{algorithmic}[1]
		\Function{block\_rmsd}{$topology$, $trajectory$, $ref$, $index$, $start$, $stop$}                       
		\State $u \gets$ Universe$(topology$, $trajectory)$\Comment{$u$ holds all the information of the system}
		\State $g \gets$ $u.$atoms[$index$]  \Comment{select AtomGroup $g$}
		\ForAll{$iframe$ in $u.$trajectory[$start:stop$]} \Comment iterate through frames, enumerated by $iframe$
		\State $results[iframe] \gets \text{RMSD}(g, ref)$ \Comment{Eq.~\protect\ref{eq:rmsd}}
		\EndFor
		\State \Return results
		\EndFunction
		\\        
		\State MPI Init
		\State $rank \gets rank\_ID$
		\State $index \gets$ indices of the mobile AtomGroup
		\State $xref0 \gets$ reference AtomGroup position
		\State $out \gets$ \Call{block\_rmsd}{$topology, trajectory, xref0, index, start, stop$}
                \\
		\State \Call{Gather}{$out, RMSD\_data, rank\_ID=0$}
		\State MPI Finalize
	\end{algorithmic}
\end{algorithm}

\subsection{MPI for Python (\package{mpi4py})}
\label{sec:methods-mpi4py}

MPI for Python (\package{mpi4py}) is a Python wrapper for the Message Passing Interface (MPI) standard and allows any Python program to employ multiple processors \cite{Dalcin:2011aa, Dalcin:2005aa}.
Performance degradation due to using \package{mpi4py} is not prohibitive \cite{Dalcin:2011aa, Dalcin:2005aa} and the overhead is far smaller than the overhead associated with the use of interpreted versus compiled languages \cite{GAiN}.
Overheads in \package{mpi4py} are small compared to C code if efficient raw memory buffers are used for communication \cite{Dalcin:2011aa}, as used in the present study.

\subsection{MPI and Parallel HDF5}
\label{sec:methods-hdf5}

HDF5 is a structured self-describing hierarchical data format which is a common mechanism for storing large quantities of numerical data in Python (\url{http://www.hdfgroup.org/HDF5}, \cite{pythonhdf5}).
Parallel HDF5 (\package{PHDF5}) typically sits on top of a MPI-IO layer (parallel I/O with MPI) and can use MPI-IO optimizations. 
In \package{PHDF5}, all file access is coordinated by the MPI library; otherwise, multiple processes would compete over accessing the same file on disk. 
MPI-based applications launch multiple parallel instances of the Python interpreter that communicate with each other via the MPI library. 
Implementation is straightforward as long as the user supplies a MPI communicator and takes into account some constraints required for data consistency \cite{pythonhdf5}.
\package{HDF5} itself handles nearly all the details involved with coordinating file access when the shared file is opened through the \emph{mpio} driver.

MPI has two flavors of operation: collective (all processes have to participate in the same order) and independent (processes can perform the operation in any order or not at all) \cite{pythonhdf5}.
With \package{PHDF5}, modifications to file metadata must be performed collectively and although all processes perform the same task, they do not need to be synchronized \cite{pythonhdf5}. 
Other tasks and any type of data operations can be performed independently by processes.
In the present study, we use independent operations.

\begin{algorithm}[ht]
	\scriptsize
	\caption{MPI-parallel Multi-frame RMSD Algorithm with HDF5 files.}
	\label{alg:RMSDhdf5}
	\hspace*{\algorithmicindent} \textbf{Input:} \emph{size}: total number of frames \\
	\hspace*{\algorithmicindent} \emph{ref}: mobile group in the initial frame which will be considered as reference \\
	\hspace*{\algorithmicindent} \emph{start \& stop}: starting and stopping frame index\\
	\hspace*{\algorithmicindent} \emph{dataset}: HDF5 Dataset of the coordinates in the trajectory \\
	\hspace*{\algorithmicindent} \textbf{Output:} calculated RMSD arrays
	\begin{algorithmic}[1]
		\Function{$block\_rmsd$}{$dataset$, $ref$, $start$, $stop$}                       
		\For{ $start \leq iframe < stop$ }
		\State $results[iframe] \gets \Call{RMSD}{dataset[iframe], ref}$ \Comment{Eq.~\protect\ref{eq:rmsd}}
		\EndFor
		\State \Return results
		\EndFunction
		\\        
		\State MPI Init
		\State $rank \gets rank\_ID$
		\State $xref0 \gets$ reference atom group position
                \State $f \gets$ open HDF5 file for parallel MPI-IO reading \Comment{use \emph{mpio} driver}
                \State $dataset \gets$ get dataset 'pos' from $f$
		\State $out \gets \Call{block\_rmsd}{dataset, xref0, start, stop}$
                \\
		\State $\Call{Gather}{out, RMSD\_data, rank\_ID=0}$
		\State MPI Finalize
	\end{algorithmic}
\end{algorithm}

MDAnalysis does not currently have a reader for HDF5 files or HDF5-based trajectories.
In order to get a sense of the performance that is possible with a HDF5 trajectory we replaced the MDAnalysis trajectory reading in Algorithm~\ref{alg:RMSD} with directly accessing a HDF5 Dataset in a HDF5 file that was opened with the \emph{mpio} parallel driver, as shown in Algorithm~\ref{alg:RMSDhdf5}.
The HDF5 file was generated from the original XTC trajectory file as described in Section~\ref{sec:data}.

\section{Benchmark Environment}
\label{sec:system}
Our benchmark environment consisted of three different XSEDE \cite{xsede} HPC resources (described in section~\ref{sec:hpcresources}), the software stack used (section~\ref{sec:software}), which had to be compiled for each resource, and the common test data set (section~\ref{sec:data}).

\subsection{HPC Resources}
\label{sec:hpcresources}

The computational experiments were executed on standard compute nodes of three XSEDE \cite{xsede} supercomputers, \emph{SDSC Comet}, \emph{PSC Bridges}, and \emph{LSU SuperMIC} (Table~\ref{tab:sys-config}).
\emph{SDSC Comet} is a 2 PFlop/s cluster with 2,020 compute nodes in total. It is optimized for running a large number of medium-size calculations (up to 1,024 cores) to support the most prevalent type of calculation on XSEDE resources.
\emph{PSC Bridges} is a 1.35 PFlop/s cluster with different types of computational nodes, including 16 GPU nodes, 8 large memory and 2 extreme memory nodes, and 752 regular nodes.
It was designed to flexibly support both traditional (medium scale calculations) and non-traditional (data analytics) HPC uses.
\emph{LSU SuperMIC} offers 360 standard compute nodes with a peak performance of 557 TFlop/s.
The parallel file system on all three machines is Lustre (\url{http://lustre.org/}) and is shared between the nodes of each cluster.

\begin{table}[ht!]
        \centering
  	\caption[Configuration of HPC resources]
	{Configuration of the HPC resources that were benchmarked. Only a subset of the total available nodes were used. IB: InfiniBand; OPA: Omni-Path Architecture.}
	\label{tab:sys-config}
	\begin{adjustbox}{max width=\textwidth}
		\begin{tabular}{c c c c c c c c c}
			\toprule
			\bfseries\thead{Name} & \bfseries\thead{Nodes} & \makecell{\bfseries\thead{Number \\of Nodes}} & \bfseries\thead{CPUs} &  \bfseries\thead{RAM} & \bfseries\thead{Network Topology} & \makecell{\bfseries\thead{Scheduler and  \\ Resource Manager}} & \makecell{\bfseries\thead{parallel\\file system}}\\
			\midrule
			\bfseries \emph{SDSC Comet} & Compute & 6400 & \makecell{2 Intel Xeon (E5-2680v3) \\ 12 cores/CPU, 2.5 GHz} &128 GB DDR4 DRAM & 56 Gbps IB & SLURM & Lustre\\
			\bfseries \emph{PSC Bridges} & RSM & 752 & \makecell{2 Intel Haswell (E5-2695 v3)  \\14 cores/CPU, 2.3 GHz} & 128 GB, DDR4-2133MHz & 12.37 Gbps OPA & SLURM & Lustre\\
			\bfseries \emph{LSU SuperMIC} & Standard & 360 & \makecell{2 Intel Ivy Bridge (E5-2680) \\10 cores/CPU, 2.8 GHz} & 64 GB, DDR3-1866MHz  & 56 Gbps IB & PBS & Lustre\\
			\bottomrule
		\end{tabular}
	\end{adjustbox}
\end{table}

\subsection{Software}
\label{sec:software}

Table~\ref{tab:version} lists the tools and libraries that were required for our computational experiments.  Many domain specific packages are not available in the standard software installation on supercomputers.
We therefore had to compile them, which in some cases required substantial effort due to non-standard building and installation procedures or lack of good documentation.
Because this is a common problem that hinders reproducibility we provide detailed version information, notes on the installation process, as well as comments on the ease of installation and the quality of the documentation in Table~\ref{tab:version}.
For the MPI implementation we used Open MPI release 1.10.7  (\url{https://www.open-mpi.org/}) consistently everywhere.
We used the \package{h5py} package for HDF5, which enables parallel HDF5 from Python because its dependencies, the HDF5 library itself and \package{mpi4py}, were both built against Open MPI.
We used Python 2.7 because it provided maximum compatibility between packages at the time when the project was started.
In principle the complete Python-dependent software stack could also be set up with Python 3.5 or higher, which is recommended because Python 2 reached end of life in January 2020.
Detailed instructions to create the computing environments together with the benchmarking code can be found in the GitHub repository as described in Section~\ref{sec:sharing}.
Carefully setting up the same software stack on the three different supercomputers allowed us to clearly demonstrate the reproducibility of our results and showed that our findings were not dependent on machine specifics.

\begin{table}[ht!]
\centering
\caption[Version of the packages used in the present study]%
{Detailed comparison on the dependencies and installation of different software packages used in the present study. Software was built from source or obtained via a package manager and installed on the multi-user HPC systems in Table~\protect\ref{tab:sys-config}. Evaluation of ease of installation and documentation uses a subjective scale with ``++'' (excellent), ``+'' (good), ``0'' (average), and ``$-$'' (difficult/lacking) and reflects the experience of a typical domain scientist at the graduate/post-graduate level in a discipline such as computational biophysics or chemistry.}
\label{tab:version}  
\begin{adjustbox}{max width=\textwidth}
\begin{tabular}{l c l c c l l}
  \toprule
            \bfseries\thead{Package} & \bfseries\thead{Version} & \bfseries\thead{Description} & \bfseries\thead{Ease of Installation} & \bfseries\thead{Documentation} & \bfseries\thead{Installation} & \bfseries\thead{Dependencies}\\
  \midrule
   \bfseries GCC & 4.9.4 & GNU Compiler Collection & 0 & ++ & \makecell[l]{via configuration \\files, environment \\or command line options, \\ minimal configuration \\ is required} &--\\
   \midrule
   \bfseries Open MPI & 1.10.7 & MPI Implementation & 0 & ++ & \makecell[l]{via configuration \\ files, environment \\or command line options, \\ minimal configuration \\ is required} &--\\
   \midrule
   \bfseries Python & 2.7.13 & Python language & + & ++ & Conda Installation & --\\
   \midrule
   \bfseries mpi4py & 3.0.0 & MPI for Python & + & ++ & Conda Installation &\makecell[l]{Python 2.7 or above, \\ MPI 1.x/2.x/3.x  \\ implementation like \\ Open MPI \\built with shared/dynamic \\libraries, Cython}\\
   \midrule
   \bfseries PHDF5 & 1.10.1 & Parallel HDF5 & $-$ & ++ & \makecell[l]{via configuration files,\\ environment \\or command line options, \\ several optional configuration\\ settings available} &\makecell[l]{MPI 1.x/2.x/3.x  \\ implementation like \\ Open MPI  \\GNU, MPIF90,  \\MPICC, MPICXX}\\
   \midrule
   \bfseries h5py &  2.7.1 & Pythonic wrapper around the HDF5 & + & ++ & Conda Installation & \makecell[l]{Python 2.7, or above,\\ PHDF5, Cython}\\    
   \midrule
   \bfseries MDAnalysis & 0.17.0 & \makecell[l]{Python library to analyze \\trajectories from MD simulations} & + & ++ & Conda Installation & \makecell[l]{Python $\ge$2.7, Cython,\\ GNU, Numpy}\\
  \bottomrule
\end{tabular}
\end{adjustbox}
\end{table}

\subsection{Data Set}
\label{sec:data}

The test system contained the protein adenylate kinase with 214 amino acid residues and 3341 atoms in total~\cite{Seyler:2014il} and the topology information (atoms types and bonds) was stored in a file in CHARMM PSF \cite{Brooks:2009pt} format.
The test trajectory was created by concatenating 600 copies of a MD trajectory with 4,187 time frames \cite{Seyler:2017aa} (saved every 240~ps for a total simulated time of 1.004~$\mu\text{s}$) in CHARMM DCD \cite{Brooks:2009pt} format and converting to Gromacs \cite{Abraham:2015aa} XTC format trajectory, as described for the ``600x'' trajectory in~\citet{Khoshlessan:2017ab}.
The trajectory had a file size of about 30 GB and contained 2,512,200 frames (corresponding to 602.4~$\mu\text{s}$ simulated time).
The file size was relatively small because water molecules that were also part of the original MD simulations were stripped to reduce the original file size by a factor of about 10; such preprocessing is a common approach if one is only interested in the protein behavior.
Thus, the trajectory represents a small to medium system size in the number of atoms and coordinates that have to be loaded into memory for each time frame.
The XTC format is a format with lossy compression \cite{Lindahl01, Spangberg:2011zr}, which also contributed to the compact file size.
XTC trades lower I/O demands for higher CPU demands during decompression and therefore performed well in our previous study~\cite{Khoshlessan:2017ab}.
In order to assess the performance of reading from an HDF5 file in parallel (see Section~\ref{sec:methods-hdf5}) we generated a trajectory-like HDF5 file with the data required to perform the RMSD calculation.
This HDF5 file was created from the XTC file by sub-selecting the atoms for which the RMSD was calculated as detailed in Section~\ref{sec:mda}; a Python script to perform the trajectory conversion can be found in the GitHub repository (see Section~\ref{sec:sharing}).
The coordinates were stored as a two-dimensional $T \times 3N$ array where the first dimension contained $T=2,512,200$ frames and the second dimension the $3N = 438$ Cartesian coordinates.
Although 2,512,200 frames represents a long simulation for current standards, such trajectories will become increasingly common due to the use of special hardware~\cite{Shaw:2009ly, Shaw:2014aa} and GPU-acceleration~\cite{Salomon-Ferrer:2013cr, Glaser:2015ys, Abraham:2015aa}.


\section{Methods}
\label{sec:methods}

In the following we define the quantities and approach used for our performance measurements, with a full summary of all definitions in Table~\ref{tab:notation}.
We evaluated MPI performance of the parallel RMSD time series algorithm~\ref{alg:RMSD} and its variation (algorithm \ref{alg:RMSDhdf5}) by timing the total time to solution as well as the execution time for different parts of the code for individual MPI ranks with the help of the Python \texttt{time.time()} function.

\begin{table}[!htb]
  \centering
  \caption[Summary of measured timing quantities.]  {Summary of measured timing quantities.  Timings are collected for the specified line numbers in the code, labeled as $t_{\text{L$n$}}$ where $\text{L$n$}$ refers to the line number in the corresponding algorithm (columns Algorithm~\ref{alg:RMSD} and \ref{alg:RMSDhdf5}), or are calculated in the same way for both algorithms from the specific quantities.  Variables in the top part of the table refer to measurements of an individual MPI rank.  Variables in the bottom part are aggregates such as averages over all ranks or the total time to solution.}
  \label{tab:notation}
\begin{threeparttable}
  \begin{tabular}{cccp{0.35\textwidth}}
    \toprule
    \bfseries\thead{Quantity} & \multicolumn{2}{c}{\bfseries\thead{Definition}} & \bfseries\thead{Description}\\
                              & Algorithm~\ref{alg:RMSD} & Algorithm~\ref{alg:RMSDhdf5} & \\
    \midrule  
    $t_{\text{opening\_trajectory}}$ &  $t_{\text{L2}}+t_{\text{L3}}$ & ---\textsuperscript{a} & file opening and data structure initialization\\
    $\tIO^{\text{frame}}$   & $t_{\text{L4}}$ & $t_{\text{L2}}$ & data reading per frame\\  
    $\tcomp^{\text{frame}}$ & $t_{\text{L5}}$ & $t_{\text{L3}}$ & compute per frame\\  
    $t_{\text{all\_frame}}$ & $t_{\text{L4}}+t_{\text{L5}}+t_{\text{L6}}$ & $t_{\text{L2}}+t_{\text{L3}}+t_{\text{L4}}$  & time to analyze one frame\\
    $t_{\text{RMSD}}$ &  $t_{\text{L1}} + ...+ t_{\text{L8}}$ & $t_{\text{L1}} + ...+ t_{\text{L6}}$ & time to execute the body of the \texttt{block\_rmsd()} function\\
    $t_{\text{end\_loop}}$ & $t_{\text{L6}} $  & $t_{\text{L4}} $ & closing of the trajectory at the end of the loop\\
    $\tcomm$  & $t_{\text{L16}}$ &  $t_{\text{L15}}$ & data communication with \texttt{MPI\_Gather()}\\
    \cmidrule(l){2-3}
    $N_{\text{frames}}^{\text{total}}$ & & & total number of trajectory frames\\
    $N$ & & & total number of MPI ranks (processes), equals the number of trajectory blocks\\
    $N_{\text{b}}$ & \multicolumn{2}{c}{$N_{\text{frames}}^{\text{total}}/N$} & number of frames per block\\
    $\tcomp$ & \multicolumn{2}{c}{$\sum_{\text{frame}=1}^{N_{\text{b}}}\tcomp^{\text{frame}}$} & total compute time for a rank (process)\\
    $\tIO$ & \multicolumn{2}{c}{$\sum_{\text{frame}=1}^{N_{\text{b}}}\tIO^{\text{frame}}$} & total read I/O time for a rank (process)\\  
    $t_{\text{Overhead1}}$ & \multicolumn{2}{c}{$t_{\text{all\_frame}}-t_{\text{I/O}}-t_{\text{comp}}-t_{\text{end\_loop}}$}  & time inside \texttt{block\_rmsd()} that was not measured explicitly\\
    $t_{\text{Overhead2}}$ & \multicolumn{2}{c}{$t_{\text{RMSD}}-t_{\text{all\_frame}}-t_{\text{opening\_trajectory}}$} & overhead for the \texttt{block\_rmsd()} function call\\
    $t_{N}$ & \multicolumn{2}{c}{$t_{\text{RMSD}}+\tcomm$} & total time to completion for a rank (process)\\
    \toprule
    $\overline{\tcomp}$ & \multicolumn{2}{c}{$\frac{1}{N}\sum_{\text{rank}=1}^{N} \tcomp$} & average compute time over all ranks\\
    $\overline{\tIO}$ & \multicolumn{2}{c}{$\frac{1}{N}\sum_{\text{rank}=1}^{N} \tIO$} & average read I/O time over all ranks\\
    $\overline{\tcomm}$ & \multicolumn{2}{c}{$\frac{1}{N}\sum_{\text{rank}=1}^{N} \tcomm$} & average communication time over all ranks\\
    $t_{\text{total}}$ & \multicolumn{2}{c}{$\max t_{N}$} & total time to solution\\
    \bottomrule
  \end{tabular}
    \begin{tablenotes}[para,flushleft]
    \item [a] Algorithm~\ref{alg:RMSDhdf5} does not need to open a trajectory inside the \texttt{block\_rmsd()} function and hence $t_{\text{opening\_trajectory}}$ only measures time to allocate empty arrays, which is not explicitly shown in Algorithm~\ref{alg:RMSDhdf5}.
      \end{tablenotes}
\end{threeparttable}
\end{table}

\subsection{Timing Observables}

We abbreviate the timings in the following as variables $t_{\text{L$n$}}$ where $\text{L$n$}$ refers to the line number in algorithm~\ref{alg:RMSD} (or algorithm \ref{alg:RMSDhdf5}, see Table~\ref{tab:notation}).
In the function \texttt{block\_rmsd()}, we measured the \emph{read I/O time} for ingesting the data of one trajectory frame from the file system into memory, $t_{\text{I/O}}^{\text{frame}} = t_{\text{L4}}$, and the \emph{compute time} per trajectory frame to perform the computation, $\tcomp^{\text{frame}} = t_{\text{L5}}$.
The \emph{total read I/O time for a MPI rank},  $\tIO = \sum_{\text{frame}=1}^{N_{\text{b}}} \tIO^{\text{frame}}$, is the sum over all I/O times for all the $N_{\text{frames}}$ frames assigned to the rank; similarly, the \emph{total compute time for a MPI rank} is $\tcomp = \sum_{\text{frame}=1}^{N_{\text{b}}} \tcomp^{\text{frame}}$. 
The time delay between the end of the last iteration and exiting the \texttt{for} loop is $t_{\text{end\_loop}} = t_{\text{L6}}$.
The time $t_{\text{opening\_trajectory}} = t_{\text{L2}}+t_{\text{L3}}$ measures the problem setup, which includes data structure initialization and opening of topology and trajectory files.
The \emph{communication time}, $\tcomm = t_{\text{L16}}$, is the time to gather all data from all processor ranks to rank zero.
The total time (for all frames) spent in \texttt{block\_rmsd()} is $t_{\text{RMSD}} = \sum_{i=1}^{8}t_{\text{L$i$}}$. 
There are parts of the code in \texttt{block\_rmsd()} that are not covered by the detailed timing information of \tcomp and \tIO.
Unaccounted time is considered as \emph{overhead}.
We define $t_{\text{Overhead1}}$ and $t_{\text{Overhead2}}$ as the overheads of the calculations (see Table \ref{tab:notation} for the definitions); both are expected to be negligible, which was the case in all our measurements. 
Finally, the \emph{total time to completion of a single MPI rank}, when utilizing $N$ cores for the execution of the overall experiment, is $t_{N}$, and as a result $t_{\text{RMSD}} + \tcomm \equiv t_{N}$.

\subsection{Performance Parameters}

We measured the \emph{total time to solution} $t_{\text{total}}(N)$ with $N$ MPI processes on $N$ cores, which is effectively
$t_{\text{total}}(N) \approx \max(t_{N})$. 
Strong scaling was quantified by the speed-up
\begin{equation}
  \label{eq:speedup}
  S(N) = \frac{t_{\text{total}}(1)}{t_{\text{total}}(N)},
\end{equation}
relative to performance on a single core ($t_{\text{total}}(1)$), and the efficiency
\begin{equation}
  \label{eq:efficiency}
  E(N) = \frac{S(N)}{N}.
\end{equation}
Averages over ranks were calculated as
\begin{equation}
  \label{eq:avg-tcomp}
  \overline{\tcomp} = \frac{1}{N}
  \sum_{\text{rank}=1}^{N}\tcomp = \frac{1}{N}\sum_{\text{rank}=1}^{N}\sum_{\text{frame}=1}^{N_\text{b}}\tcomp^{\text{frame}},
\end{equation}
\begin{equation}
  \label{eq:avg-tIO}
  \overline{\tIO} = \frac{1}{N}\sum_{\text{rank}=1}^{N}\tIO = \frac{1}{N}\sum_{\text{rank}=1}^{N}\sum_{\text{frame}=1}^{N_{\text{b}}}\tIO^{\text{frame}},
\end{equation}
and
\begin{equation}
  \label{eq:avg-tcomm}
  \overline{\tcomm} = \frac{1}{N}\sum_{\text{rank}=1}^{N}\tcomm.
\end{equation}

Additionally, we introduced two performance parameters that we found to be indicative of the occurrence of stragglers.
We defined the ratio of compute time to read I/O time for the serial code as
\begin{equation}
  \label{eq:Compute-IO}  
  \RcompIO = \frac{\tcomp}{\tIO} = %
  \frac{\tcomp/N_{\text{frames}}^{\text{total}}}{\tIO/N_{\text{frames}}^{\text{total}}}  = %
  \frac{\overline{\tcomp^{\text{frame}}}}{\overline{\tIO^{\text{frame}}}}  
\end{equation}
where the last equality shows that the ratio can also be computed from the average times per frame, $\overline{\tcomp^{\text{frame}}}$ and $\overline{\tIO^{\text{frame}}}$.
\RcompIO was calculated with the serial versions of our algorithms (on a single CPU core) in order to characterize the computational problem in the absence of parallelization.
The ratio of compute to communication time was defined by the ratio of average total compute time to the average total communication time   
\begin{equation}
  \label{eq:Compute-comm}
  \Rcompcomm = \frac{\overline{\tcomp}}{\overline{\tcomm}}.
\end{equation}
Because \tcomm cannot be measured for a serial code, we estimated \Rcompcomm from the rank-averages (Eqs.~\ref{eq:avg-tcomp} and \ref{eq:avg-tcomm}) for a given number of MPI ranks.

\subsection{Data sharing}
\label{sec:sharing}

Documentation and benchmark/trajectory conversion scripts are made available in the repository \url{https://github.com/hpcanalytics/supplement-hpc-py-parallel-mdanalysis} under the GNU General Public License v3.0 (code) and the Creative Commons Attribution-ShareAlike (documentation).
All materials are archived under DOI \href{https://doi.org/10.5281/zenodo.3351616}{10.5281/zenodo.3351616}.
These materials should enable users to recreate the computational environment on the tested XSEDE HPC resources (\emph{SDSC Comet}, \emph{PSC Bridges}, \emph{LSU SuperMIC}), prepare data files, and run the computational experiments.


\section{Computational Experiments}
\label{sec:experiments}

We had previously measured the performance of the MPI-parallelized RMSD analysis task on two different HPC resources (\emph{SDSC Comet} and \emph{TACC Stampede}) and had found that it only scaled well up to a single node due to high variance in the runtime of the MPI ranks, similar to the straggler phenomenon observed in big-data analytics \cite{Khoshlessan:2017ab}.
However, the ultimate cause for this high variance could not be ascertained.
We therefore performed more measurements with more detailed timing information (see section \ref{sec:methods}) on \emph{SDSC Comet} (described in this section) and two other supercomputers (summarized in section \ref{sec:clusters}) in order to better understand the origin of the stragglers and find solutions to overcome them. 

\subsection{RMSD Benchmark}
\label{sec:RMSD}

\begin{figure}[!htb]
  \centering
  \begin{subfigure}{.4\textwidth}
    \includegraphics[width=\linewidth]{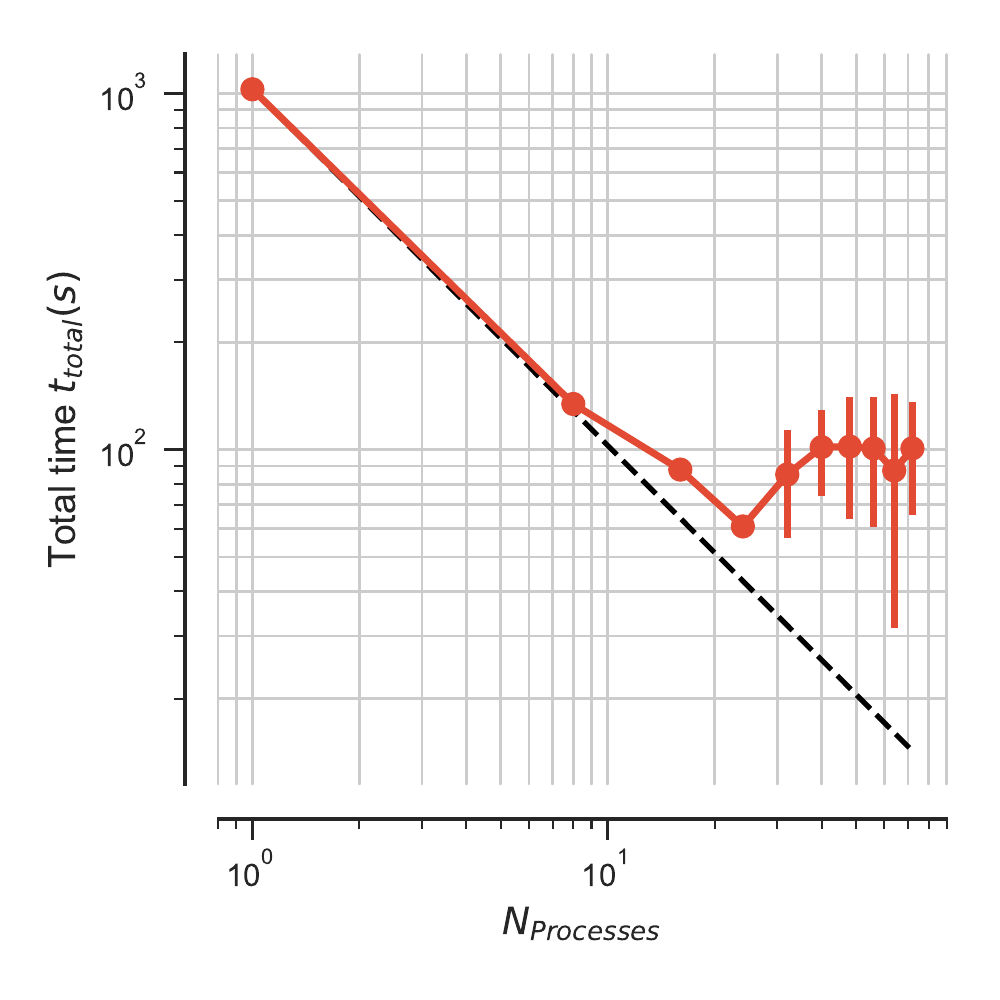}
    \captionsetup{format=hang}
    \caption{Scaling total (five repeats)}
    \label{fig:MPIscaling}
  \end{subfigure}
  \hfill
  \begin{subfigure}{.4\textwidth}
    \includegraphics[width=\linewidth]{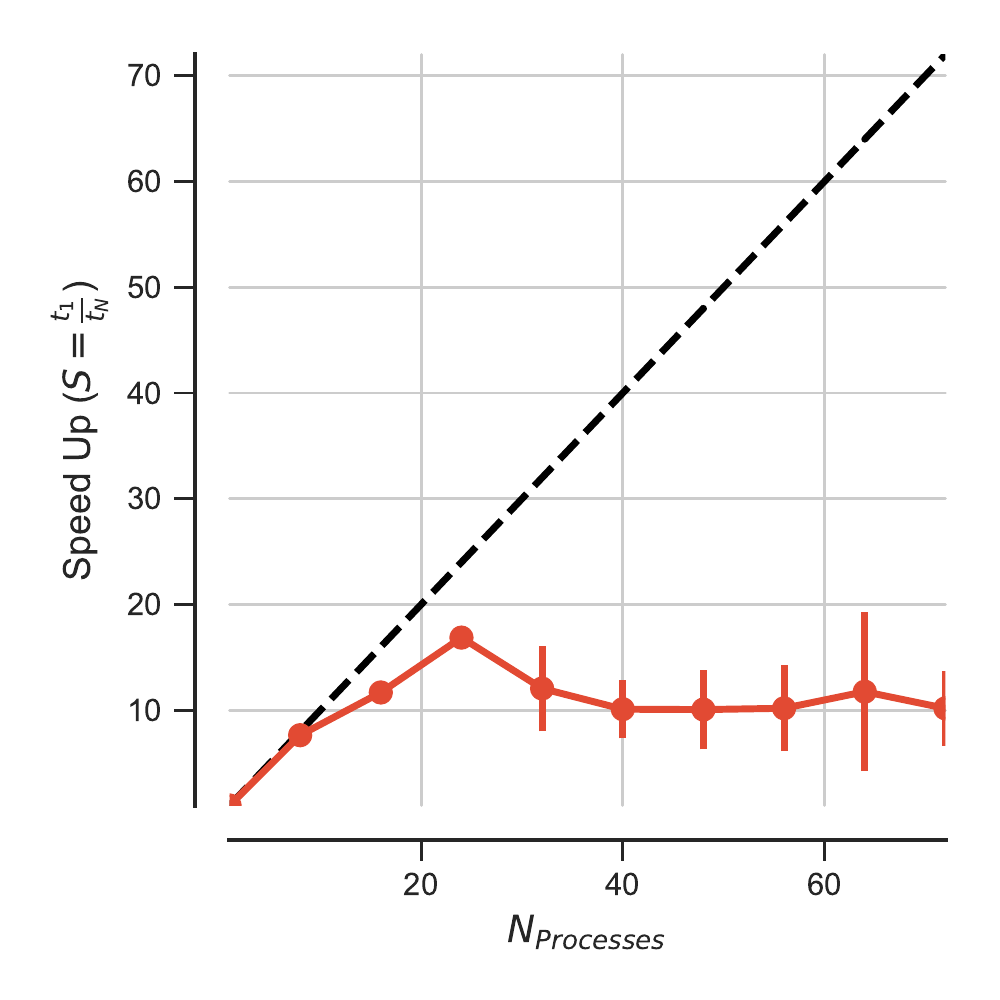}
    \captionsetup{format=hang}
    \caption{Speed-up (five repeats)}
    \label{fig:MPIspeedup}
  \end{subfigure}
  \bigskip
  \begin{subfigure}{.45\textwidth}
    \includegraphics[width=\linewidth]{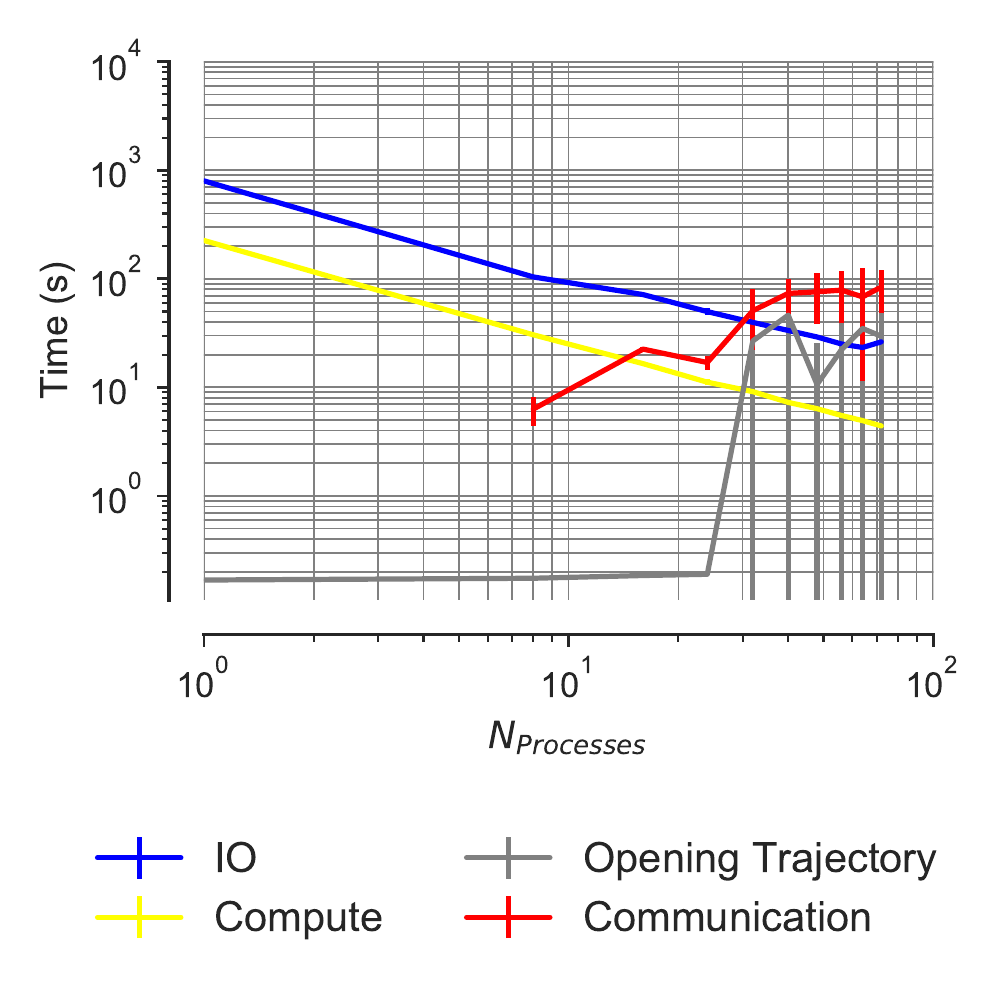}
    \captionsetup{format=hang}
    \caption{Scaling for different components (five repeats)}
    \label{fig:ScalingComputeIO}
  \end{subfigure}
  \hfill
  \begin{subfigure} {.5\textwidth}
    \includegraphics[width=\linewidth]{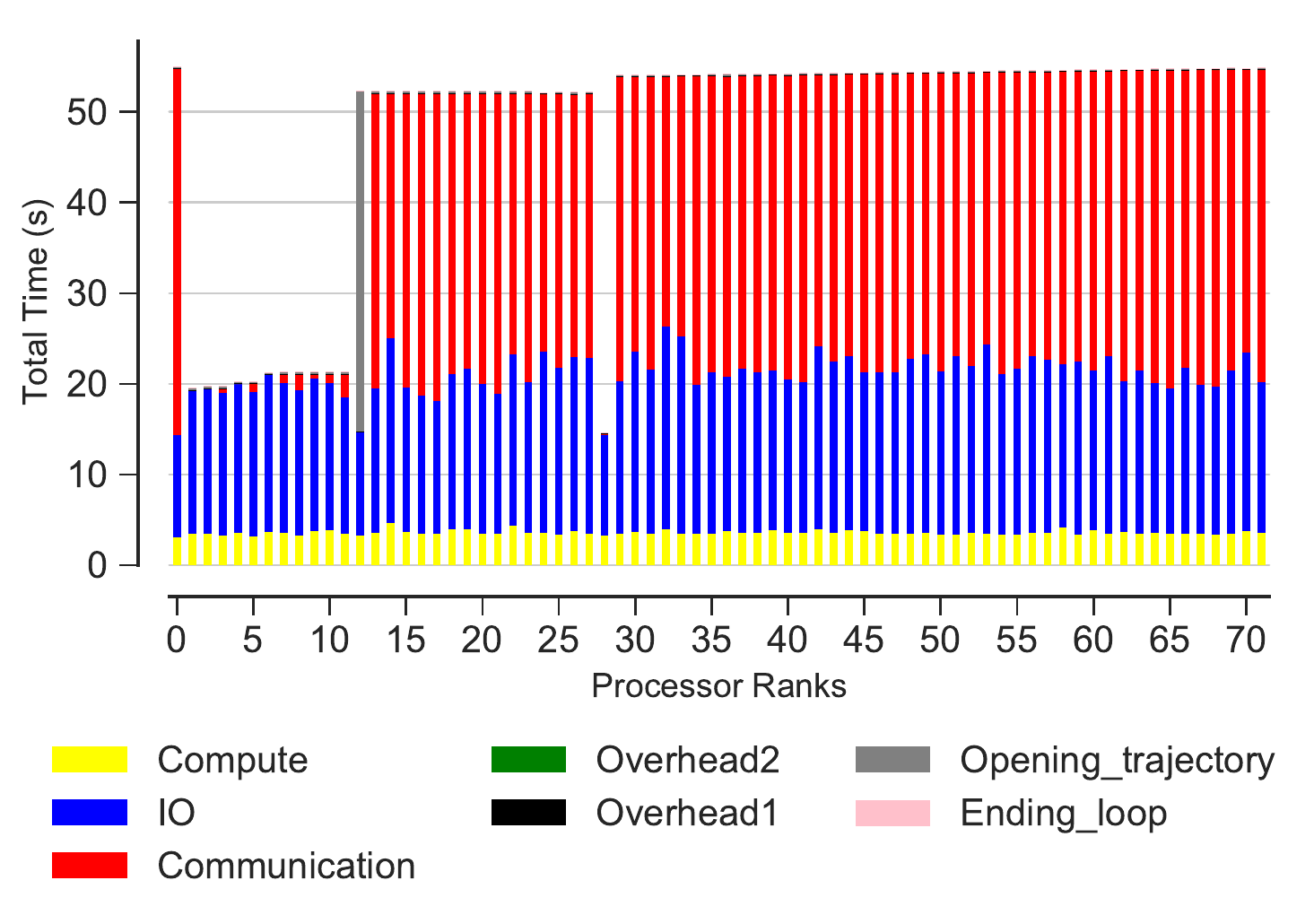}
    \captionsetup{format=hang}
    \caption{Time comparison on different parts of the calculations per MPI rank (example)}
    \label{fig:MPIranks}
  \end{subfigure}
  \caption{Performance of the RMSD task parallelized with MPI on \emph{SDSC Comet}.
    Results were communicated back to rank 0.
    Five independent repeats were performed to collect statistics.
    (a-c) The error bars show standard deviation with respect to the mean.
    In serial, there is no communication and no data points are shown for $N=1$ in (c).
    (d) Compute \tcomp, read I/O \tIO, communication \tcomm, ending the for loop $t_{\text{end\_loop}}$, opening the trajectory $t_{\text{opening\_trajectory}}$, and overheads $t_{\text{overhead1}}$, $t_{\text{overhead2}}$ per MPI rank; see Table \ref{tab:notation} for definitions.
    These are data from one run of the five repeats.
    MPI ranks 0, 12--27 and 29--72 are stragglers.
  }
  \label{fig:MPIwithIO}
\end{figure}

We measured strong scaling for the RMSD analysis task (Algorithm \ref{alg:RMSD}) with the 2,512,200 frame test trajectory (section \ref{sec:data}) on 1 to 72 cores (one to three nodes) of \emph{SDSC Comet} (Figures~\ref{fig:MPIscaling} and \ref{fig:MPIspeedup}). 
We observed poor strong scaling performance beyond a single node (24 cores), comparable to our previous results \cite{Khoshlessan:2017ab}.
A more detailed analysis showed that the RMSD computation, and to a lesser degree the read I/O, considered on their own, scaled well beyond 50 cores (yellow and blue lines in Figure~\ref{fig:ScalingComputeIO}). 
But communication (sending results back to MPI rank 0 with \texttt{MPI\_Gather()}; red line in Figure~\ref{fig:ScalingComputeIO}) and the initial file opening (loading the system information into the \texttt{MDAnalysis.Universe} data structure from a shared topology file and opening the shared trajectory file; gray line in Figure~\ref{fig:ScalingComputeIO}) started to dominate beyond 50 cores.
Communication cost and initial time for opening the trajectory were distributed unevenly across MPI ranks, as shown in Figure~\ref{fig:MPIranks}. 
The ranks that took much longer to complete than the typical execution time of the other ranks were the stragglers that hurt performance.

We qualitatively denoted by \emph{straggler} any MPI rank that took at least about twice as long as the group of ranks that finished fastest, roughly following the original description of a straggler as a task that took an ``unusually long time to complete'' \cite{Dean2008}.
The fast-finishing ranks were generally clearly distinguishable in the per-rank timings such as in Figures~\ref{fig:MPIranks} and \ref{fig:MPIranks-Bridges}.
Such a qualitative definition of stragglers was sufficient for our purpose of identifying scalability bottlenecks, as shown in the following discussion.

\subsubsection*{Identification of Scalability Bottlenecks}

In the example shown in Figure~\ref{fig:MPIranks}, 62 ranks out of 72 took about 60~s (the stragglers) whereas the remaining ranks only took about 20~s.
In other instances, far fewer ranks were stragglers, as shown, for example, in Figure~\ref{fig:MPIranks-Bridges}.
The detailed breakdown of the time spent on each rank (Figure~\ref{fig:MPIranks}) showed that the computation, \tcomp, was relatively constant across ranks. 
The time spent on reading data from the shared trajectory file on the Lustre file system into memory, \tIO, showed variability across different ranks. 
The stragglers, however, appeared to be defined by occasionally much larger \emph{communication} times, \tcomm (line 16 in Algorithm~\ref{alg:RMSD}), which were on the order of 30~s, and by larger times to initially open the trajectory (line 2 in Algorithm~\ref{alg:RMSD}).
\tcomm varied across different ranks and was barely measurable for a few of them.
Although the data in Figure~\ref{fig:MPIranks} represented one run and in other instances different number of ranks were stragglers, the averages over all ranks in five independent repeats (Figure~\ref{fig:ScalingComputeIO}) showed that increased \tcomm were generally the reason for large variations in the run time for each rank.
This initial analysis indicated that communication was a major issue that prevented good scaling beyond a single node but the problems related to file I/O also played an important role in limiting scaling performance.

\subsubsection*{Influence of Hardware}
We ran the same benchmarks on multiple HPC systems that were equipped with a Lustre parallel file system [XSEDE's \emph{PSC Bridges} (Figure~\ref{fig:MPIwithIO-Bridges}) and \emph{LSU SuperMIC} (Figure~\ref{fig:MPIwithIO-SuperMIC})], and observed the occurrence of stragglers, in a manner very similar to the results described for \emph{SDSC Comet}.
There was no clear pattern in which certain MPI ranks would always be a straggler, and neither could we trace stragglers to specific cores or nodes.
Therefore, the phenomenon of stragglers in the RMSD case was reproducible on different clusters and thus appeared to be independent from the underlying hardware.

\subsection{Effect of Compute to I/O Ratio on Performance}
\label{sec:bound}

The results in section~\ref{sec:RMSD} indicated that opening the trajectory, communication, and read I/O were important factors that appeared to correlate with stragglers. 
In order to better characterize the RMSD task, we computed the ratio between the time to complete the computation and the time spent on I/O per frame, \RcompIO (Eq.~\ref{eq:Compute-IO}).
The average values were $\overline{\tcomp^{\text{frame}}} = 0.09\ \text{ms}$, $\overline{t_{\text{IO}}^{\text{frame}}} = 0.3\ \text{ms}$, resulting in a compute-to-I/O ratio $\RcompIO \approx 0.3$.
With $\RcompIO \ll 1$, the RMSD analysis task was characterized as I/O bound.

In other studies, better scaling was observed for analysis tasks that were more compute-intensive than the RMSD calculation, such as a radial distribution function calculation \cite{Roe:2018aa, Fan:2019aa}, i.e., analysis tasks that could be characterized as compute-bound.
Such behavior is expected, as the contribution from the parallel part of the program that requires neither I/O nor communication is increased.
From a practical point of view, it is of interest to understand the size of the effect of increasing the computational load on strong scaling, and in our case, we were interested in seeing if changes in the compute part (namely the RMSD calculation on coordinates held in memory) would have an effect on the execution time of other parts of the program.
In Appendix~\ref{sec:shiftload} we set out to analyze compute bound tasks, i.e.\ ones with $\RcompIO \gg 1$.
To assess the effect of the $\RcompIO$ ratio on performance while leaving other parameters the same, we artificially increased the computational load by repeating the RMSD calculation and measured strong scaling (Figure \ref{fig:tcomp_tIO_effect}).
With increasing $\RcompIO$, the impact of stragglers appeared to lessen (although they did not disappear) and scaling performance improved, as expected (see Appendix \ref{sec:increasedworkload}).
Better scaling also went together with a higher ratio of compute to communication ($\Rcompcomm$, Eq.~\ref{eq:Compute-comm}) as shown in Appendix \ref{sec:tcomm} but ultimately I/O seemed to be the key determinant for performance.

 \begin{figure}[!htb]
   \centering
   \begin{subfigure}{.25\textwidth}
     \includegraphics[width=\linewidth]{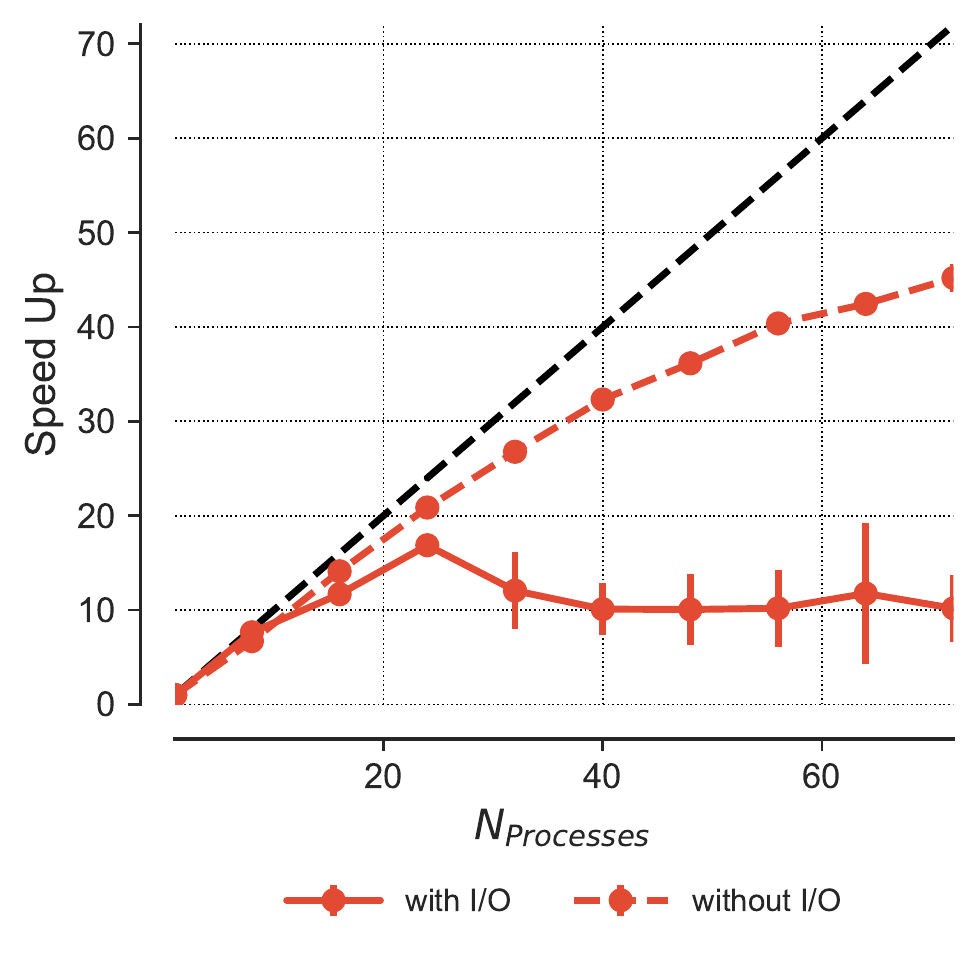}
     \caption{Speed up comparison}
     \label{fig:MPIspeedup-no-IO}
   \end{subfigure}
   \begin{subfigure}{0.3\textwidth}
     \includegraphics[width=\linewidth]{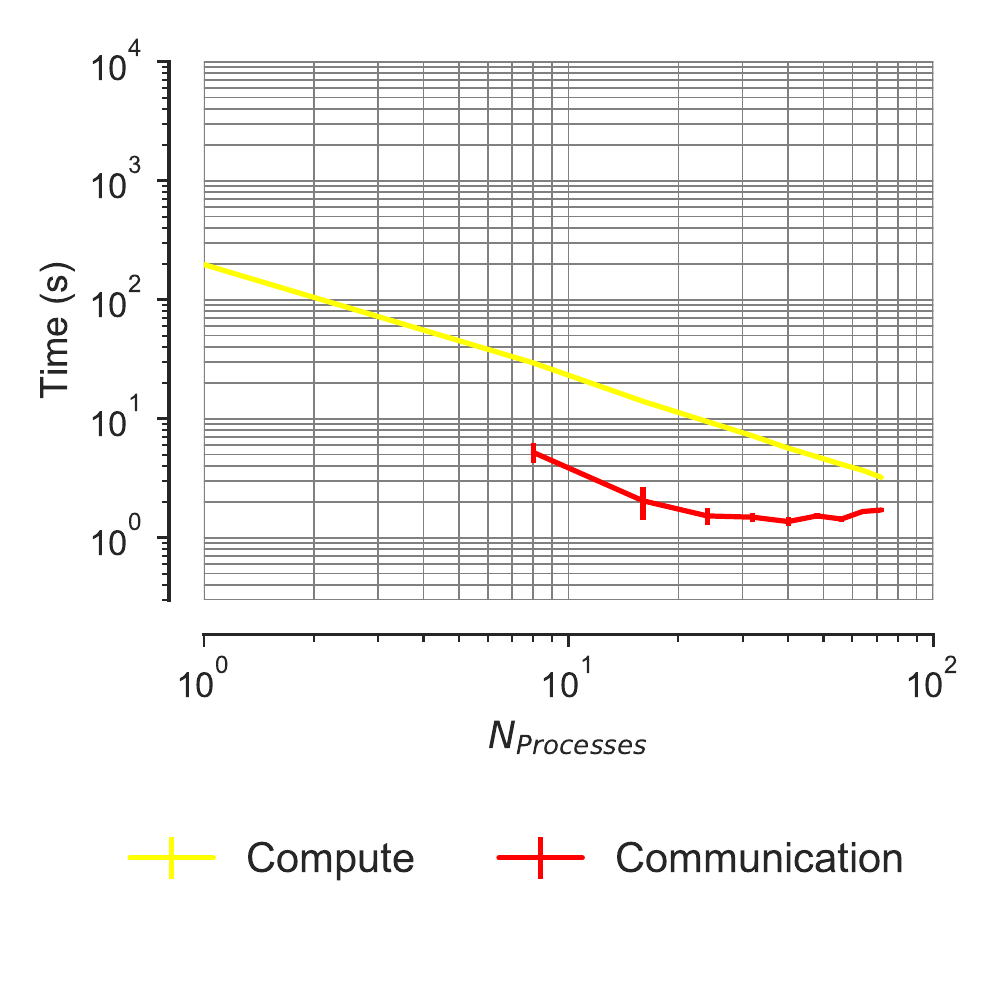}
     \caption{Scaling for different components}
     \label{fig:MPIScalingCompute-Comm-no-IO}     
   \end{subfigure}
   \begin{subfigure}{.4\textwidth}
     \includegraphics[width=\linewidth]{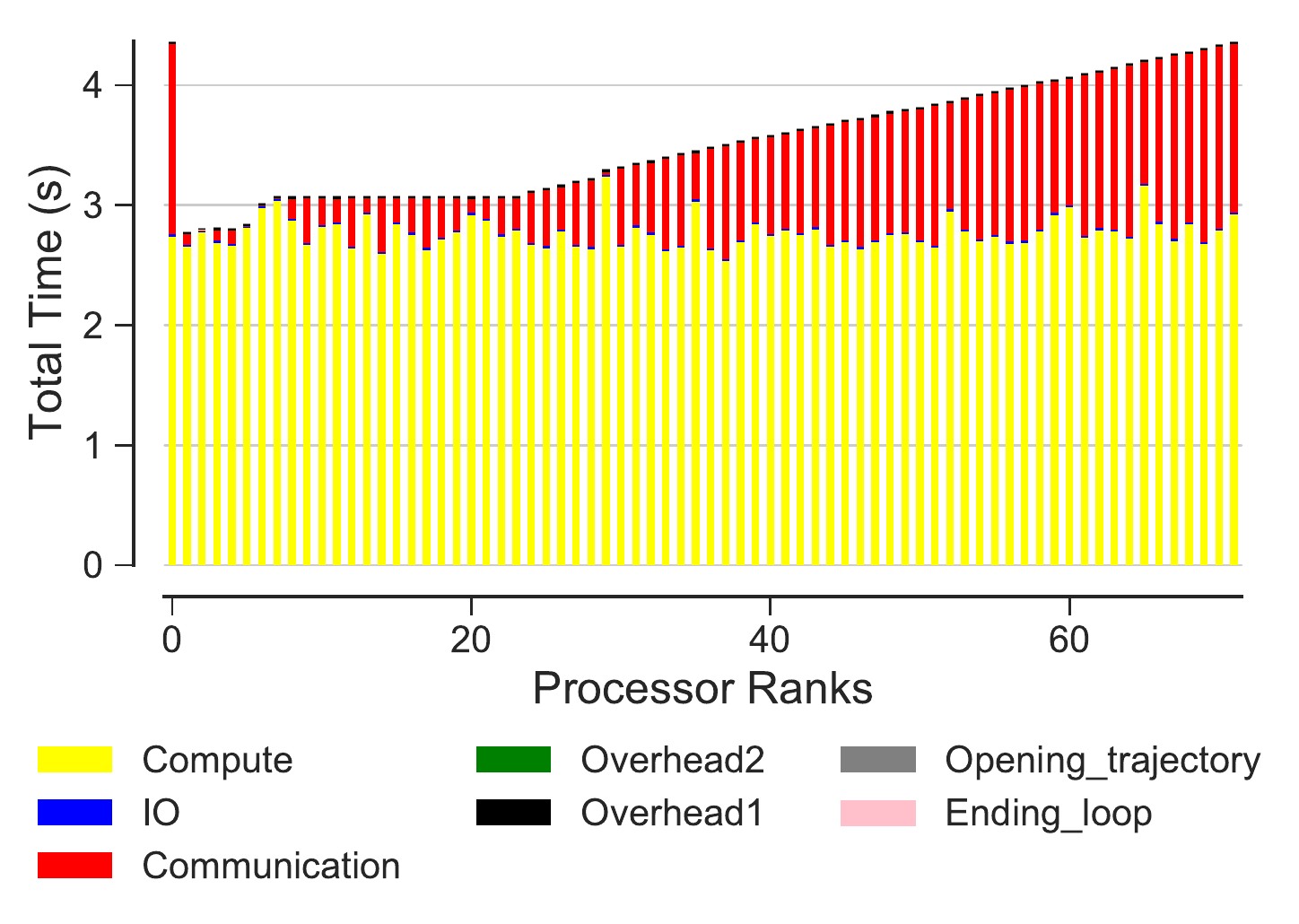}
     \captionsetup{format=hang}
     \caption{Time comparison on different parts of the calculations per MPI rank when I/O is removed}
     \label{fig:MPIranks-no-IO}
   \end{subfigure}
   \caption{Comparison of the performance of the RMSD task with I/O ($\RcompIO \approx 0.3$) and without I/O ($\RcompIO = +\infty$) on \emph{SDSC Comet}.
     Five repeats were performed to collect statistics.
     (a-b) The error bars show standard deviation with respect to the mean.
     (b) Only compute \tcomp and communication \tcomm are included; there are no timings related to I/O (\tIO, $t_{\text{opening\_trajectory}}$) as in Figure~\protect{\ref{fig:ScalingComputeIO}}.
     (c) Compute \tcomp, read I/O $\tIO=0$, communication \tcomm, ending the for loop \text{$t_{\text{end\_loop}}=0$},
     opening the trajectory \text{$t_{\text{opening\_trajectory}}=0$}, and overheads \text{$t_{\text{overhead1}}$}, \text{$t_{\text{overhead2}}$} per MPI rank.
     (See Table \ref{tab:notation} for definitions.)}
   \label{fig:MPIwithoutIO}
\end{figure}

In order to study an extreme case of a compute-bound task that would demonstrate the effect of ``ideal'' read I/O, we eliminated all I/O from the RMSD task by randomly generating artificial trajectory data in memory; the data had the same size as if they had been obtained from the trajectory file.
The time for the data generation was excluded and no file access was necessary. 
Without any I/O, performance improved markedly (Figure~\ref{fig:MPIwithoutIO}), with reasonable scaling up to 72 cores (3 nodes).
No stragglers were observed because overall communication time decreased dramatically by more than a factor of ten and showed less variability (Figure~\ref{fig:MPIScalingCompute-Comm-no-IO}) compared to the same runs with I/O (Figure~\ref{fig:ScalingComputeIO}), even though exactly the same amount of data were communicated.
The scaling performance suffered somewhat for more than 40 processes only because the cost of communication \tcomm became comparable to the compute time \tcomp and would not decrease further.
Although in practice I/O cannot be avoided, this experiment demonstrated that the way how the trajectory file was accessed on the Lustre file system was at least one cause for the observed stragglers.
It also showed that the communication cost for the \emph{same amount of data transfer} could be dramatically higher in the presence of I/O than in its absence.

\subsection{Reducing I/O Cost}
\label{sec:I/O}
In order to improve performance we needed to employ strategies to avoid the competition over file access across different ranks when the $\RcompIO$ ratio was small.
One obvious approach when using the Lustre parallel file system is to increase the number of stripes, i.e., the number of copies of a file that are stored on different object storage targets (OSTs).
But because in our previous work we did not see scaling performance improvement with varying the stripe count \cite{Khoshlessan:2017ab} we decided to just use the system default, i.e., one stripe per file.
Instead we experimented with two different ways for reducing the I/O cost:
\begin{inparaenum}[1)]
	\item splitting the trajectory file into as many segments as the number of processes (subfiling), thus using file-per-process access, and
	\item using the HDF5 file format together with MPI-IO parallel reads instead of the XTC trajectory format.
\end{inparaenum}
We discuss these two approaches and their performance improvements in detail in the following sections.

\subsubsection{Splitting the Trajectories (subfiling)}
\label{sec:splitting-traj}
\emph{Subfiling} is a mechanism previously used for splitting a large multi-dimensional global array to a number of smaller subarrays in which each smaller array is saved in a separate file. Subfiling reduces the file system control overhead by decreasing the number of processes concurrently accessing a shared file~\cite{scalable-IO, scalable-IO1}.
Because subfiling is known to improve performance of parallel shared-file I/O, we investigated splitting our trajectory file into as many trajectory segments as the number of processes.
The trajectory file was split into $N$ segments, one for each process, with each segment having $N_{b}$ frames. 
This way, each process would access its own trajectory segment file without competing for file accesses. 

\begin{figure}[!htb]
  \centering
  \begin{subfigure}{.4\textwidth}
    \includegraphics[width=\linewidth]{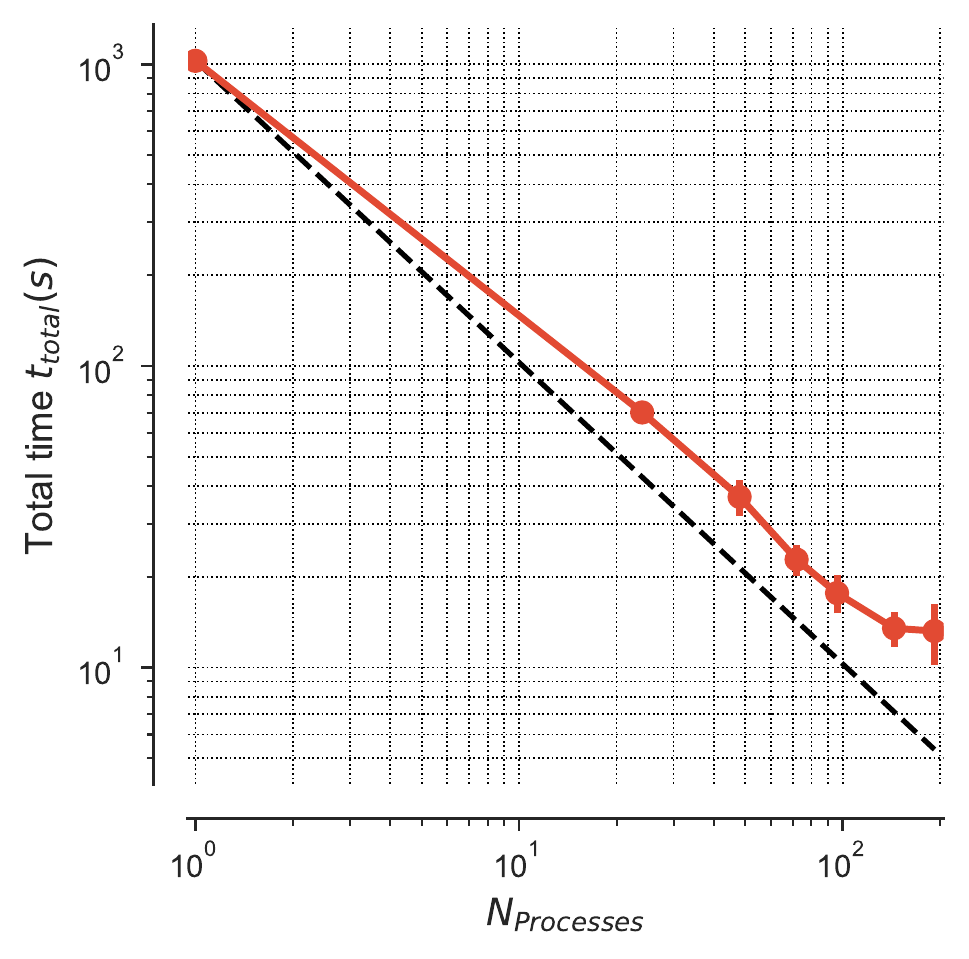}
    \caption{Scaling total}
    \label{fig:MPIscaling-split}
  \end{subfigure}
  \hfill
  \begin{subfigure}{.4\textwidth}
    \includegraphics[width=\linewidth]{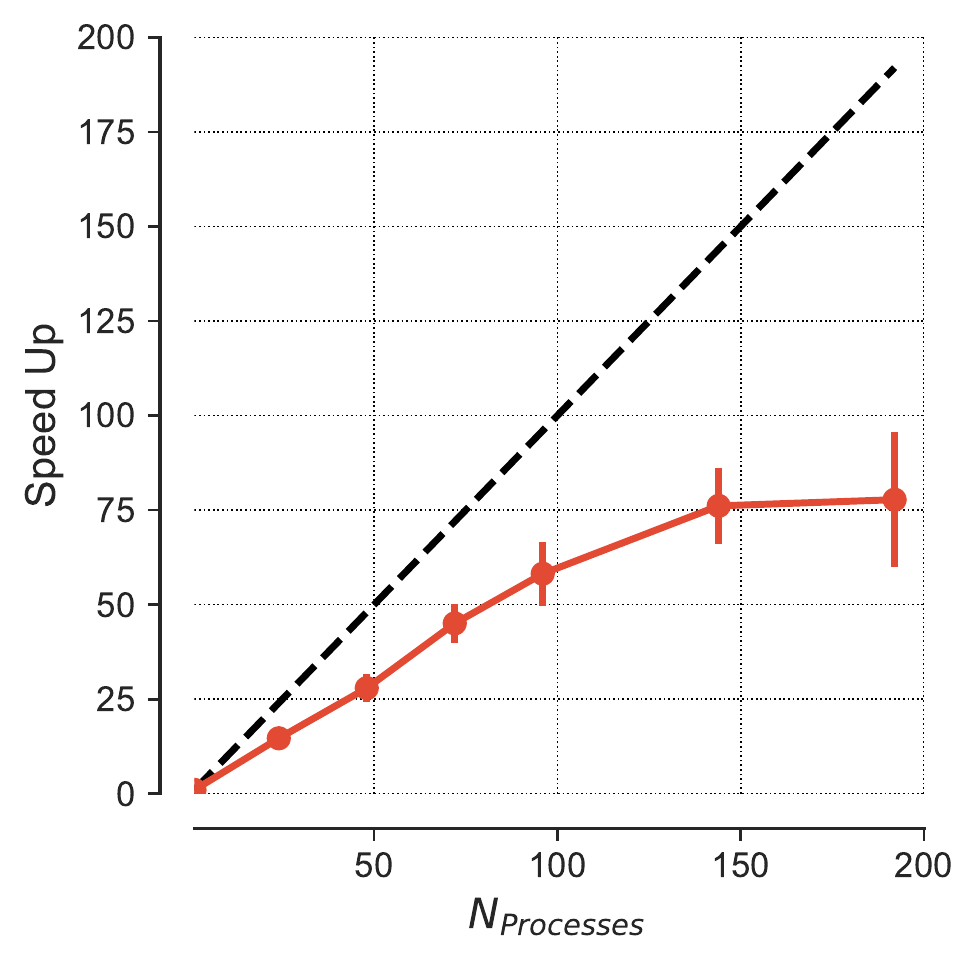}
    \captionsetup{format=hang}
    \caption{Speed-up}
    \label{fig:MPIspeedup-split}
  \end{subfigure}
  \bigskip
  \begin{subfigure}{.45\textwidth}
    \includegraphics[width=\linewidth]{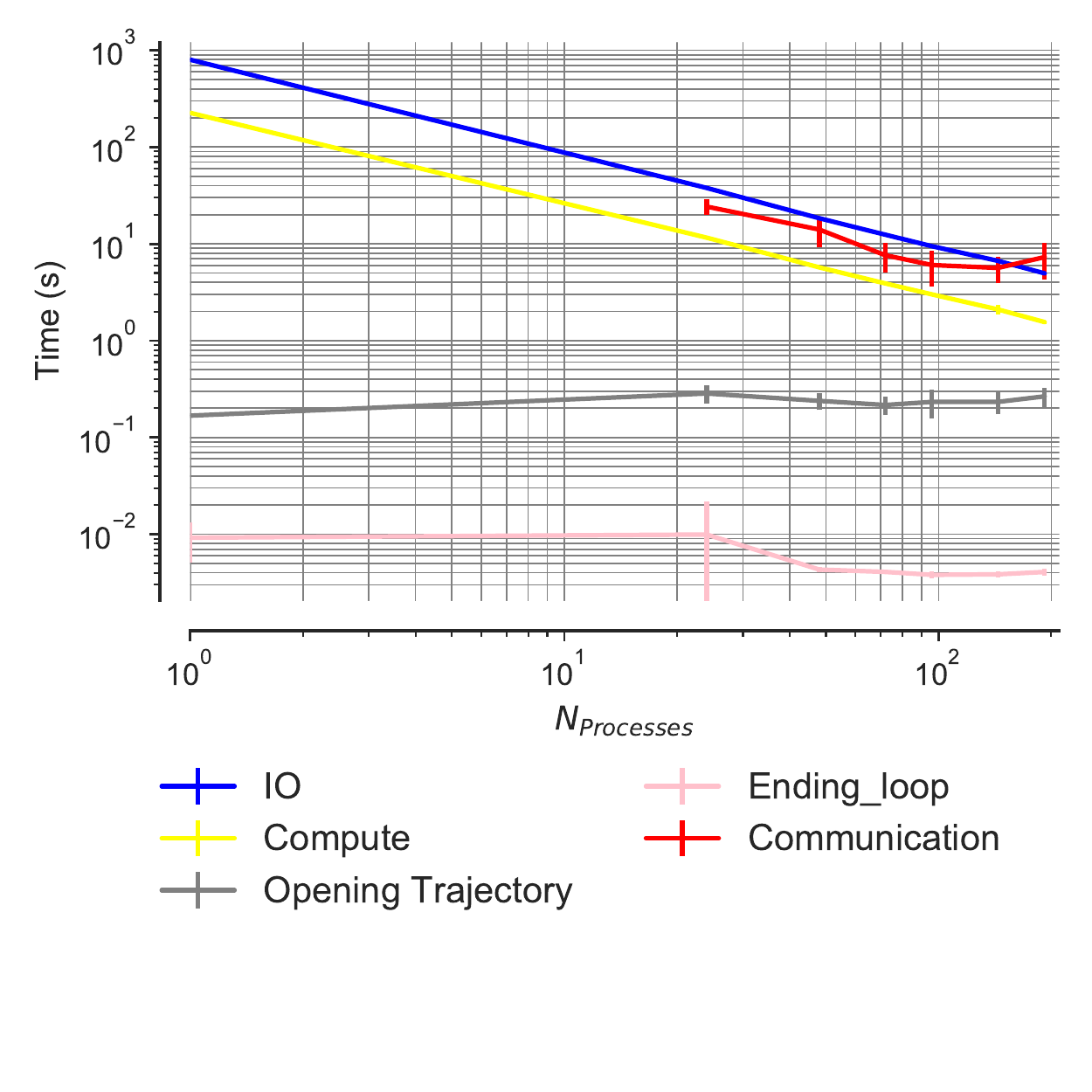}
    \captionsetup{format=hang}
    \caption{Scaling for different components}
    \label{fig:ScalingComputeIO-split}
  \end{subfigure}
  \hfill
  \begin{subfigure}{.5\textwidth}
    \includegraphics[width=\linewidth]{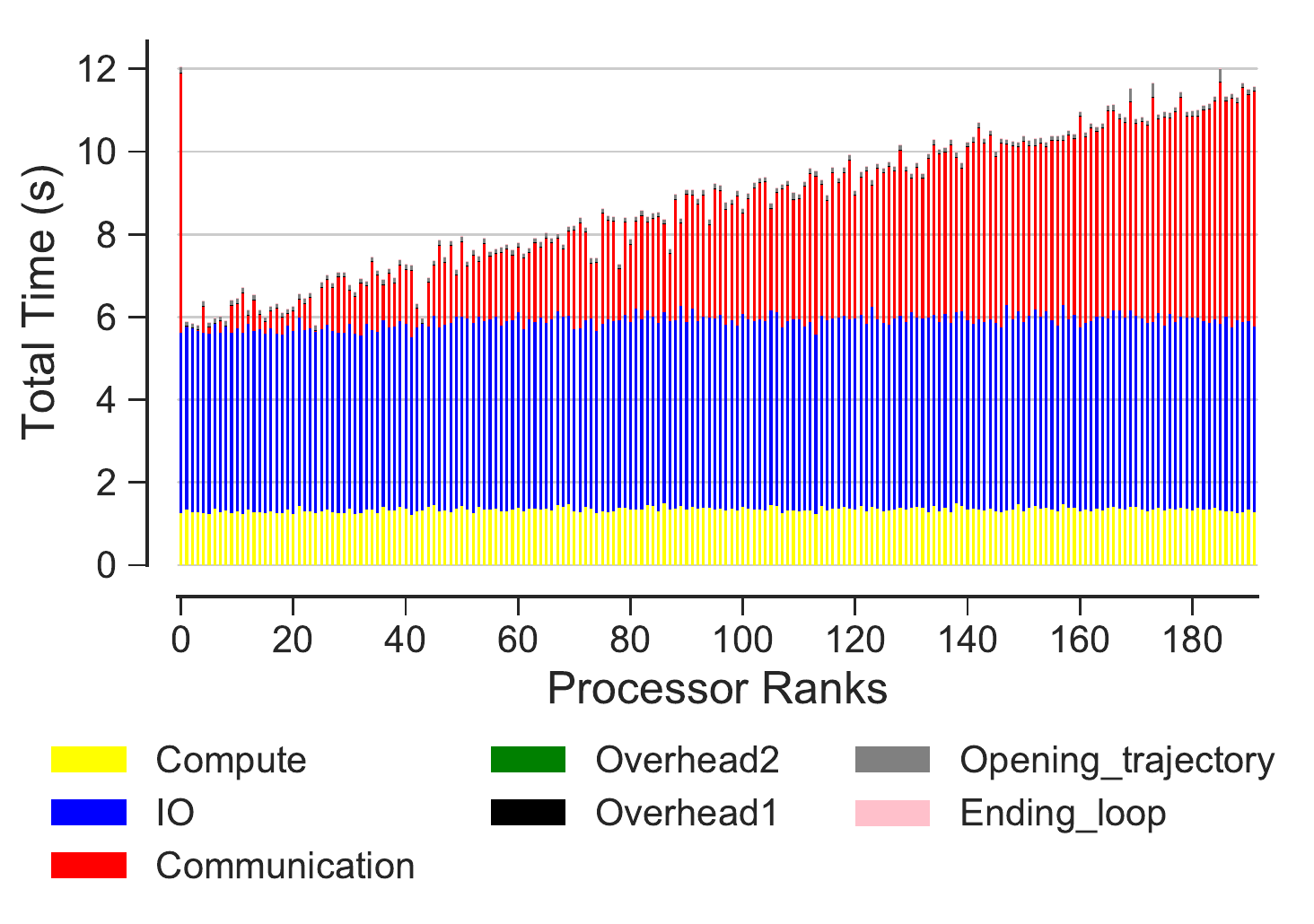}
    \captionsetup{format=hang}
    \caption{Time comparison on different parts of the calculations per MPI rank.}
    \label{fig:MPIranks-split}
  \end{subfigure}
  \caption{Performance of the RMSD task on \emph{SDSC Comet} when the trajectories are split into one trajectory segment per process (\emph{subfiling}).
    Five repeats were performed to collect statistics.
    In serial, there is no communication and no data points are shown for $N=1$ in (c).
    (a-c) The error bars show standard deviation with respect to the mean.
    (d) Compute \tcomp, read I/O \tIO, communication \tcomm, opening the trajectory $t_{\text{opening\_trajectory}}$, ending the for loop  $t_{\text{end\_loop}}$ (includes closing the trajectory), and overheads $t_{\text{overhead1}}$, $t_{\text{overhead2}}$ per MPI rank; see Table \protect\ref{tab:notation} for the definitions.
}
\label{fig:MPIwithIO-split}
\end{figure}

We ran a benchmark up to 8 nodes (192 cores) and observed rather better scaling behavior with efficiencies above 0.6 (Figure~\ref{fig:MPIscaling-split} and~\ref{fig:MPIspeedup-split}) with the delay time for stragglers reduced from 65~s to about 10~s for 72 processes. 
However, scaling was still far from ideal due to the MPI communication costs. 
Although the delay due to communication was much smaller than compared to parallel RMSD with shared-file I/O [compare Figure~\ref{fig:MPIranks-split} ($\tcomm^{\text{Straggler}} \gg \tcomp+\tIO$) to Figure~\ref{fig:MPIranks} ($\tcomm^{\text{Straggler}} \approx \tcomp+\tIO$)], it was still delaying several processes and resulted in longer job completion times (Figure~\ref{fig:MPIranks-split}). 
These delayed tasks impacted performance so that speed-up remained far from ideal (Figure~\ref{fig:MPIspeedup-split}).

The subfiling approach appeared promising but it required preprocessing of trajectory files and additional storage space for the segments.
We benchmarked the necessary time for splitting the trajectory in parallel using different number of MPI processes (Table~\ref{tab:timing-splitting}); in general the operation scaled well, with efficiencies $> 0.8$ although performance fluctuated, as seen for the case on six nodes where the efficiency dropped to $0.34$ for the run.
These preprocessing times were not included in the estimates because we focused on better understanding the principal causes of stragglers and we wanted to make the results directly comparable to the results of the previous sections.
Nevertheless, from an end user perspective, preprocessing of trajectories can be integrated in workflows (especially as the data in Table~\ref{tab:timing-splitting} indicated good scaling) and the preprocessing time can be quickly amortized if the trajectories are analyzed repeatedly.
However, the requirement of needing as many segments as processes makes the approach somewhat inflexible as a new set of trajectory segments must be produced when a different level of parallelization is needed.
Finally, the performance of parallel file systems generally suffers when too many files are processed and so there exists a limit as to how far the subfiling approach can be pushed.

\begin{SCtable}[1.0][!htb]
  \centering
  \caption[Time necessary for writing the trajectory segments]
  {The wall-clock time spent for writing $N_{\text{seg}}$ trajectory segments using $N_{\text{p}}$ processes using MPI on \emph{SDSC Comet}.
    One set of runs was performed for the timings.
    Scaling $S$ and efficiency $E$ are relative to the 1 node case (24 MPI processes).}
  \label{tab:timing-splitting}  
  \begin{tabular}{rrrrrr}
    \toprule
    \thead{$N_{\text{seg}}$} & \thead{$N_{\text{p}}$} & \thead{nodes}
    & \thead{time (s)} & \thead{$S$} & \thead{$E$}\\
    \midrule
    24 &  24 & 1 & 89.9 & 1.0 & 1.0\\
    48 &  48 & 2 & 46.8 & 1.9 & 0.96\\
    72 &  72 & 3 & 33.7 & 2.7 & 0.89\\
    96 &  96 & 4 & 25.1 & 3.6 & 0.89\\
    144 & 144 & 6 & 43.7 & 2.1 & 0.34\\
    192 & 192 & 8 & 13.5 & 6.7 & 0.83\\  
    \bottomrule
  \end{tabular}
\end{SCtable}

Often trajectories from MD simulations on HPC machines are produced and kept in smaller files or segments that can be concatenated to form a full continuous trajectory file.
These trajectory segments could be used for the subfiling approach.
However, it might not be feasible to have exactly one segment per MPI rank, with all segments of equal size, which constitutes the ideal case for subfiling.
MDAnalysis can create virtual trajectories from separate trajectory segment files that appear to the user as a single trajectory.
In Appendix~\ref{sec:chainreader} we investigated if this so-called \emph{ChainReader} functionality could benefit from the subfiling approach.
We found some improvements in performance but discovered limitations in the design of the ChainReader (namely that all segment files are initially opened) that will have to be addressed before equivalent performance can be reached.
 
\subsubsection{MPI-based Parallel HDF5}
\label{sec:HDF5}

In the HPC community, parallel I/O with MPI-IO is widely used in order to address I/O limitations.
We investigated MPI-based Parallel HDF5 to improve I/O scaling. 
We converted our XTC trajectory file into a simple custom HDF5 format so that we could test the performance of parallel I/O with the HDF5 file format.
The time it took to convert our XTC file with $2,512,200$ frames into HDF5 format was about 5,400~s on a local workstation with network file system (NFS).

\begin{figure}[!htb]
  \centering
  \begin{subfigure}{.4\textwidth}
    \includegraphics[width=\linewidth]{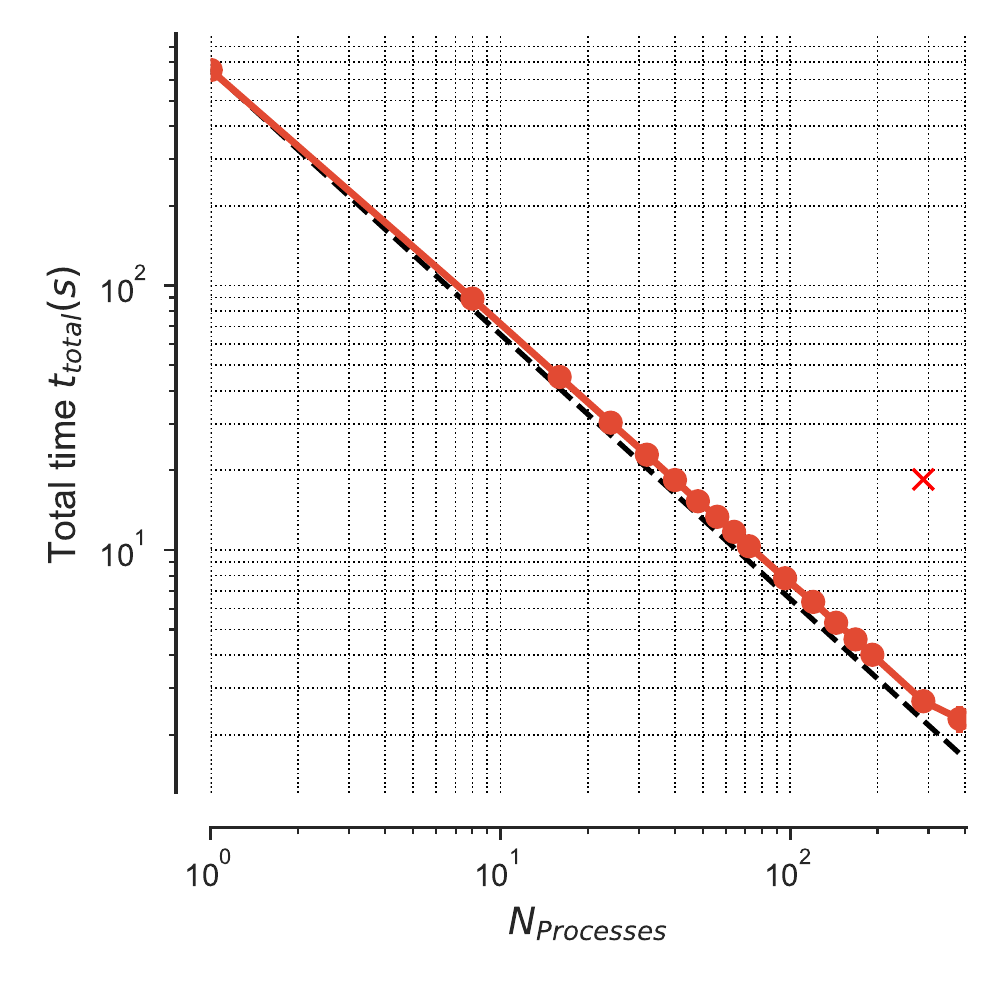}
    \caption{Scaling total}
    \label{fig:MPIscaling-hdf5}
  \end{subfigure}
  \hfill
  \begin{subfigure}{.4\textwidth}
    \includegraphics[width=\linewidth]{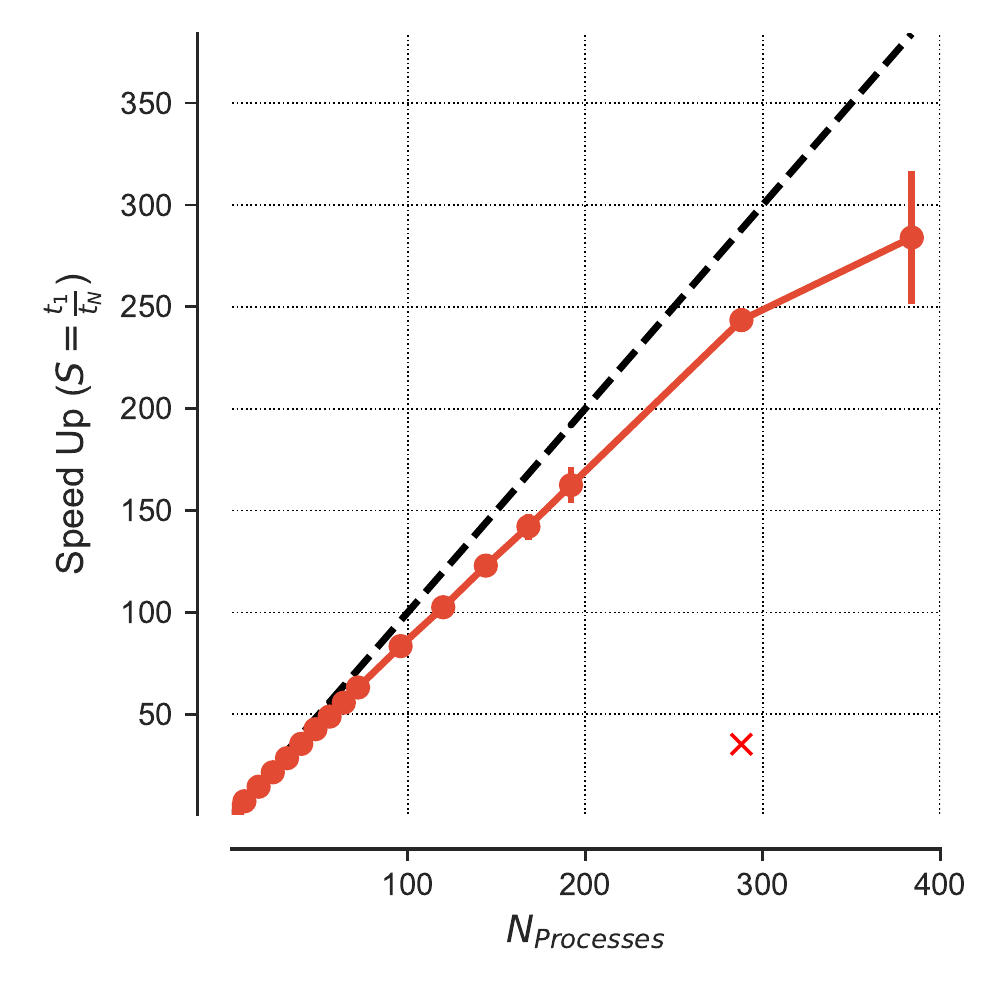}
    \caption{Speed-up}
    \label{fig:MPIspeedup-hdf5}
  \end{subfigure}
  \bigskip

  \begin{subfigure}{.45\textwidth}
    \centering
    \includegraphics[width=\linewidth]{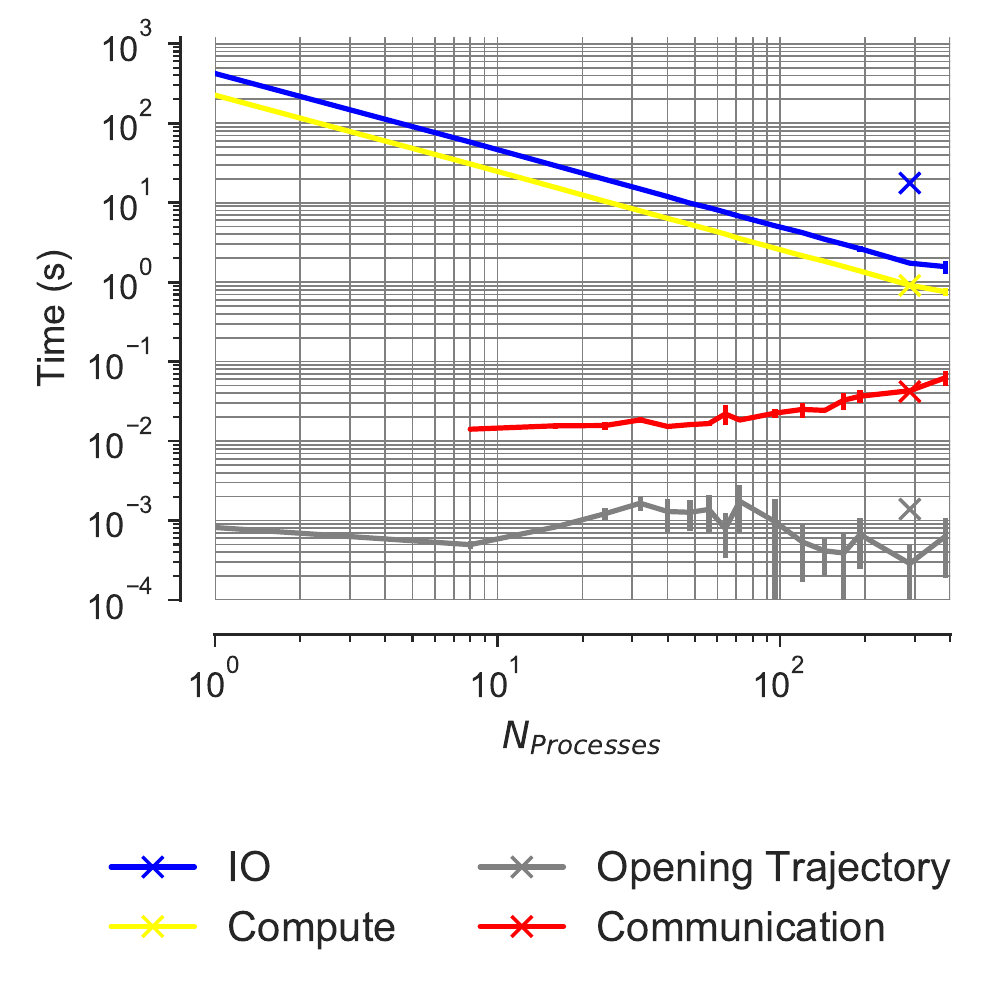}
    \captionsetup{format=hang}
    \caption{Scaling for different components}
    \label{fig:ScalingComputeIO-hdf5}
  \end{subfigure}
  \hfill
  \begin{subfigure} {.5\textwidth}
    \includegraphics[width=\linewidth]{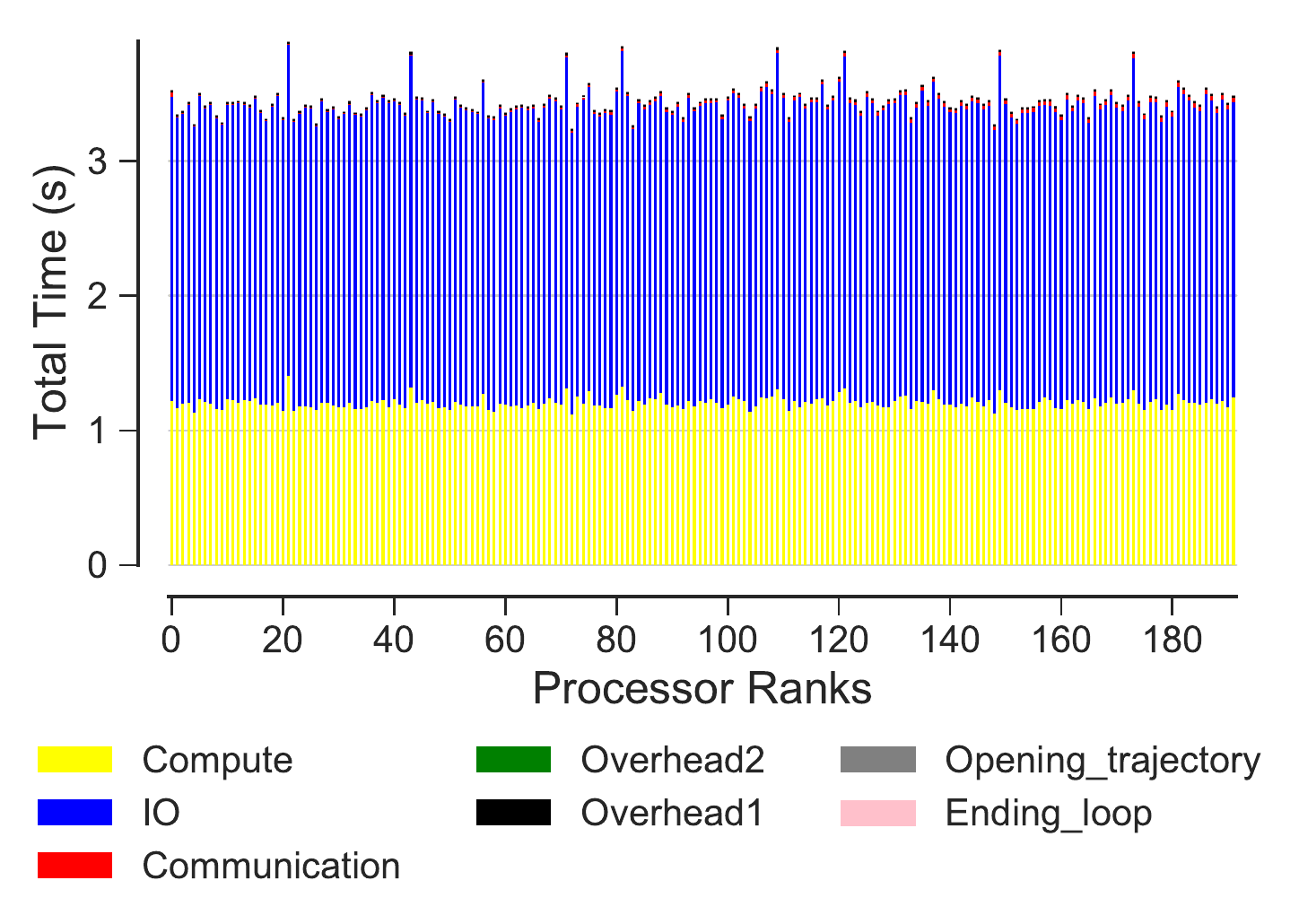}
    \captionsetup{format=hang}
    \caption{Time comparison on different parts of the calculations per MPI rank}
    \label{fig:MPIranks-hdf5}
  \end{subfigure}
  \caption{Performance of the RMSD task with MPI-based parallel HDF5 (MPI-IO) on \emph{SDSC Comet}.
    Data are read from the file system from a shared HDF5 file format instead of XTC format (independent I/O) and results are communicated back to rank 0. 
    Five repeats were performed to collect statistics; one repeat for 288 processes had abnormally high \tIO and was treated as an outlier and excluded from the averages but is shown as ``x'' in the graphs.
    (a-c) The error bars show standard deviation with respect to the mean.
    In serial, there is no communication and no data points are shown for $N=1$ in (c).
    (d) Compute \tcomp, read I/O \tIO, communication \tcomm, ending the for loop $t_{\text{end\_loop}}$,
    opening the trajectory $t_{\text{opening\_trajectory}}$, and overheads $t_{\text{overhead1}}$, $t_{\text{overhead2}}$ per MPI rank; see Table \ref{tab:notation} for definitions.
    These are typical data from one run of the five repeats.
  }
  \label{fig:MPIwithIO-hdf5}
\end{figure}

We ran our benchmark on up to 16 nodes (384 cores) on \emph{SDSC Comet} and we observed near ideal scaling behavior (Figures~\ref{fig:MPIscaling-hdf5} and~\ref{fig:MPIspeedup-hdf5}) with parallel efficiencies above 0.8 on up to 8 nodes (Figure~\ref{fig:comparison_efficiency}) with no straggler tasks (Figure~\ref{fig:MPIranks-hdf5}).
The trajectory reading I/O (\tIO) was the dominant contribution, followed by compute (\tcomp), but because both contributions scaled well, overall scaling performance remained good, even for 384 cores.
Amongst the five repeats for 12 nodes (288 cores) we observed one run with much slower I/O than typical (Figure~\ref{fig:ScalingComputeIO-hdf5}) but we did not further investigate this spurious case and classified it as an outlier that was excluded from the statistics.
Importantly, the trajectory opening cost remained negligible (in the millisecond range) and the cost for MPI communication also remained small (below 0.1 s, even for 16 nodes).
Overall, parallel MPI with HDF5 appeared to be a robust approach to obtain good scaling, even for I/O-bound tasks.

\subsection{Potential Causes of Stragglers}
\label{sec:likelycauses}

The data indicated that an increase in the duration of both MPI communication and trajectory file access lead to large variability in the run time of individual MPI processes and ultimately poor scaling performance beyond a single node.
A discussion of likely causes for stragglers begins with the observation that opening and reading a single trajectory file from multiple MPI processes appeared to be at the center of the problem. 

In MDAnalysis, individual trajectory frames are loaded into memory, which ensures that even systems with tens of millions of atoms can be efficiently analyzed on resources with moderate RAM sizes.
The test trajectory (file size 30 GB) had $2,512,200$ frames in total so each frame was about 0.011 MB in size.
With $\tIO \approx 0.3~\text{ms}$ per frame, the data were ingested at a rate of about $40$~MB/s for a single process.
For 24 MPI ranks (one \emph{SDSC Comet} node), the aggregated reading rate would have been about 1 GB/s, well within the available bandwidth of 56 Gb/s of the InfiniBand network interface that served the Lustre file system, but nevertheless sufficient to produce substantial constant network traffic.

Furthermore, in our study the default Lustre stripe size value was 1~MB, i.e., the amount of contiguous data stored on a single Lustre OST.
Each I/O request read a single Lustre stripe but because the I/O size (0.011~MB) was smaller than the stripe size, many of these I/O requests were likely just accessing the same stripe on the same OST but nevertheless had to acquire a new reading lock for each request.
The reason for this behavior is related to ensuring POSIX consistency that relates to POSIX I/O API and POSIX I/O semantics, which can have adverse effects on scalability and performance.
Parallel file systems like Lustre implement sophisticated distributed locking mechanisms to ensure consistency.
For example, locking mechanisms ensures that a node can not read from a file or part of a file while it might be being modified by another node. 
When the application I/O is not designed in a way to avoid scenarios where multiple nodes are fighting over locks for overlapping extents, Lustre can suffer from scalability limitations~\cite{optimize_lustre}.
Continuously keeping metadata updated in order to have fully consistent reads and writes (POSIX metadata management), requires writing a new value for the file's last-accessed time (POSIX atime) every time a file is read, imposing a significant burden on parallel file system~\cite{POSIX2017}. 
Mache \textit{et al.} observed that contention for the interconnect between OSTs and compute nodes due to MPI communication may lead to variable performance in I/O measurements~\cite{Mache:2005aa}.
Conversely, our data suggest that single-shared-file I/O on Lustre can negatively affect MPI communication as well, even at moderate numbers (tens to hundreds) of concurrent requests, similar to the results from recent network simulations that predicted interference between MPI and I/O traffic~\cite{Brown:2018ab}.
Brown \textit{et al.}'s  work \cite{Brown:2018ab} indicated that MPI traffic (inter-process communication) could be affected by increasing I/O.
In particular, a few MPI processes were always delayed by one to two orders of magnitude more than the median time, which we would classify as stragglers.
In summary, our observations, seen in the context of the work by \citet{Brown:2018ab}, suggest that our observed stragglers with large variance in the communication step could be due to interference of MPI communications with the I/O requests on the same network.
Further detailed work will be needed to test this hypothesis.

Our results clearly showed that reading a single shared file is an inefficient way to use the Lustre parallel file system; instead, parallel I/O via MPI-IO and HDF5 emerged as the most promising approach to avoid stragglers and obtain good strong scaling behavior on hundreds of cores, even for I/O bound analysis tasks.


\section{Reproducibility and Performance Comparison on Different Clusters}
\label{sec:clusters}

In this section we compare the performance of the RMSD task on different HPC resources (Table \ref{tab:sys-config}) to examine the robustness of the methods we used for our performance study and to ensure that the results are general and independent from the specific HPC system.
Scripts and instructions to set up the computational environments and reproduce our computational experiments are provided in a git repository as described in section \ref{sec:methods}.

In Appendix~\ref{sec:supplement} we demonstrated that stragglers occured on \emph{PSC Bridges} (Figure \ref{fig:MPIwithIO-Bridges}) and \emph{LSU SuperMIC} (Figure \ref{fig:MPIwithIO-SuperMIC}) in a manner similar to the one observed on \emph{SDSC Comet} (section \ref{sec:RMSD}).
We performed additional comparisons for several cases discussed previously, namely (1) splitting the trajectories with blocking collective communications in MPI and (2) MPI-based parallel HDF5.

\subsection{Splitting the Trajectories}
\begin{figure}[!htb]
  \centering
  \begin{subfigure}{.49\textwidth}
    \includegraphics[width=\linewidth]{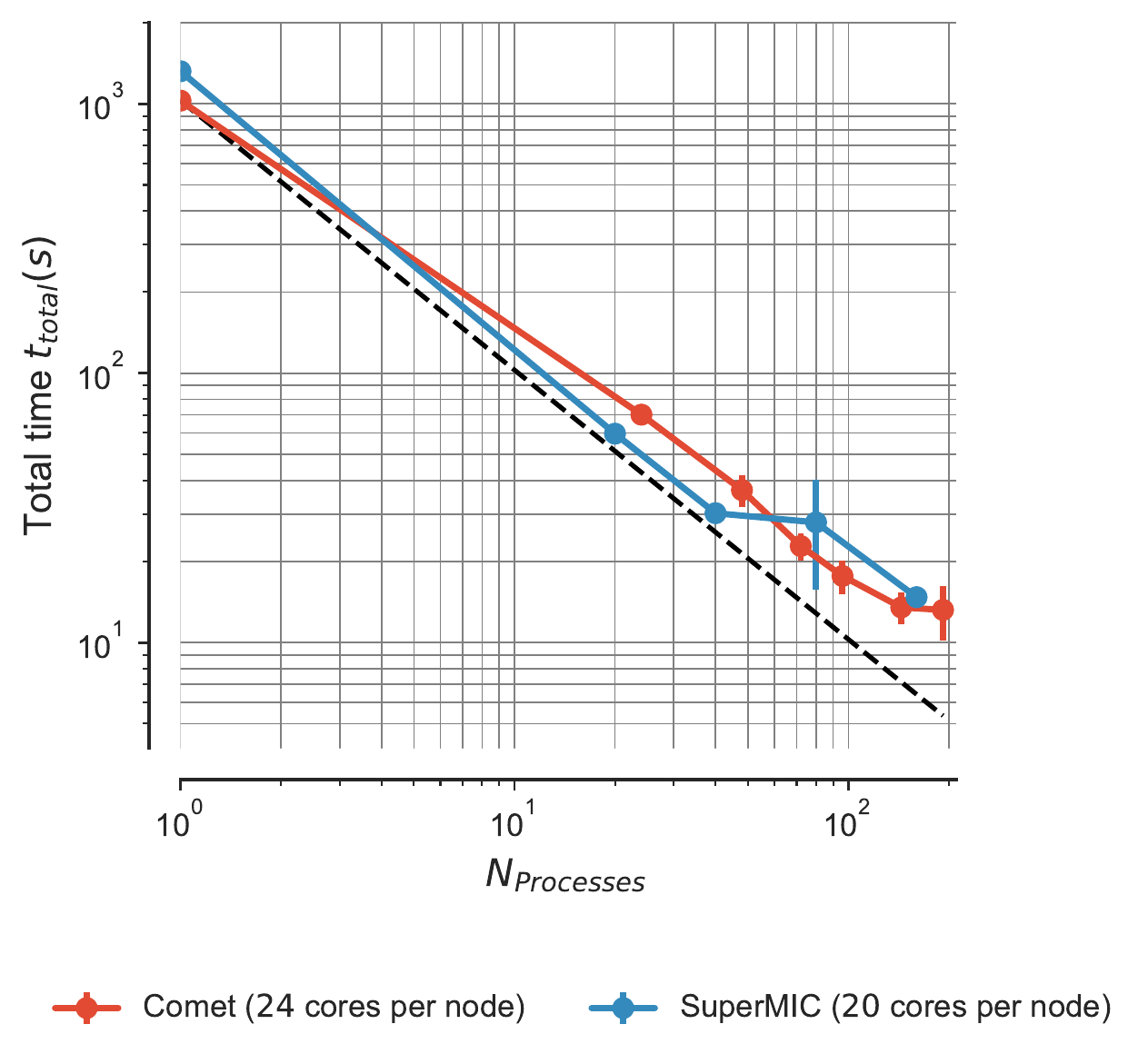}
    \caption{Scaling total}
    \label{fig:MPIscaling-clusters-splitting}
  \end{subfigure}
  \hfill
  \begin{subfigure}{.49\textwidth}
    \includegraphics[width=\linewidth]{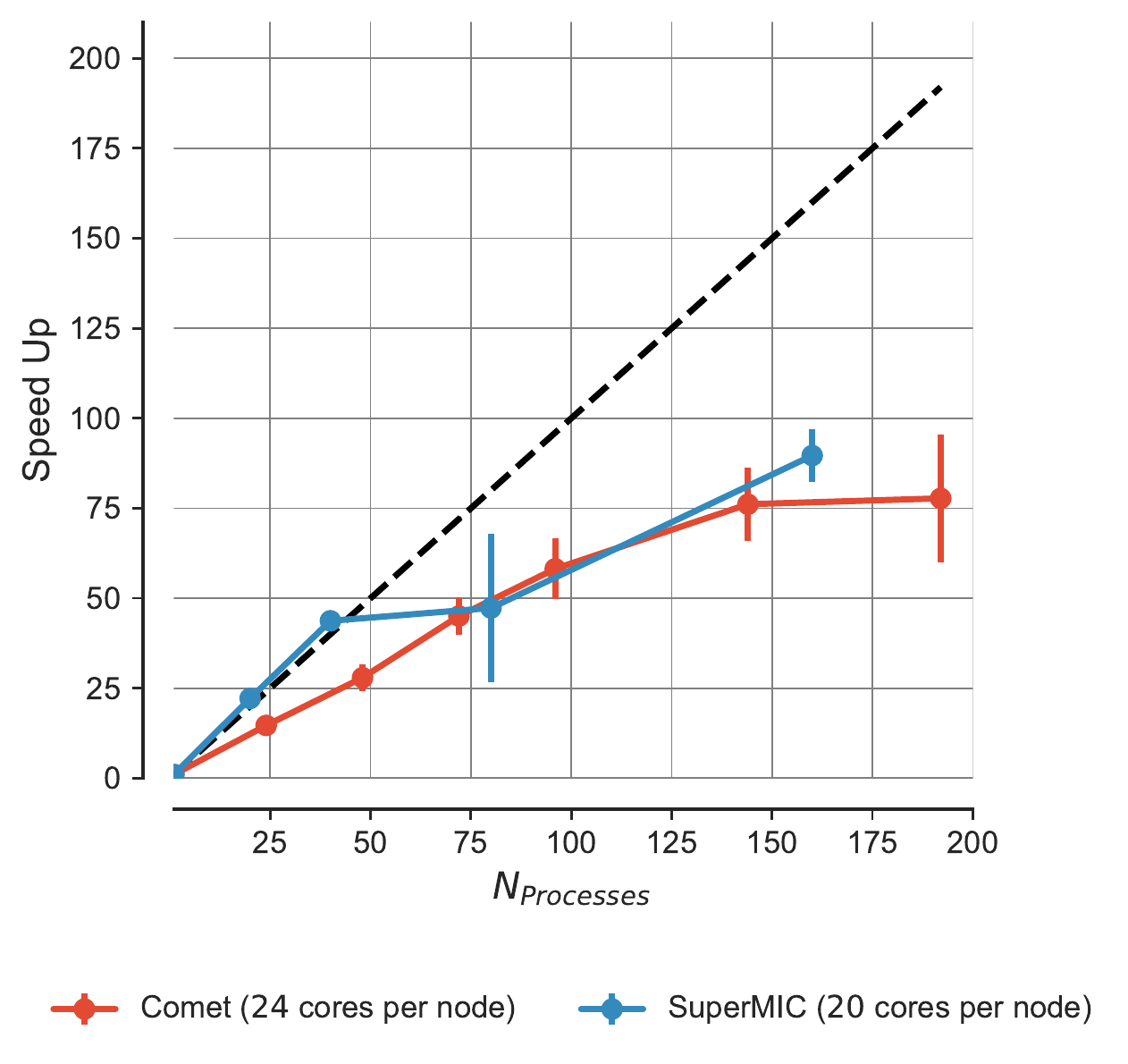}
    \caption{Speed-up}
    \label{fig:MPIspeedup-clusters-splitting}
  \end{subfigure}
  \bigskip

  \begin{subfigure}{0.7\textwidth}
    \includegraphics[width=\linewidth]{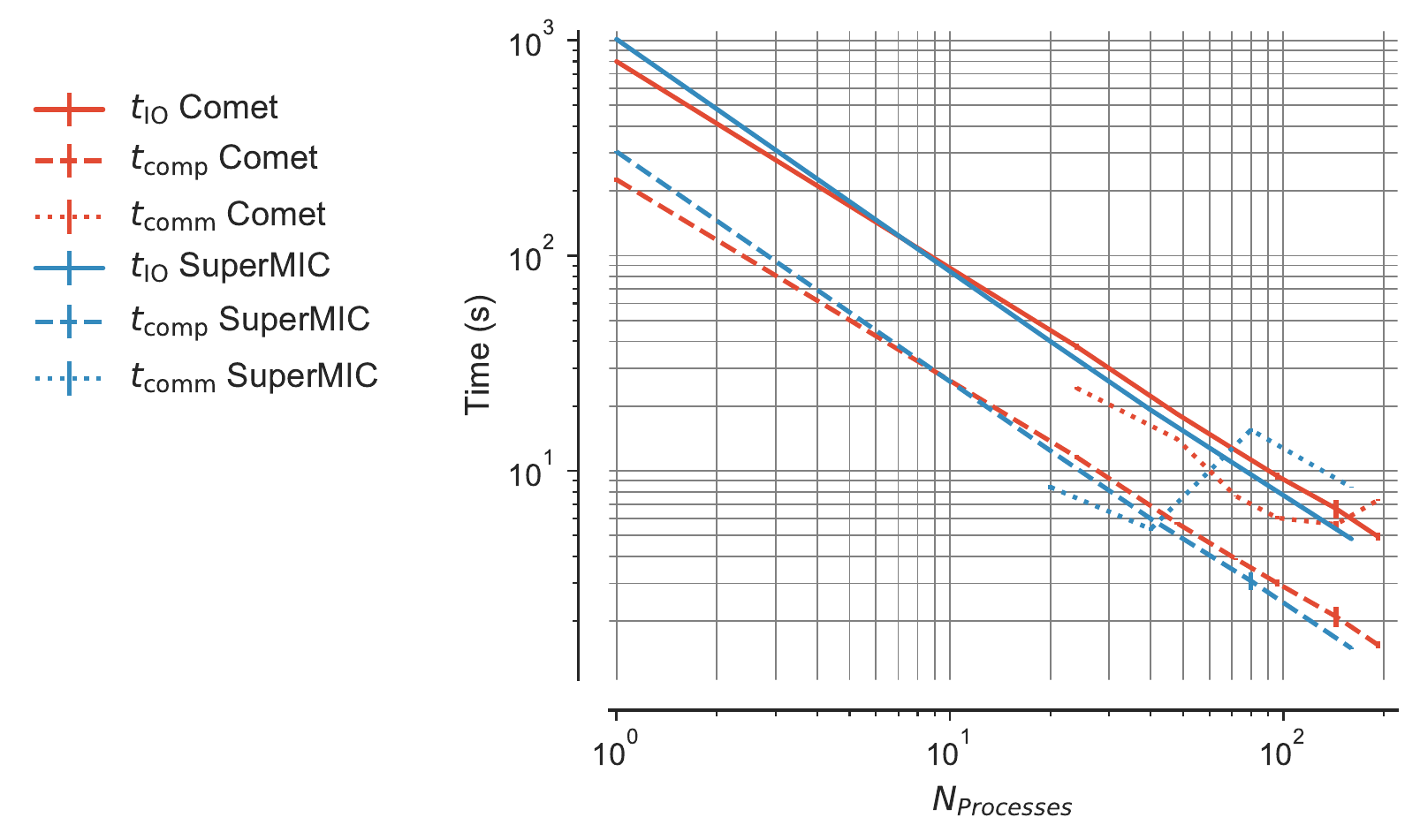}
    \captionsetup{format=hang}
    \caption{Scaling of \tcomp and \tIO.}
    \label{fig:compute-IO-scaling-clusters-splitting}
  \end{subfigure}
  \caption{Comparison of the performance of the RMSD task across different clusters [\emph{SDSC Comet} (24 cores per node), \emph{LSU SuperMIC} (20 cores per node)] when the trajectories are split (\emph{subfiling}).
    Five repeats were performed to collect statistics.
    The error bars show the standard deviation with respect to the mean.}
\label{fig:MPI-splitting-clusters}
\end{figure}

Figure \ref{fig:MPI-splitting-clusters} shows the strong scaling of the RMSD task on \emph{LSU SuperMIC} and \emph{SDSC Comet} with subfiling.
The results were comparable between the two clusters, with scaling far from ideal due to the communication cost (see section \ref{sec:splitting-traj} and Figure \ref{fig:compute-IO-scaling-clusters-splitting}).
On \emph{SuperMIC}, scaling was excellent on up to two nodes (40 cores) but beyond two nodes the communication cost increased markedly for unknown reasons and thus leading to reduced scaling behavior.
Overall, the scaling of the RMSD task was slighly better on \emph{LSU SuperMIC} than on \emph{SDSC Comet} but qualitatively similar.

\subsection{MPI-based Parallel HDF5}

\begin{figure}[!htb]
  \centering
  \begin{subfigure}{.49\textwidth}
    \includegraphics[width=\linewidth]{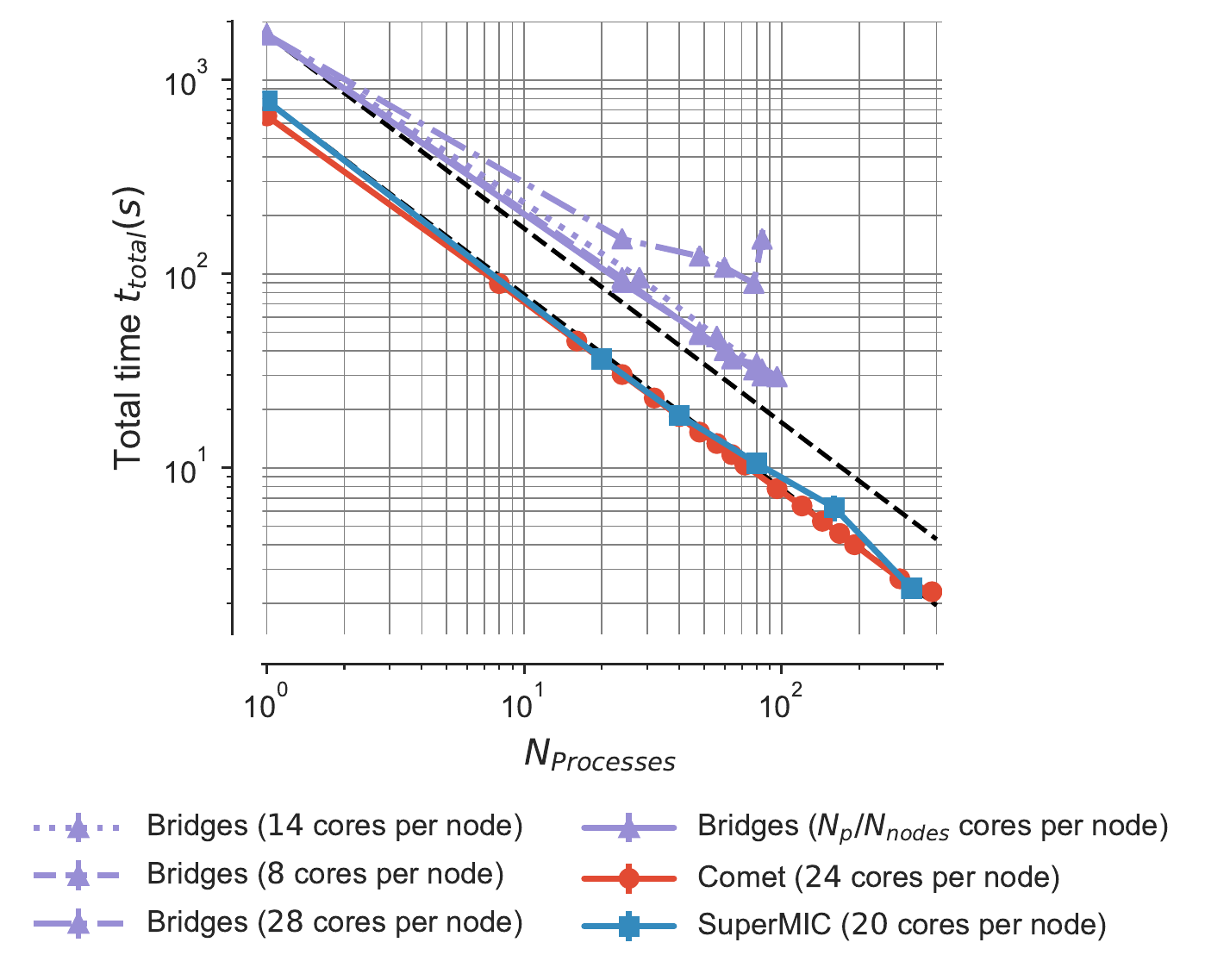}
    \caption{Scaling total}
    \label{fig:MPIscaling-clusters}
  \end{subfigure}
  \hfill
  \begin{subfigure}{.49\textwidth}
    \includegraphics[width=\linewidth]{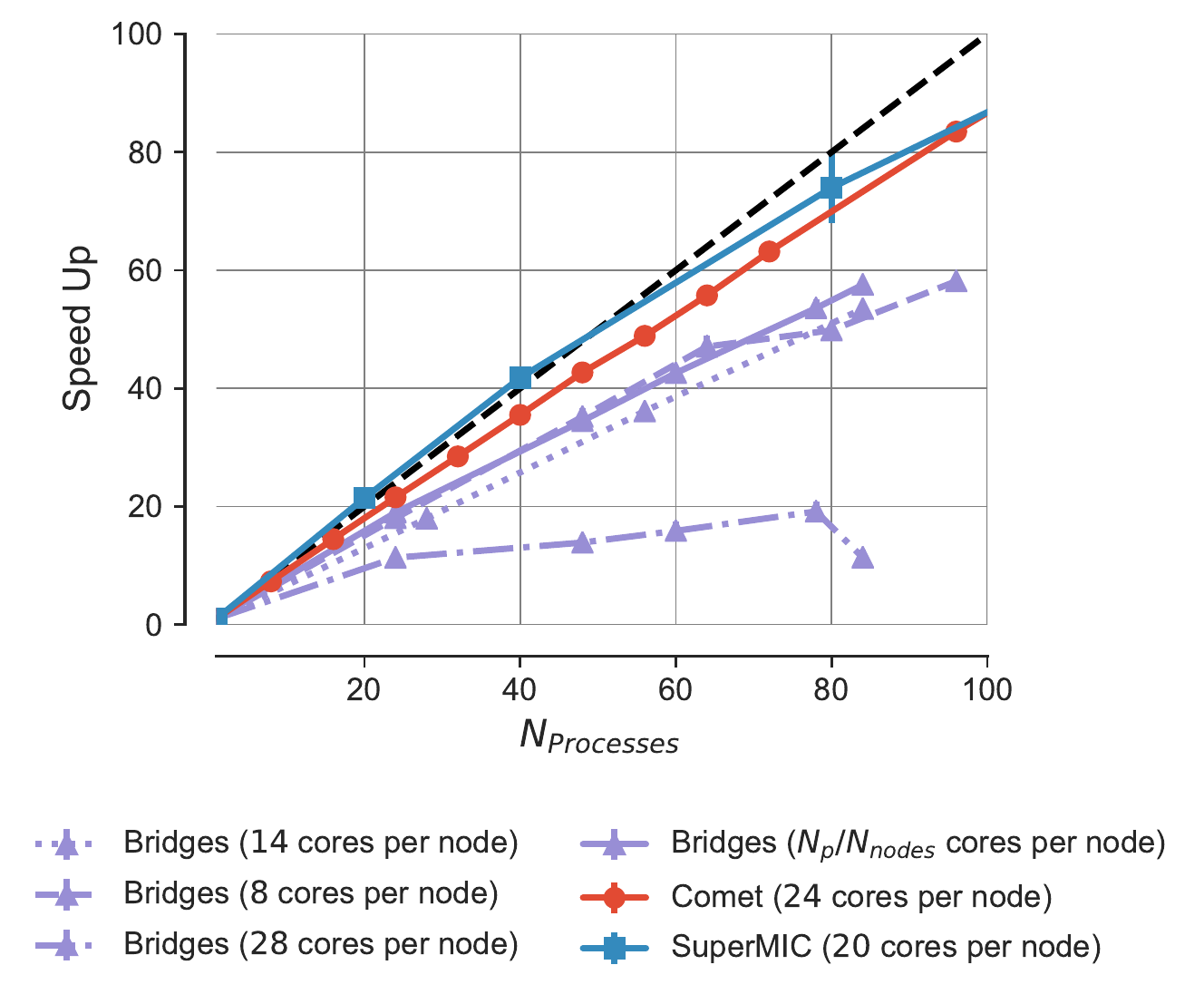}
    \caption{Speed-up}
    \label{fig:MPIspeedup-clusters}
  \end{subfigure}
  \bigskip
  
  \begin{subfigure} {\textwidth}
    \includegraphics[width=\linewidth]{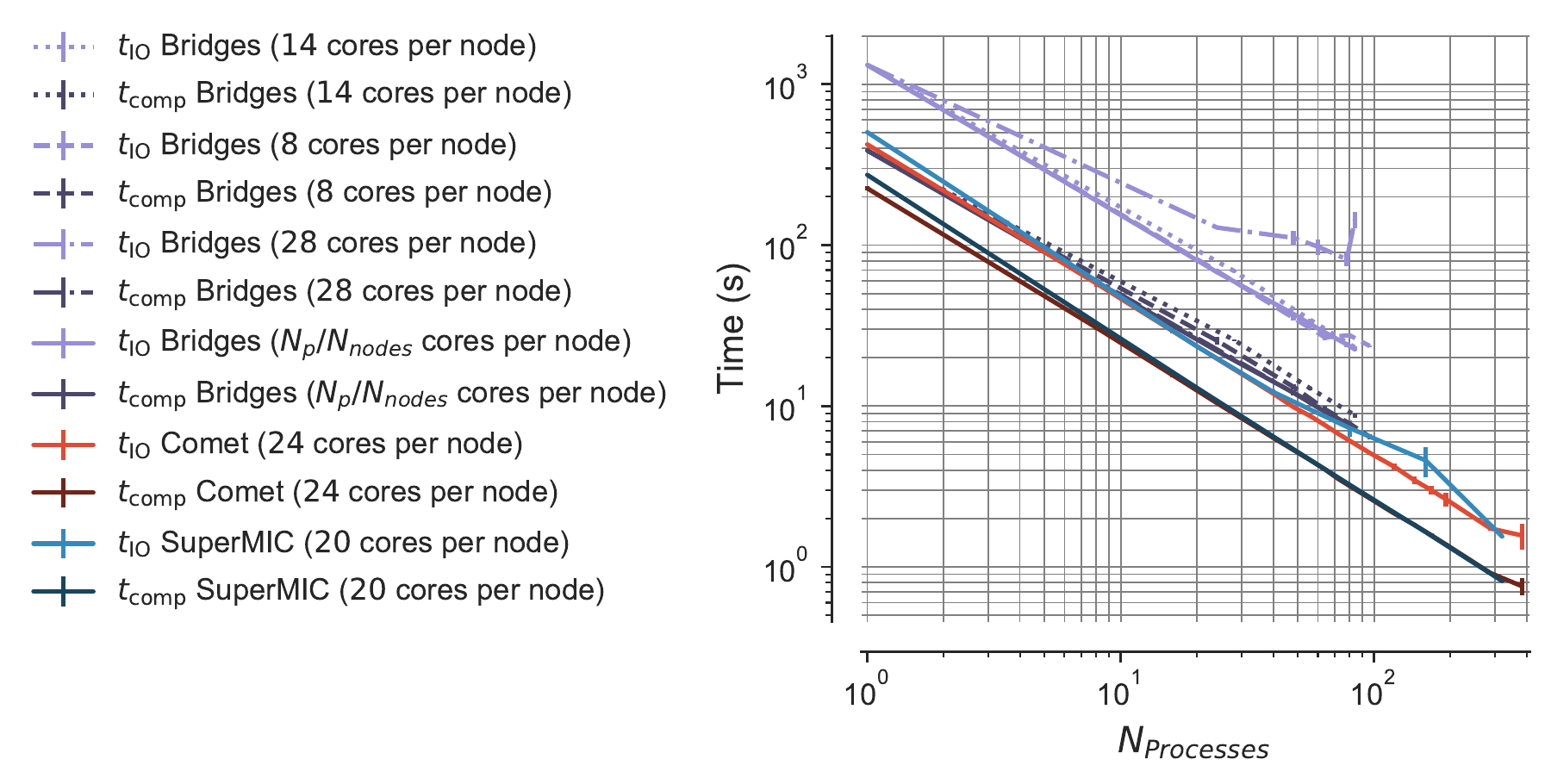}
    \captionsetup{format=hang}
    \caption{Scaling of \tcomp and \tIO}
    \label{fig:compute-IO-scaling-clusters}
  \end{subfigure}
  \caption{Comparison of the performance of the RMSD task across different clusters (\emph{SDSC Comet}, \emph{PSC Bridges}, \emph{LSU SuperMIC}) with MPI-IO.
    Data were read from a shared HDF5 file instead of an XTC file, using MPI independent I/O in the PHDF5 library.
    $N_{\text{P}}/N_{\text{nodes}}$ indicates that number of processes used for the task were equally distributed over all compute nodes.
    Five repeats were performed to collect statistics (except for \emph{SDSC Comet} at 288 cores, where only four repeats were included, as described in Figure~\protect\ref{fig:MPIwithIO-hdf5}).
    The error bars show standard deviation with respect to mean.
    In (b) only results up to 100 cores are shown to simplify the comparison; see Figure~\protect\ref{fig:MPIspeedup-hdf5} for \emph{SDSC Comet} and Figure~\protect\ref{fig:comparison_efficiency_SuperMIC} for \emph{LSU SuperMic} all data.
  }
\label{fig:MPIwithIO-clusters}
\end{figure}

Figure \ref{fig:MPIwithIO-clusters} shows the scaling on \emph{SDSC Comet}, \emph{LSU SuperMIC}, and \emph{PSC Bridges} using MPI-based parallel HDF5.  
Performance on \emph{SDSC Comet} and \emph{LSU SuperMIC} was very good with near ideal linear strong scaling.
The performance on \emph{PSC Bridges} was sensitive to how many cores per node were used.
Using all 28 cores in a node resulted in poor performance but decreasing the number of cores per node and equally distributing processes over nodes improved the scaling (Figure \ref{fig:MPIwithIO-clusters}), mainly by reducing variation in the I/O times.

\begin{figure}[!htb]
  \centering
  \begin{subfigure}{.49\textwidth}
    \includegraphics[width=\linewidth]{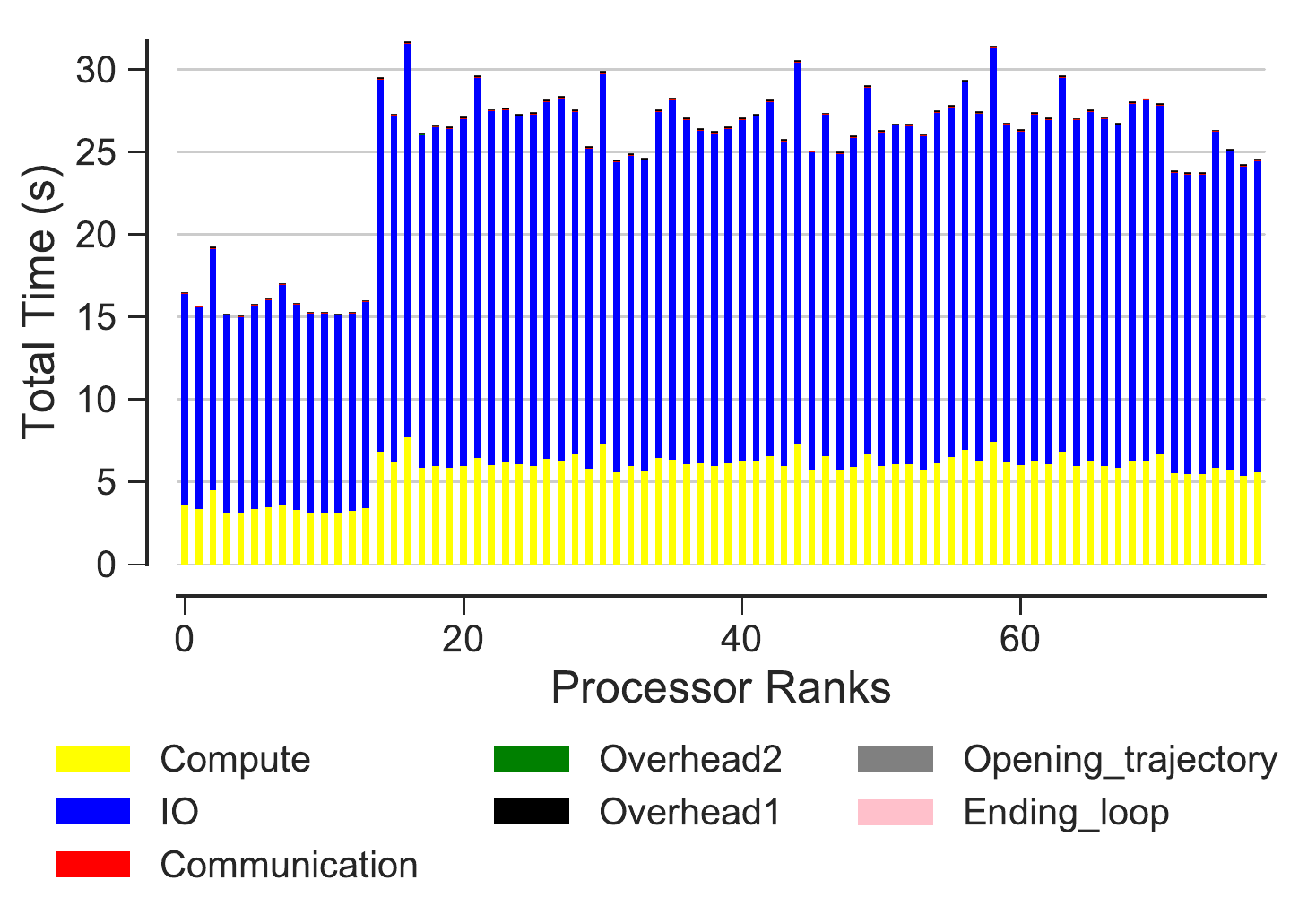}
    \caption{\emph{PSC Bridges}}
    \label{fig:hdf5-bridge}
  \end{subfigure}
  \bigskip
  \begin{subfigure} {.49\textwidth}
    \includegraphics[width=\linewidth]{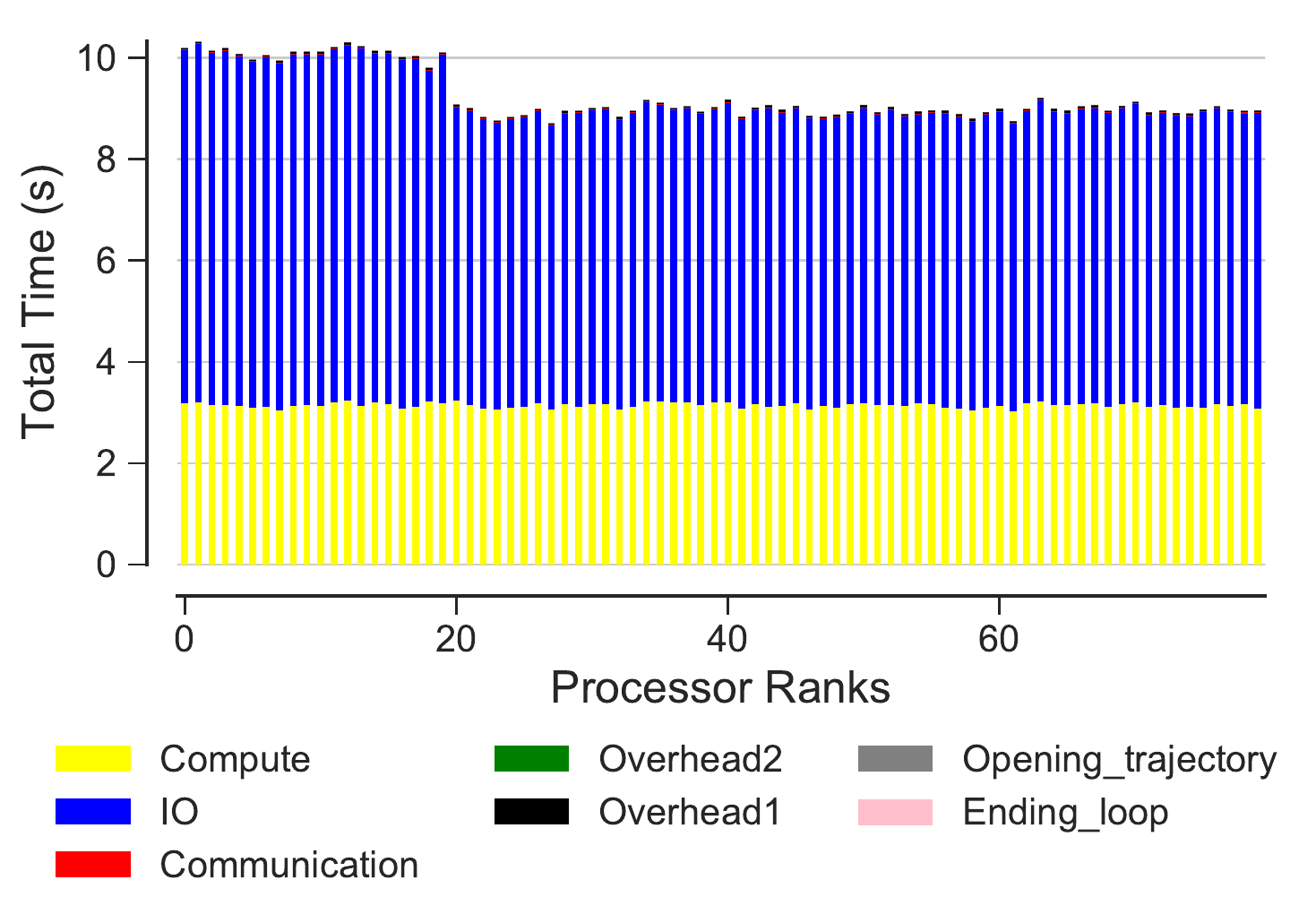}
    \caption{\emph{LSU SuperMIC}}
    \label{fig:hdf5-SuperMIC}
  \end{subfigure}
  \caption{Examples of timing per MPI rank for the RMSD task with MPI-based parallel HDF5 on (a) \emph{PSC Bridges} and (b) \emph{LSU SuperMIC}.
    Five repeats were performed to collect statistics and these were typical data from one run of the five repeats. Compute \tcomp, read I/O \tIO, communication \tcomm, ending the for loop $t_{\text{end\_loop}}$,  opening the trajectory $t_{\text{opening\_trajectory}}$, and overheads $t_{\text{overhead1}}$, $t_{\text{overhead2}}$ per MPI rank; see Table \ref{tab:notation} for definitions.}
  \label{fig:MPIwithIO-clusters-rank}
\end{figure}

The main difference between the runs on \emph{PSC Bridges} and \emph{SDSC Comet}/\emph{LSU SuperMIC} appeared to be the variance in \tIO (Figure \ref{fig:compute-IO-scaling-clusters}).
The I/O time distribution was fairly small and uniform across all ranks on \emph{SDSC Comet} and \emph{LSU SuperMIC} (Figures \ref{fig:hdf5-SuperMIC} and \ref{fig:MPIranks-hdf5}).
However, on \emph{PSC Bridges} the I/O time was on average about two and a half times larger and the I/O time distribution was also more variable across different ranks (Figure \ref{fig:hdf5-bridge}).  

\subsection{Comparison of Compute and I/O Scaling Across Different Clusters}
A full comparison of compute and I/O scaling across different clusters for different test cases and algorithms is shown in Table \ref{tab:comp-IO-scaling}. 
For MPI-based parallel HDF5, both the compute and I/O time on \emph{Bridges} were consistently larger than their corresponding values on \emph{SDSC Comet} and \emph{LSU SuperMIC}.
For example, with one core the corresponding compute and I/O time were $\tcomp = 387~\text{s}$, $\tIO = 1318~\text{s}$ versus $225~\text{s}$, $423~\text{s}$ on \emph{SDSC Comet} and $273~\text{s}$, $503~\text{s}$ on \emph{LSU SuperMIC}.
This performance difference became larger with increasing core number.

Overall, the results from \emph{SDSC Comet} and \emph{LSU SuperMIC} were consistent with each other.
Performance on \emph{PSC Bridges} seemed sensitive to the exact allocation of cores on each node but nevertheless the approaches that decreased the occurrence of stragglers on \emph{SDSC Comet} and \emph{LSU SuperMIC} also improved performance on \emph{PSC Bridges}.
Thus, the findings described in the previous sections were valid for a range of different HPC clusters with Lustre file systems.

\begin{sidewaystable}[hp]
\centering
\caption
{Comparison of selected compute and I/O timings for different clusters, test cases, and number of processes.
  Five repeats were performed to collect statistics.
  The mean value and the standard deviation with respect to mean are reported for each case.}
\label{tab:comp-IO-scaling}  
\begin{adjustbox}{max width=\textwidth}
\begin{tabular}{c c c c c c c c c c c c}
  \toprule
 \multicolumn{10}{r}{\bfseries $N_{Processes}$} \\
\cmidrule(r){5-12}
           \bfseries\thead{Cluster} & \bfseries\thead{Gather} & \bfseries\thead{File Access} & \bfseries\thead{Time} & \bfseries\thead{Serial} & \begin{tabular}{c} \bfseries\thead{Comet:} 24 \\ \bfseries\thead{Bridges:} 24 \\ \bfseries\thead{SuperMIC:} 20 \end{tabular}  & \begin{tabular}{c} \bfseries\thead{Comet:} 48 \\ \bfseries\thead{Bridges:} 48 \\ \bfseries\thead{SuperMIC:} 40 \end{tabular} & \begin{tabular}{c} \bfseries\thead{Comet:} 72 \\ \bfseries\thead{Bridges:} 60 \\ \bfseries\thead{SuperMIC:} 80 \end{tabular} & \begin{tabular}{c} \bfseries\thead{Comet:} 96 \\ \bfseries\thead{Bridges:} 78 \end{tabular} & \begin{tabular}{c} \bfseries\thead{Comet:} 144 \\ \bfseries\thead{Bridges:} 84 \\ \bfseries\thead{SuperMIC:} 160\end{tabular} & \bfseries\thead{Comet: 192} & \begin{tabular}{c} \bfseries\thead{Comet:} 384 \\  \bfseries\thead{SuperMIC:} 320\end{tabular}\\
  \midrule
    Comet & MPI & Single & \begin{tabular}{c} \tIO \\ \tcomp  \end{tabular} & \begin{tabular}{c} $791 \pm 5.22$ \\ $225 \pm 5.4$ \end{tabular} & \begin{tabular}{c} $49 \pm 3.45$ \\ $11 \pm 0.75$ \end{tabular} & \begin{tabular}{c} $29 \pm 1.3$ \\ $6 \pm 0.35$ \end{tabular} & \begin{tabular}{c} $26 \pm 9.19$ \\ $4 \pm 0.48$ \end{tabular} & -- & -- & -- & --\\  
  \midrule
    Bridges & MPI & Single & \begin{tabular}{c} \tIO \\ \tcomp  \end{tabular} & \begin{tabular}{c} $770 \pm 10.8$ \\ $221 \pm 3.9$ \end{tabular} &  \begin{tabular}{c} $38 \pm 0.84$ \\ $11 \pm 0.43$ \end{tabular} & \begin{tabular}{c} $33 \pm 19.4$ \\ $6 \pm 0.32$ \end{tabular} & \begin{tabular}{c} $15 \pm 1.6$ \\ $4 \pm 0.18$ \end{tabular} & -- & -- & -- & --\\  
  \midrule
    SuperMIC & MPI & Single & \begin{tabular}{c} \tIO \\ \tcomp  \end{tabular} & \begin{tabular}{c} $1014.51 \pm 2.94$ \\ $303.85 \pm2.3$ \end{tabular} & \begin{tabular}{c} $48.08 \pm 0.35$ \\ $14.56 \pm 0.14$ \end{tabular} & \begin{tabular}{c} $24.5 \pm 0.79$ \\ $7.4 \pm 0.25$ \end{tabular} & \begin{tabular}{c} $12 \pm 0.31$ \\ $3.7 \pm 0.12$ \end{tabular} & -- & \begin{tabular}{c} $6.24 \pm 0.38$ \\ $1.8 \pm 0.04$ \end{tabular} & -- & --\\  
  \midrule
    Comet & MPI & Splitting & \begin{tabular}{c} \tIO \\ \tcomp \end{tabular} & \begin{tabular}{c} $799 \pm 5.22$ \\ $225 \pm 5.4$ \end{tabular} & \begin{tabular}{c} $37 \pm 1.22$ \\ $11 \pm 0.31$ \end{tabular} & \begin{tabular}{c} $18 \pm 0.18$ \\ $5 \pm 0.07$ \end{tabular} & \begin{tabular}{c} $12 \pm 0.14$ \\ $3 \pm 0.04$ \end{tabular} & \begin{tabular}{c} $9 \pm 0.3$ \\ $3 \pm 0.11$ \end{tabular} & \begin{tabular}{c} $6  \pm 0.66$ \\ $2 \pm 0.23$ \end{tabular} & \begin{tabular}{c} $4 \pm 0.23$ \\ $1 \pm 0.07$ \end{tabular} & --\\
  \midrule
    SuperMIC &  MPI & Splitting & \begin{tabular}{c} \tIO \\ \tcomp \end{tabular} & \begin{tabular}{c} $1013.75 \pm 2.8$ \\ $304.26 \pm 2.55$ \end{tabular} & \begin{tabular}{c} $39.99 \pm 0.36$ \\ $12.41 \pm 0.22$ \end{tabular} & \begin{tabular}{c} $19.18 \pm 0.25$ \\ $5.99\pm 0.09$ \end{tabular} & \begin{tabular}{c} $9.61 \pm 0.28$ \\ $3.08 \pm 0.13$ \end{tabular} & -- & \begin{tabular}{c} $4.83 \pm 0.06$ \\ $1.5 \pm 0.01$ \end{tabular} & --& --\\ 
  \midrule 
    Comet & MPI & PHDF5 & \begin{tabular}{c} \tIO \\ \tcomp \end{tabular} & \begin{tabular}{c} $423 \pm 5.88$ \\ $225 \pm 6.55$ \end{tabular} & \begin{tabular}{c} $19 \pm 0.3$ \\ $10 \pm 0.12$ \end{tabular} & \begin{tabular}{c} $9 \pm 0.13$ \\ $5 \pm 0.1$ \end{tabular} & \begin{tabular}{c} $6 \pm 0.06$ \\ $3 \pm 0.04$ \end{tabular} & \begin{tabular}{c} $5 \pm 0.12$ \\ $2 \pm 0.05$ \end{tabular} & \begin{tabular}{c} $3 \pm 0.2$ \\ $1 \pm 0.04$ \end{tabular} & \begin{tabular}{c} $3 \pm 0.25$\\ $1 \pm 0.03$ \end{tabular} & \begin{tabular}{c} $1.57 \pm 0.29$\\ $0.76 \pm 0.09$ \end{tabular}\\
  \midrule 
    Bridges & MPI & PHDF5 & \begin{tabular}{c} \tIO \\ \tcomp \end{tabular} & \begin{tabular}{c} $1318.87 \pm 10.42$ \\ $387.8 \pm 5.51$ \end{tabular} & \begin{tabular}{c} $67.93 \pm 0.52$ \\ $21.97 \pm 0.38$ \end{tabular} & \begin{tabular}{c} $37.37 \pm 0.2$ \\ $12.12 \pm 0.34$ \end{tabular} & \begin{tabular}{c} $30.35 \pm 0.15$ \\ $9.79 \pm 0.24$ \end{tabular} & \begin{tabular}{c} $24.16 \pm 0.89$ \\ $7.72 \pm 0.03$ \end{tabular} & \begin{tabular}{c} $22.5 \pm 0.17$ \\ $7.18 \pm 0.08$ \end{tabular} & -- & --\\ 
  \midrule
    SuperMIC & MPI & PHDF5 & \begin{tabular}{c} \tIO \\ \tcomp \end{tabular} & \begin{tabular}{c} $503.69 \pm 2.57$ \\ $273.54 \pm 4.7$ \end{tabular} & \begin{tabular}{c} $12.96 \pm 0.06$ \\ $23.44 \pm 0.29$ \end{tabular} & \begin{tabular}{c} $6.46 \pm 0.02$ \\ $12.22 \pm 0.43$ \end{tabular} & \begin{tabular}{c} $3.2 \pm 0.01$ \\ $7.3 \pm 0.85$ \end{tabular} & -- & \begin{tabular}{c} $1.64 \pm 0.01$ \\ $4.59 \pm 0.96$ \end{tabular} & --& \begin{tabular}{c} $0.82 \pm 0.004$ \\ $1.55 \pm 0.009$ \end{tabular} \\
  \bottomrule
\end{tabular}
\end{adjustbox}
\end{sidewaystable}


\section{Guidelines for Improving Parallel Trajectory Analysis Performance}
\label{sec:guidelines}

Although the performance measurements were performed with \package{MDAnalysis} and therefore capture some details of this library such as the specific timings for file reading, we believe that the broad picture is fairly general and applies to any Python-based approach that uses MPI for parallelizing trajectory access with a split-apply-combine approach.
Based on the lessons that we learned, we suggest the following guidelines to improve strong scaling performance:

Calculate the compute to I/O ratio ($\RcompIO$, Eq.~\ref{eq:Compute-IO}). As discussed in Section \ref{sec:bound}, for I/O bound problems the performance of the task will be affected by stragglers that delay job completion time.
\begin{description}
\item[\textbf{Heuristic 1}] For $\RcompIO \gg 1$, single-shared-file I/O can be used and one can expect reasonable scaling up to about 50 cores; for better scaling, one of the strategies under Heuristic 2 needs to be employed.
\item[\textbf{Heuristic 2}] For $\RcompIO \le 1$ the task is I/O bound and single-shared-file I/O should be avoided.
  One might want to consider the following steps:  
  \begin{description}
  \item[\textbf{Heuristic 2.1}] If there is access to the HDF5 format, use \textbf{MPI-based Parallel HDF5} (Section \ref{sec:HDF5}). This approach may scale well to hundreds of cores.
  \item[\textbf{Heuristic 2.2}] If the trajectory file is not in HDF5 format then one can consider \textbf{subfiling} and split the single trajectory file into as many trajectory segments as the number of processes. This approach may scale reasonably well to less than 200 cores.
  \end{description}
\end{description}

The better solution is the use of parallel I/O (\textbf{Heuristic 2.1}) as it makes best use of the parallel file system and scales well to hundreds of cores, regardless of $\RcompIO$. 
Splitting the trajectories will not scale as well as parallel I/O but it can be easily performed in parallel and trajectory conversion may be integrated into the beginning of standard workflows for MD simulations.  Alternatively, trajectories may already be kept in smaller chunks if they are already produced in batches; for instance, when running simulations with \package{Gromacs} \cite{Abraham:2015aa}, the \texttt{gmx mdrun -noappend} option produces individual trajectory segments instead of extending an existing trajectory file.


\section{Conclusions}
\label{sec:conclusions}

We analyzed the strong scaling performance of a typical task when analyzing MD trajectories, the calculation of the time series of the RMSD of a protein, with the widely used Python-based \package{MDAnalysis} library.
All benchmarks were performed in five replicates on three different XSEDE supercomputers to demonstrate that our results were independent from the specifics of the hardware and local environment.

The RMSD task was parallelized with MPI following the \emph{split-apply-combine} approach by having each MPI process analyze a contiguous segment of the trajectory.
This approach did not scale beyond a single node because straggler MPI processes exhibited large upward variations in runtime.
Stragglers were primarily caused by either increased MPI communication costs or increased time to open the single shared trajectory file whereas both the computation and the ingestion of data exhibited close to ideal strong scaling behavior.
Stragglers were less prevalent for compute-bound workloads (i.e., $\RcompIO \gg 1$), suggesting that file read I/O was responsible for poor MPI communication.
In particular, artificially removing all I/O substantially improved performance of the communication step and thus brought overall performance close to ideal (i.e., linear increase in speed-up with processor count with slope one), despite the fact that the amount of data to be communicated did not depend on I/O.
Our results suggested that stragglers might be due to the competition between MPI messages and the Lustre file system on the shared InfiniBand interconnect, which would be consistent with other similar observations \cite{VMD2013} and theoretical predictions by \citet{Brown:2018ab}, but further work would be needed to validate this specific hypothesis.
One possible interpretation of our results was that for a sufficiently large per-frame compute workload, read I/O interfered much less with communication than for an I/O bound task that almost continuously accesses the file system.
This interpretation suggested that the poor scaling performance was the result of inefficient use of the Lustre file system and that we needed to improve read I/O to reduce interference.

We investigated subfiling (splitting of the trajectories into separate files, one for each MPI rank) and MPI-based parallel I/O.
Subfiling improved scaling up to about 150 cores.
However, subfiling, at least in the form described here, is not an ideal solution because creating and accessing many small files on a parallel file system such as Lustre can negatively impact the overall performance of the file system.
Furthermore, managing a large number of files can become cumbersome and inflexible, given that the number of files determines the number of processes.
When we used MPI-based parallel I/O through HDF5 together with MPI for communications we achieved nearly ideal performance up to 384 cores (16 nodes on \emph{SDSC Comet}) and speed-ups of two orders of magnitude compared to the serial execution.
The latter approach appears to be a promising way forward as it directly builds on very widely used technology (MPI-IO and HDF5) and echoes the experience of the wider HPC community that parallel file I/O is necessary for efficient data handling.

The biomolecular simulation community suffers from a large number of trajectory file formats with very few being based on HDF5, with the exception of the H5MD format \cite{Buyl:2014aa} and the MDTraj HDF5 format \cite{McGibbon:2015aa}.
Our work suggests that HDF5-based formats should be seriously considered for MD simulations if users want to make efficient use of their HPC systems for analysis. 
Alternatively, enabling MPI-IO for trajectory readers in libraries such as \package{MDAnalysis} might also provide a path forward to better read performance.

We summarized our findings in a number of guidelines for improving the scaling of parallel analysis of MD trajectory data.
We showed that it is feasible to run an I/O bound analysis task on HPC resources with a Lustre parallel file system and achieve good scaling behavior up to 384 CPU cores with an almost 300-fold speed-up compared to serial execution.

\paragraph{Future Directions}

Future work might look into testing different MPI implementations, especially in combination with parallel HDF5.
Choosing the best performing MPI implementation for a specific system and optimizing the parallel file system parameters might also lead to further improvements.
Although our results showed qualitatively similar behavior on three different HPC resources, unexplained differences in performances remained.
Deeper insights into the system-level network traffic and parallel file system access would be necessary to approach performance tuning for different HPC systems in a rational manner.

Our HDF5 results are encouraging but lack a convenient and widely available implementation.
Therefore, a HDF5-based trajectory reader needs to be implemented in MDAnalysis for an existing HDF5 trajectory format.
The algorithm for the analysis task could be optimized by reducing file access to the shared system topology file (and any other data common to all ranks, such as the reference coordinates in the RMSD analysis) by using \texttt{MPI\_Scatter} and \texttt{MPI\_Gather} to efficiently communicate the static data.
In this case, only rank 0 would read these data from the file system and then scatter them to all other ranks.
Each rank would then build their own MDAnalysis \texttt{Universe} from those data (either by gathering a serialized \texttt{Universe} data structure or by using Python \texttt{StringIO} to read the scattered text buffer containing the topology file) and their own parallel file access to an HDF5 trajectory (with the \texttt{Universe.load\_new()} method to attach a trajectory).

In summary, the encouraging finding of this work is that by using parallel file reading (here tested with HDF5), the simple split-apply-combine single trajectory parallelization approach can work on current HPC systems up to a few hundred cores, even for I/O-bound tasks.
The major advantage of the approach is its simplicity as users can directly use their serial code and apply it to blocks of a trajectory, without having to rewrite their algorithms or having to consider hybrid parallelization schemes. 
Although we focused on the \package{MDAnalysis} library, similar strategies are likely to be more generally applicable and useful to the wider biomolecular simulation community.

\section*{Acknowledgements}
\label{acknowledgements}
\ack

We are grateful to Sarp Oral for insightful comments on this manuscript.
This work was supported by the National Science Foundation under grant numbers ACI-1443054 and ACI-1440677.
This work used the Extreme Science and Engineering Discovery Environment (XSEDE), which is supported by National Science Foundation grant number ACI-1548562.
\emph{SDSC Comet} at the San Diego Supercomputer Center, \emph{LSU SuperMic} at Louisiana State University, and \emph{PSC Bridges} at the Pittsburgh Supercomputing Center were used under allocations TG-MCB090174 and TG-MCB130177.

\bibliography{main}

\clearpage

\appendix

\section{Additional Data}
\label{sec:supplement}

\begin{figure}[!htb]
  \centering
  \begin{subfigure}{.4\textwidth}
    \includegraphics[width=\linewidth]{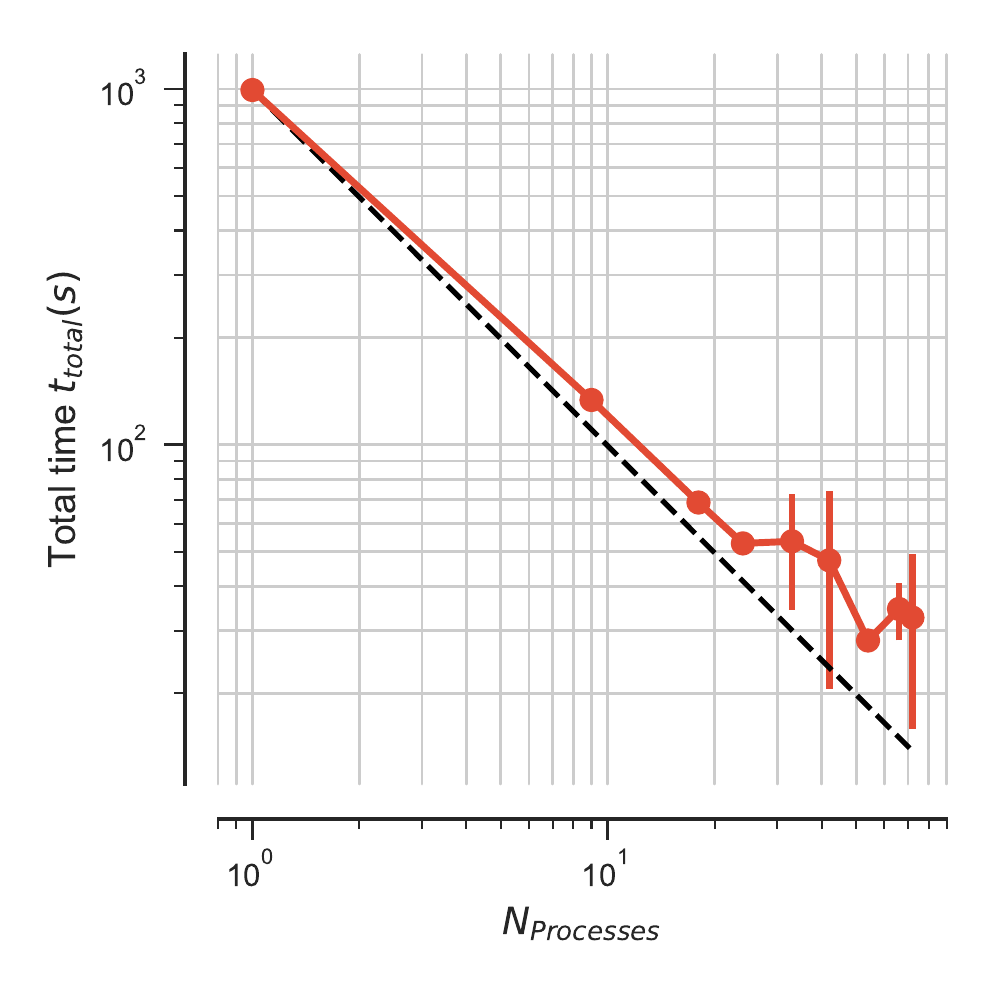}
    \caption{Scaling total}
    \label{fig:MPIscaling-Bridges}
  \end{subfigure}
  \hfill
  \begin{subfigure}{.4\textwidth}
    \includegraphics[width=\linewidth]{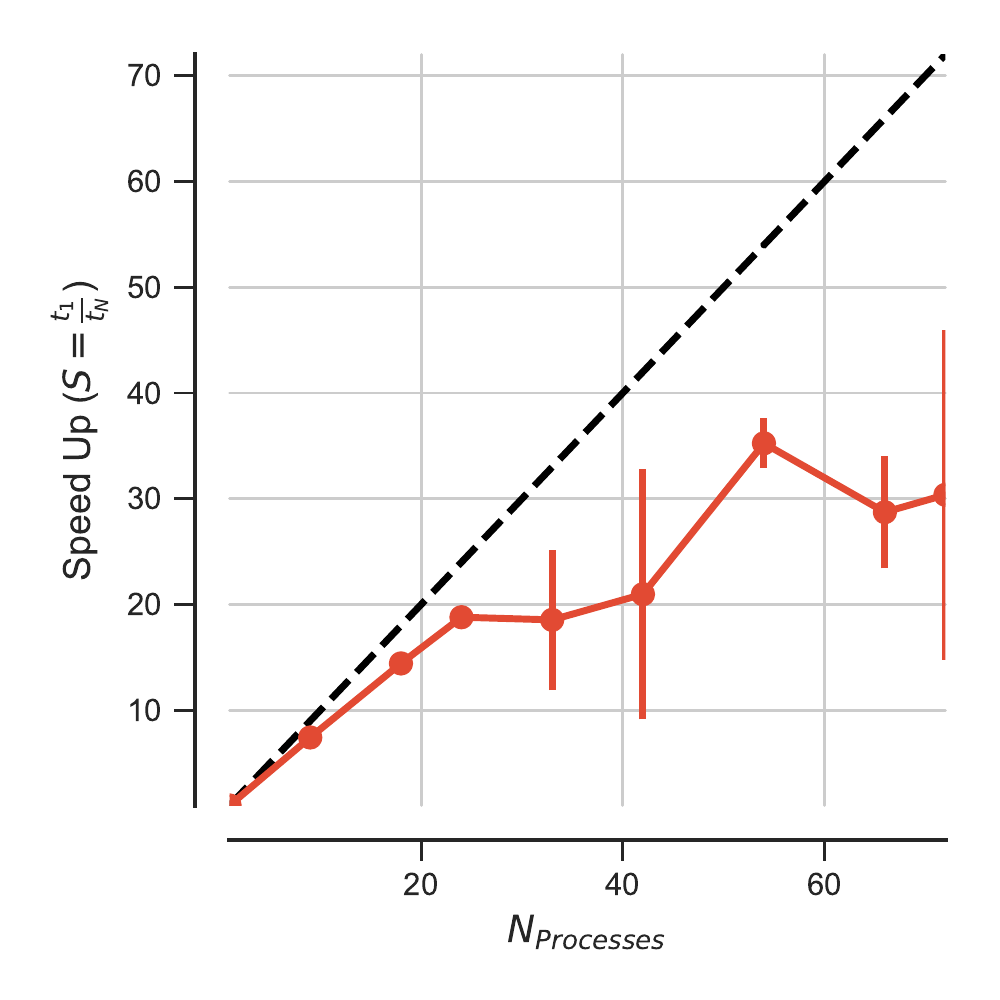}
    \caption{Speed-up}
    \label{fig:MPIspeedup-Bridges}
  \end{subfigure}
  \bigskip

  \begin{subfigure}{.45\textwidth}
    \includegraphics[width=\linewidth]{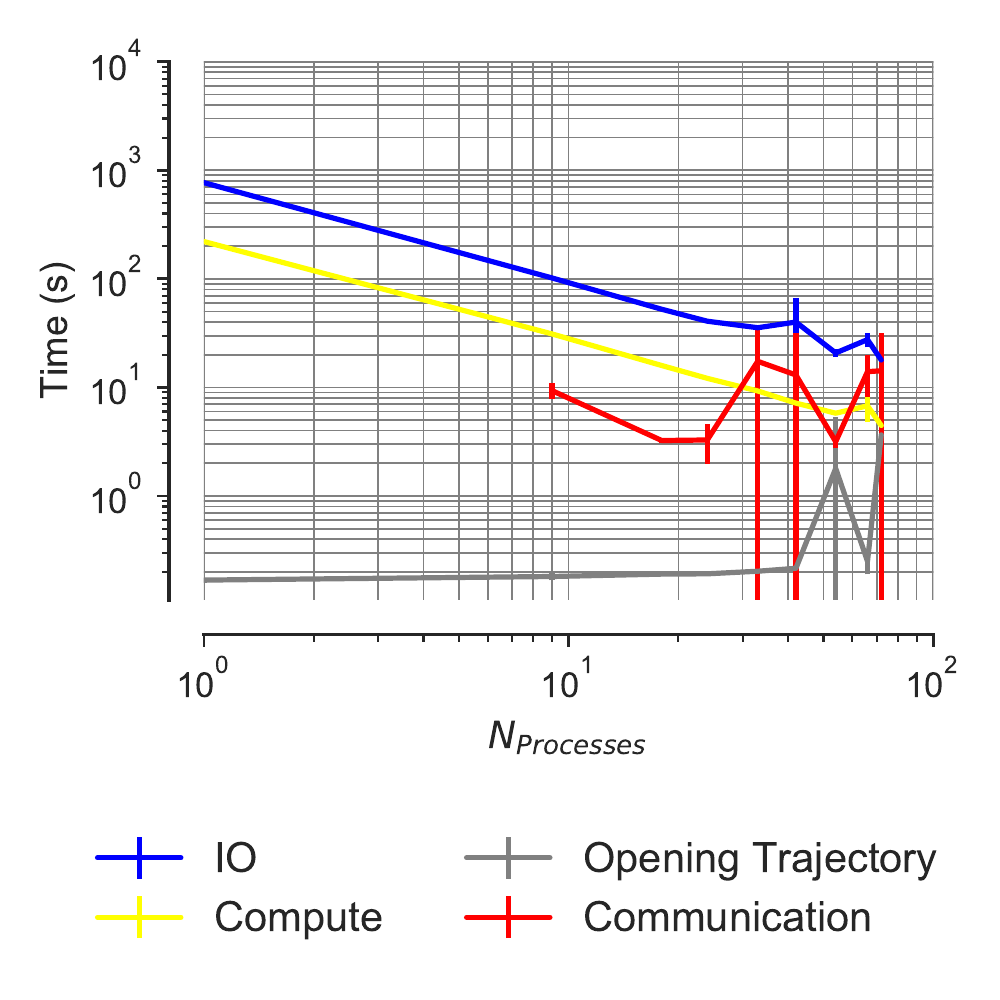}
    \captionsetup{format=hang}
    \caption{Scaling for different components}
    \label{fig:ScalingComputeIO-Bridges}
  \end{subfigure}
  \hfill
  \begin{subfigure} {.5\textwidth}
    \includegraphics[width=\linewidth]{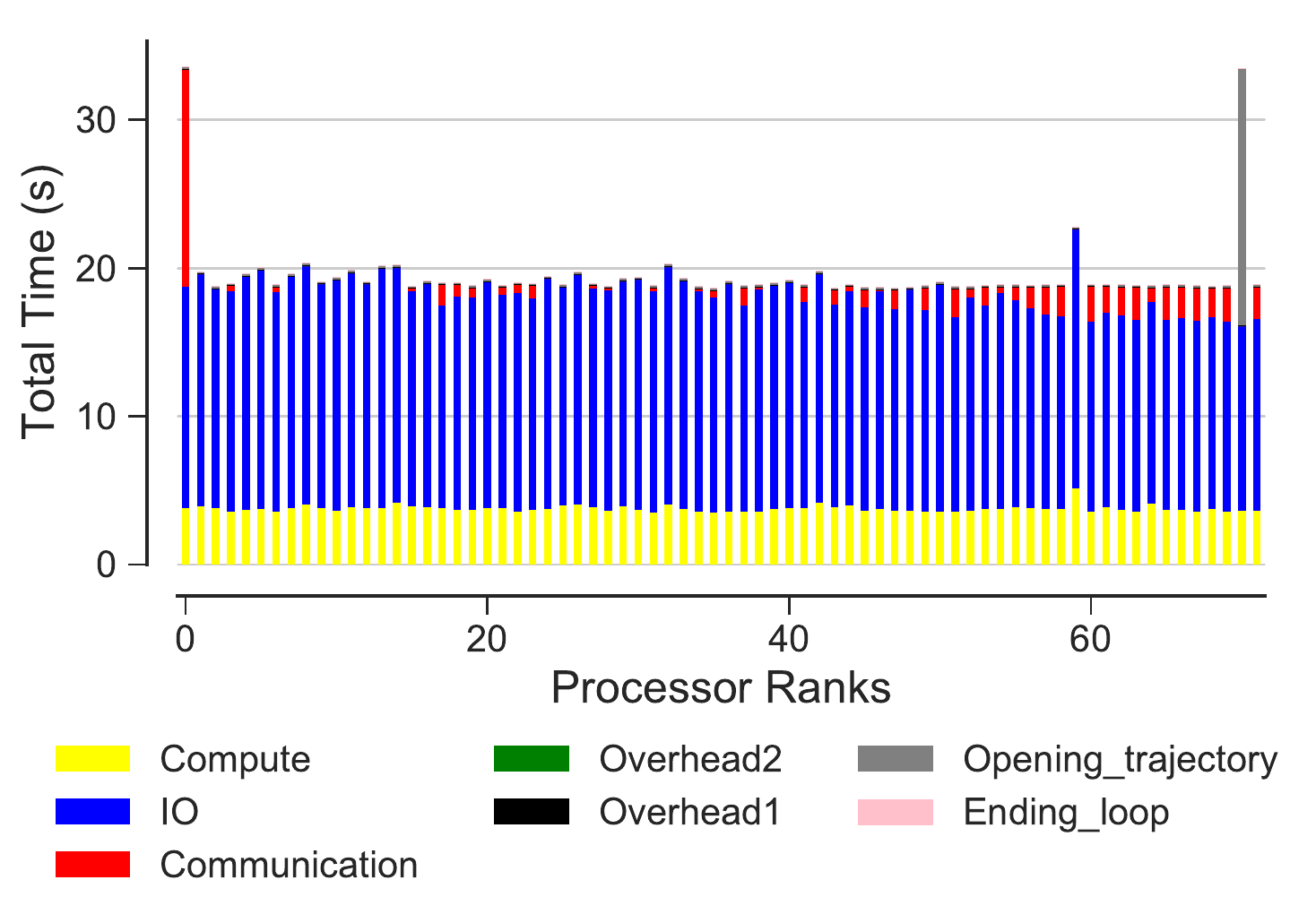}
    \captionsetup{format=hang}
    \caption{Time comparison on different parts of the calculations per MPI rank (example)}
    \label{fig:MPIranks-Bridges}
  \end{subfigure}
  \caption{\emph{PSC Bridges}: Performance of the RMSD task.
    Results are communicated back to rank 0.
    Five independent repeats were performed to collect statistics.
    (a-c) The error bars show standard deviation with respect to the mean.
    In serial, there is no communication and hence no data point is shown for $N=1$ in (c).
    (d) Compute \tcomp, read I/O \tIO, communication \tcomm, ending the for loop $t_{\text{end\_loop}}$, opening the trajectory $t_{\text{opening\_trajectory}}$, and overheads $t_{\text{overhead1}}$, $t_{\text{overhead2}}$ per MPI rank; see Table \ref{tab:notation} for definitions.
    These are data from one run of the five repeats.
    MPI ranks 0 and 70 are stragglers.
  }
\label{fig:MPIwithIO-Bridges}
\end{figure}

Figure \ref{fig:MPIwithIO-Bridges} shows performance of the RMSD task on \emph{PSC Bridges}. 

\begin{figure}[!htb]
  \centering
  \begin{subfigure}{.4\textwidth}
    \includegraphics[width=\linewidth]{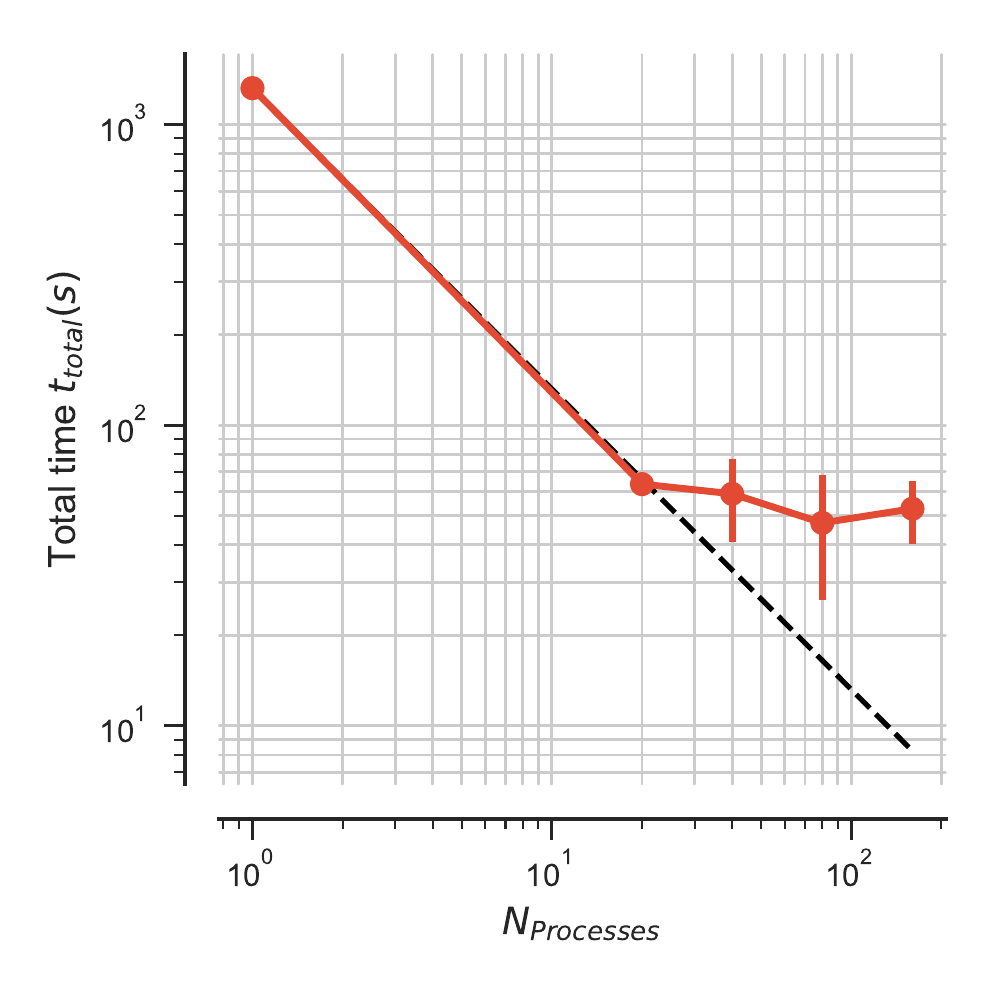}
    \caption{Scaling total}
    \label{fig:MPIscaling-SuperMIC}
  \end{subfigure}
  \hfill
  \begin{subfigure}{.4\textwidth}
    \includegraphics[width=\linewidth]{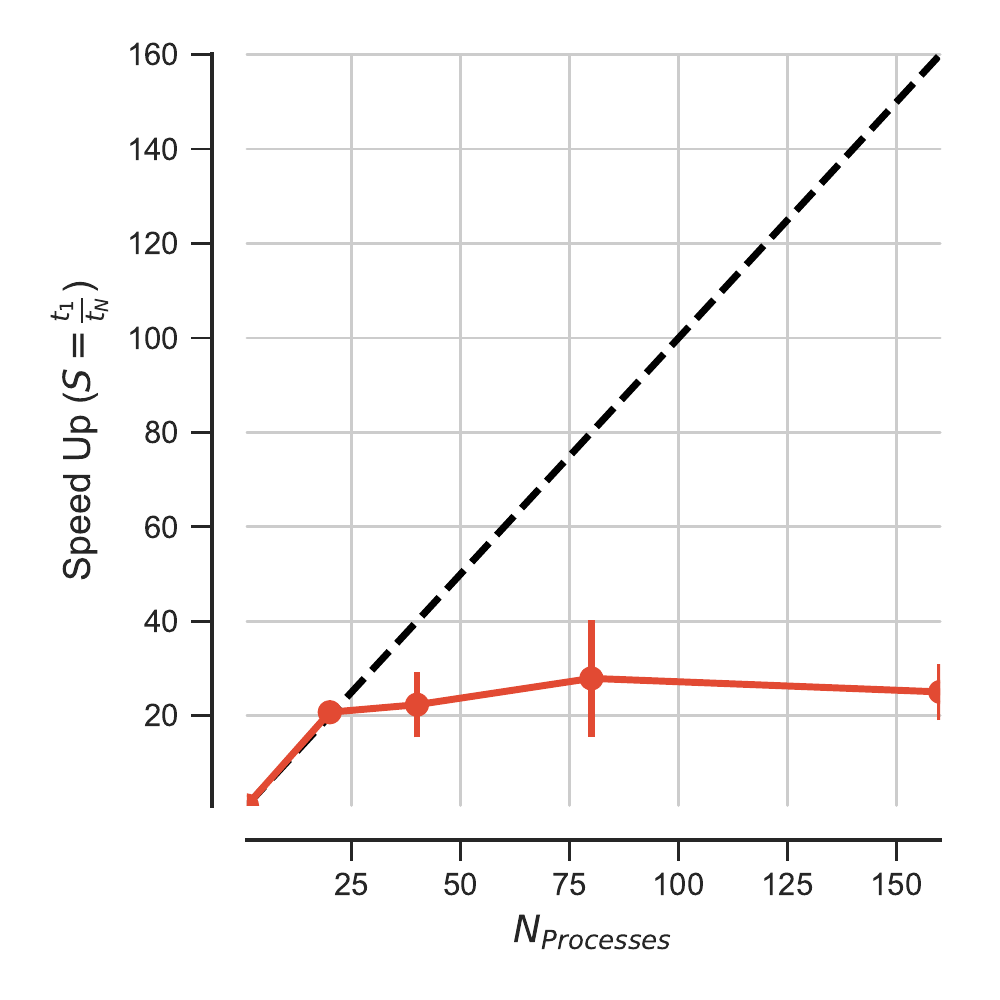}
    \caption{Speed-up}
    \label{fig:MPIspeedup-SuperMIC}
  \end{subfigure}
  \bigskip

  \begin{subfigure}{.45\textwidth}
    \includegraphics[width=\linewidth]{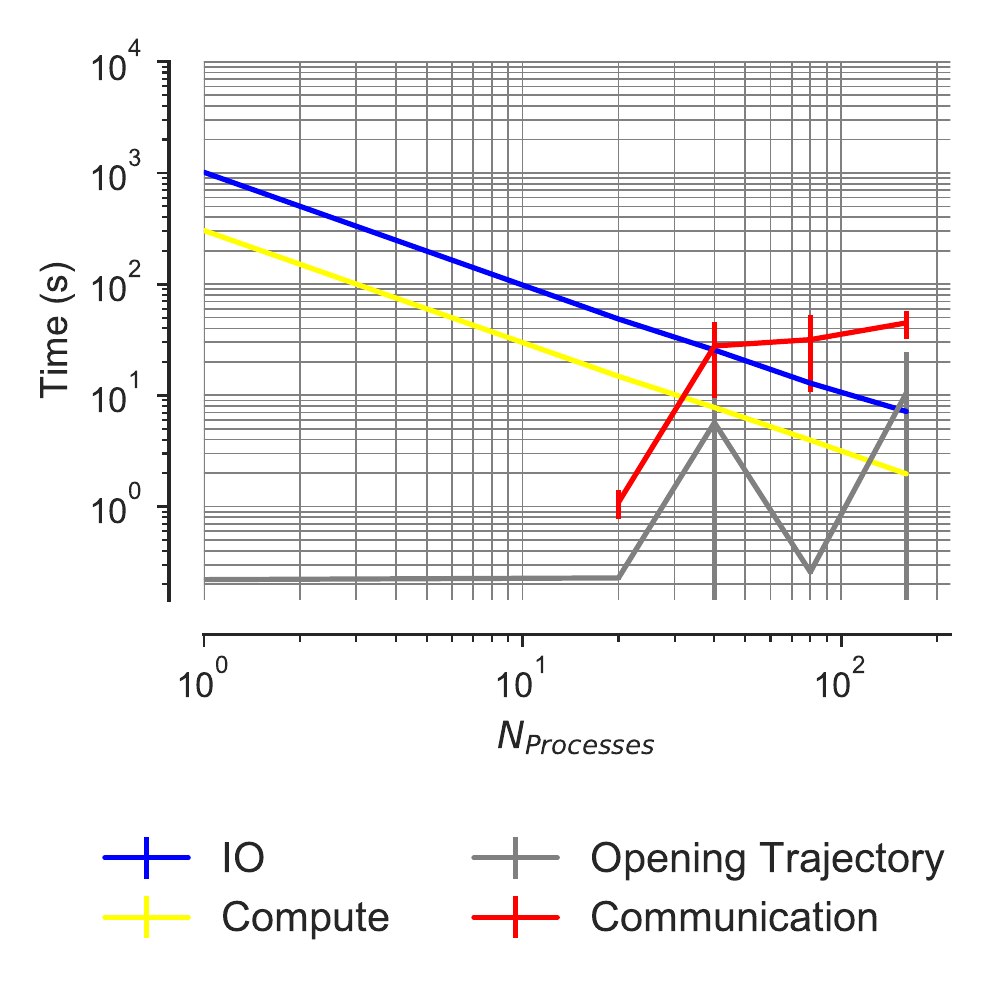}
    \captionsetup{format=hang}
    \caption{Scaling for different components}
    \label{fig:ScalingComputeIO-SuperMIC}
  \end{subfigure}
  \hfill
  \begin{subfigure} {.5\textwidth}
    \includegraphics[width=\linewidth]{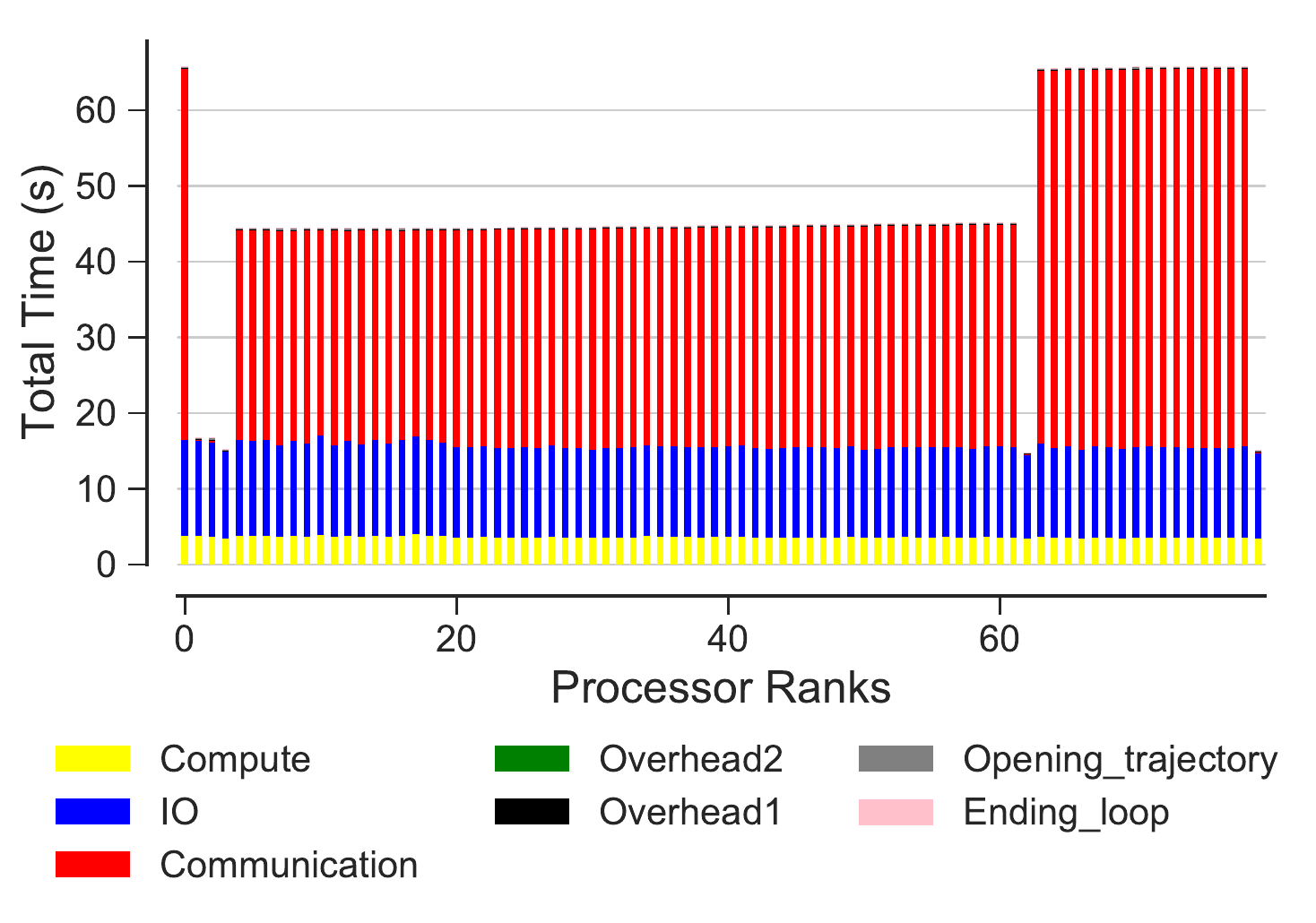}
    \captionsetup{format=hang}
    \caption{Time comparison on different parts of the calculations per MPI rank (example)}
    \label{fig:MPIranks-SuperMIC}
  \end{subfigure}
  \caption{\emph{LSU SuperMIC}: Performance of the RMSD task with MPI.
    Results are communicated back to rank 0.
    Five independent repeats were performed to collect statistics.
    (a-c) The error bars show standard deviation with respect to mean.
    In serial, there is no communication and hence the data points for $N=1$ are not shown in (c).
    (d) Compute \tcomp, read I/O \tIO, communication \tcomm, ending the for loop $t_{\text{end\_loop}}$,  opening the trajectory $t_{\text{opening\_trajectory}}$, and overheads $t_{\text{overhead1}}$, $t_{\text{overhead2}}$ per MPI rank; see Table \ref{tab:notation} for definitions.
    These are data from one run of the five repeats.
  }
  \label{fig:MPIwithIO-SuperMIC}
\end{figure}

Figure \ref{fig:MPIwithIO-SuperMIC} shows performance of the RMSD task on \emph{LSU SuperMIC}. 

\begin{figure}[!htb]
  \centering
  \begin{subfigure}{.3\textwidth}
    \includegraphics[width=\linewidth]{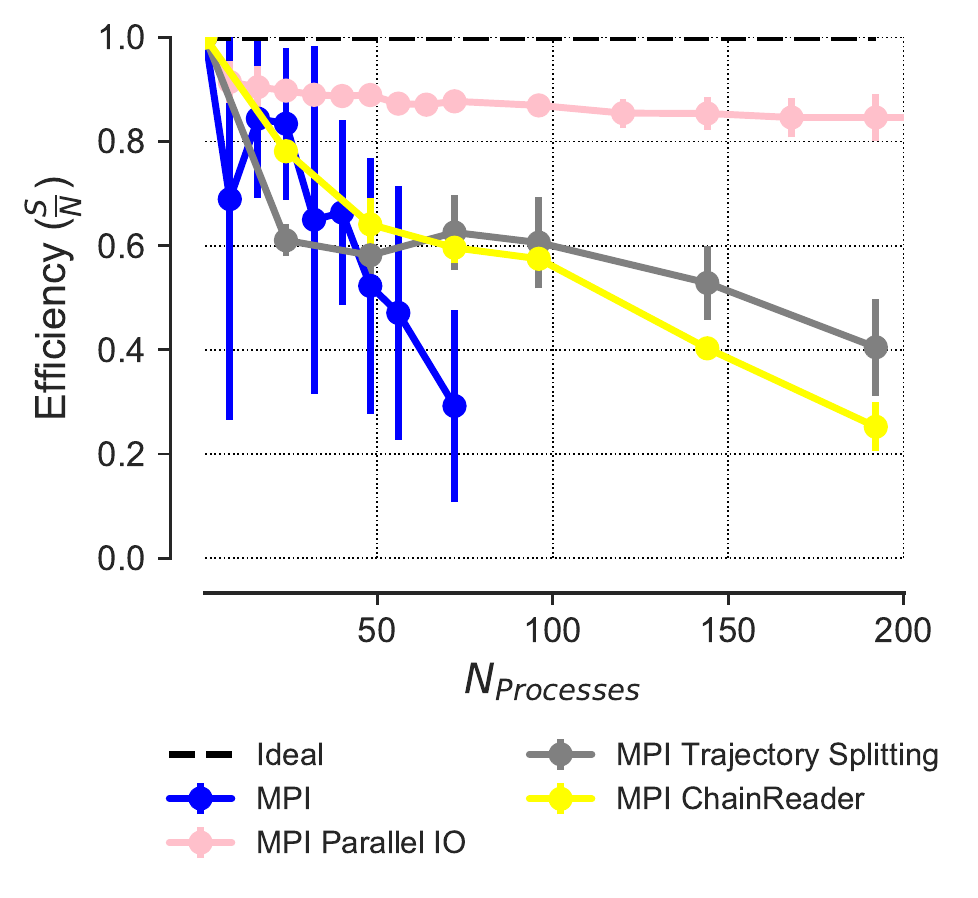}
    \caption{\emph{SDSC Comet}}
    \label{fig:comparison_efficiency}
  \end{subfigure}
  \hfill
  \begin{subfigure}{.35\textwidth}
    \includegraphics[width=\linewidth]{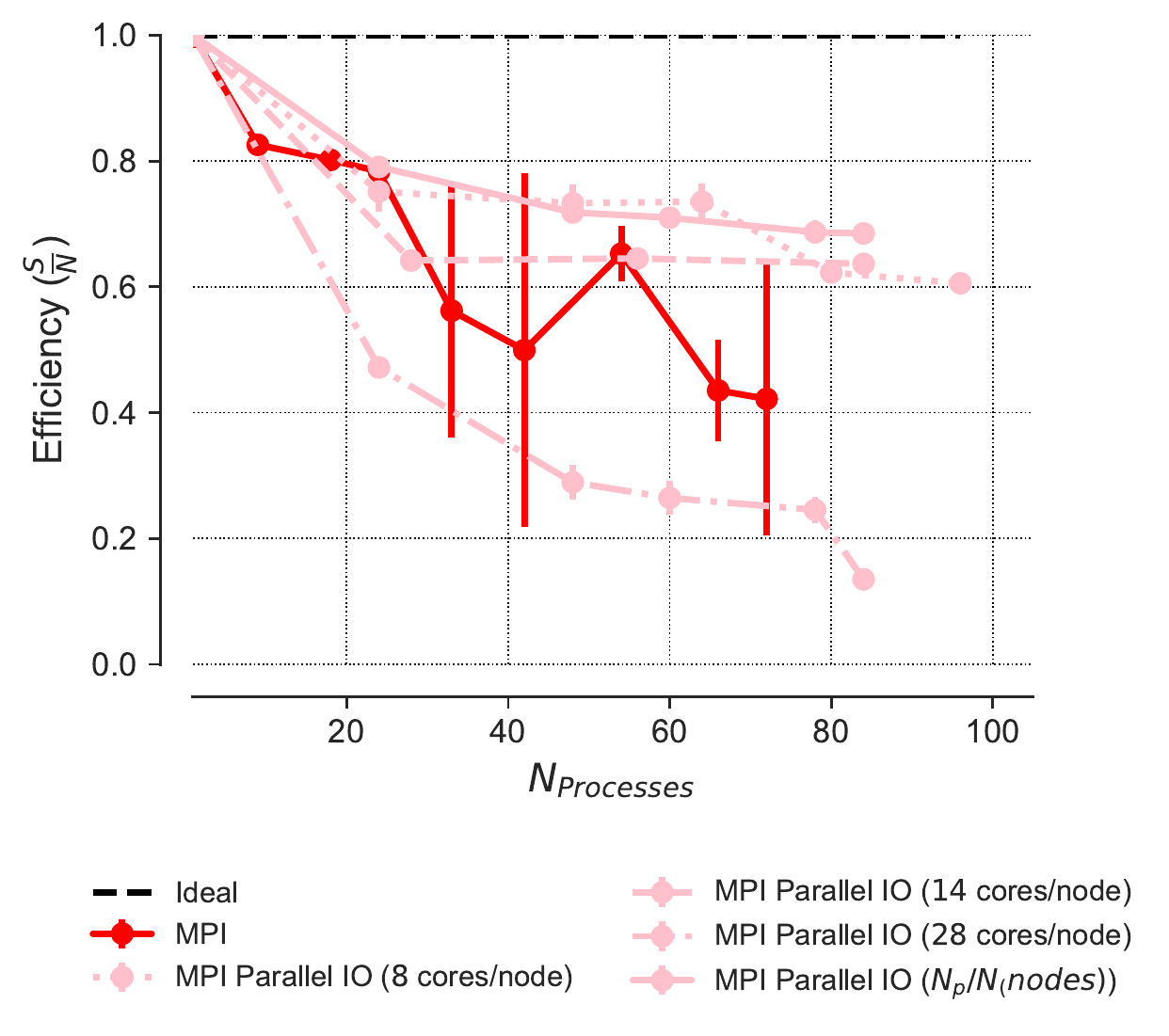}
    \caption{\emph{PSC Bridges}}
    \label{fig:comparison_efficiency_Bridges}
  \end{subfigure}
  \hfill
  \begin{subfigure}{.3\textwidth}
    \includegraphics[width=\linewidth]{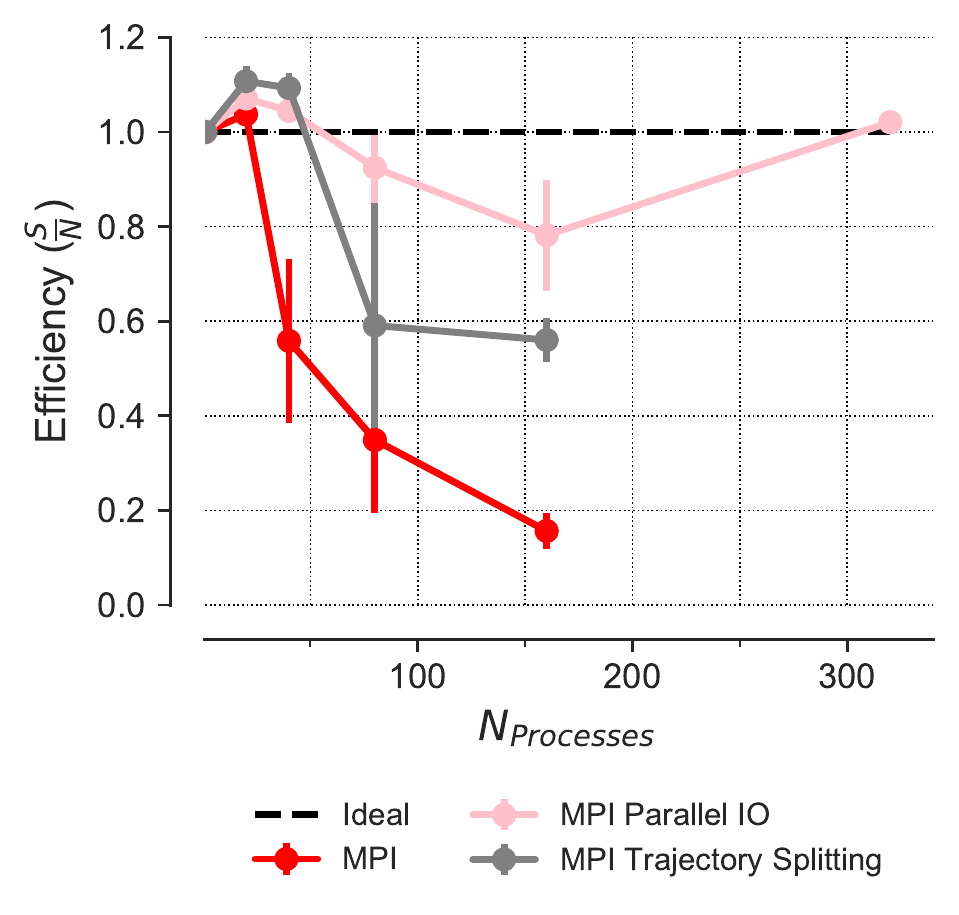}
    \caption{\emph{LSU SuperMIC}}
    \label{fig:comparison_efficiency_SuperMIC}
  \end{subfigure}
  \caption{Comparison of the parallel efficiency between different test cases on (a) \emph{SDSC Comet} (data for ``MPI Parallel IO'' are only shown up to 192 cores for better comparison across different scenarios, see Fig.~\protect\ref{fig:MPIspeedup-hdf5} for equivalent scaling data up to 384 cores), (b) \emph{PSC Bridges}, and (c) \emph{LSU SuperMIC}.
    Five repeats were performed to collect statistics and error bars show standard deviation with respect to mean.}
  \label{fig:comparison_efficiency_clusters}
\end{figure}

Figure \ref{fig:comparison_efficiency_clusters} shows comparison of the parallel efficiency of the RMSD task between different test cases on \emph{SDSC Comet}, \emph{PSC Bridges}, and \emph{LSU SuperMIC} when reading from a HDF5 file.

\section{Effect of Increasing the Computational Load on Scaling
  Performance}
\label{sec:shiftload}

We quantified the effect of increasing the computational load of the analysis task on scaling performance in order to obtain practical insights into the question under which circumstances the simple single-trajectory split-apply-combine MPI-based parallelization approach might be feasible without any other considerations such as optimization of the file read I/O.
To this end, we quantified strong scaling performance as a function of the compute-to-I/O ratio \RcompIO (Eq.~\ref{eq:Compute-IO}) and the compute-to-communication ratio \Rcompcomm (Eq.~\ref{eq:Compute-comm}).

To measure the effect of an increased compute load on performance while leaving other parameters the same, we artificially increased the computational load by repeating the same RMSD calculation (line 5, algorithm \ref{alg:RMSD}) 40, 70 and 100 times in a loop, resulting in forty-fold (``$40\times$''), seventy-fold (``$70\times$''), and one hundred-fold (``$100\times$'') load increases.


\subsection{Effect of \RcompIO on Scaling Performance}
\label{sec:increasedworkload}

For an $X$-fold increase in workload, we expected the workload for the computation to scale with $X$ as $\tcomp(X) =  N_{\text{frames}}^{\text{total}} X \overline{\tcomp^{\text{frame}}}$ while the read I/O workload $\tIO(X) = N_{\text{frames}}^{\text{total}} \overline{\tIO^{\text{frame}}}$ (number of frames times the average time to read a frame) should remain independent of $X$.
Therefore, the ratio for any $X$ should be $\RcompIO(X) = \tcomp(X)/\tIO(X) = X \RcompIO(X=1)$, i.e.,  $\RcompIO$ should just linearly scale with the workload factor $X$.
The measured $\RcompIO$ ratios of 11, 19, 27 for the increased computational workloads agreed with this theoretical analysis, as shown in Table \ref{tab:load-ratio}.

\begin{SCtable}[1.0][!htb]
\centering
\caption[Change in load-ratio with RMSD workload]{Change in $\RcompIO$ ratio with change in the RMSD workload $X$.
  The RMSD workload was artificially increased in order to examine the effect of compute to I/O ratio on the performance.
  The reported compute and I/O time were measured based on the serial version using one core.
  The theoretical $\RcompIO$ (see text) is provided for comparison.}
\label{tab:load-ratio}
\begin{tabular}{rrrrr}
  \toprule
  \bfseries\thead{Workload $X$} &  \bfseries\thead{$\tcomp$ (s)} &  \bfseries\thead{$\tIO$ (s)}
  & \multicolumn{2}{c}{\bfseries\thead{$\RcompIO$}}\\
  & & & \thead{measured} & \thead{theoretical}\\
  \midrule
    $1\times$   &   226 & 791 &  0.29 &   \\  
    $40\times$  &  8655 & 791 & 11   & 11\\    
    $70\times$  & 15148 & 791 & 19   & 20\\  
    $100\times$ & 21639 & 791 & 27   & 29\\  
  \bottomrule
\end{tabular}
\end{SCtable}

\begin{figure}[!htb]
  \centering
  \begin{subfigure} {.3\textwidth}
    \includegraphics[width=\linewidth]{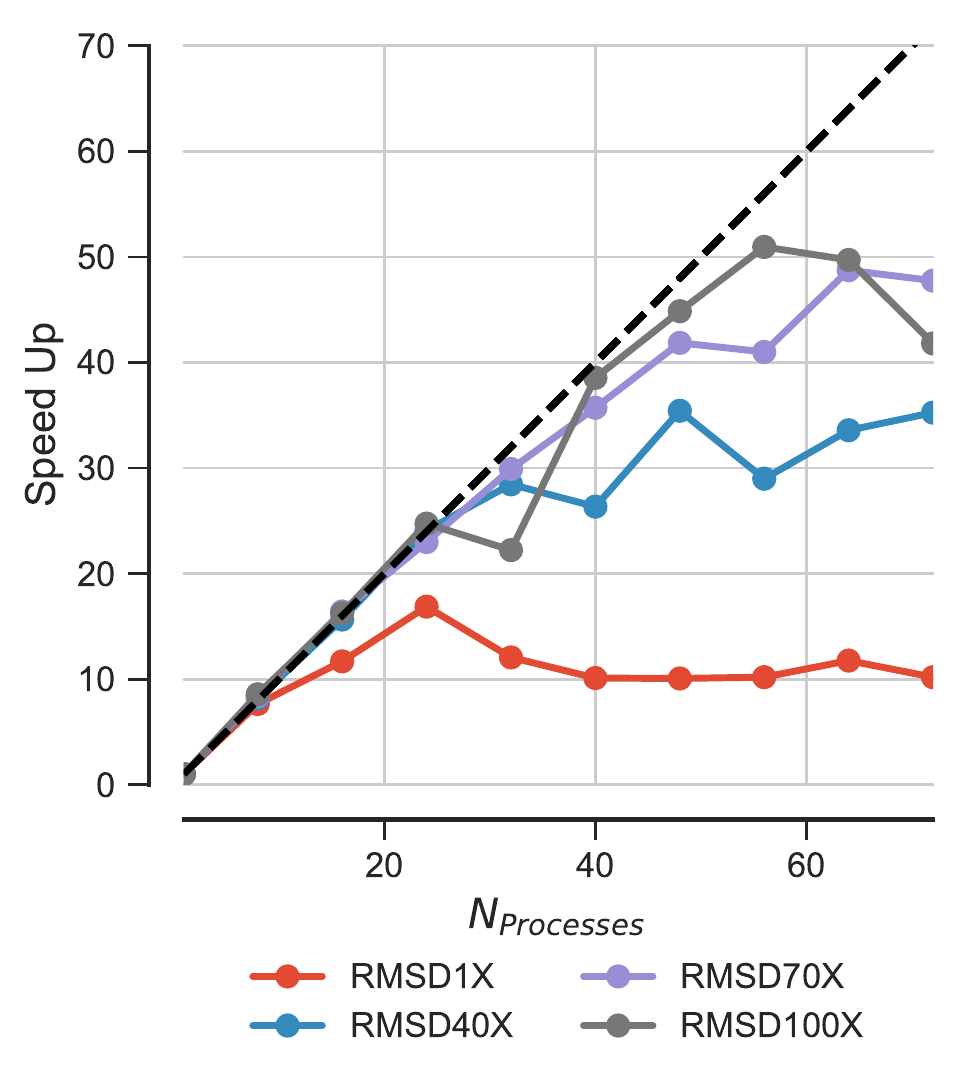}
    \caption{Speed-Up}
    \label{fig:S1_tcomp_tIO_effect}
  \end{subfigure}
  \hfill
  \begin{subfigure}{.3\textwidth}
    \includegraphics[width=\linewidth]{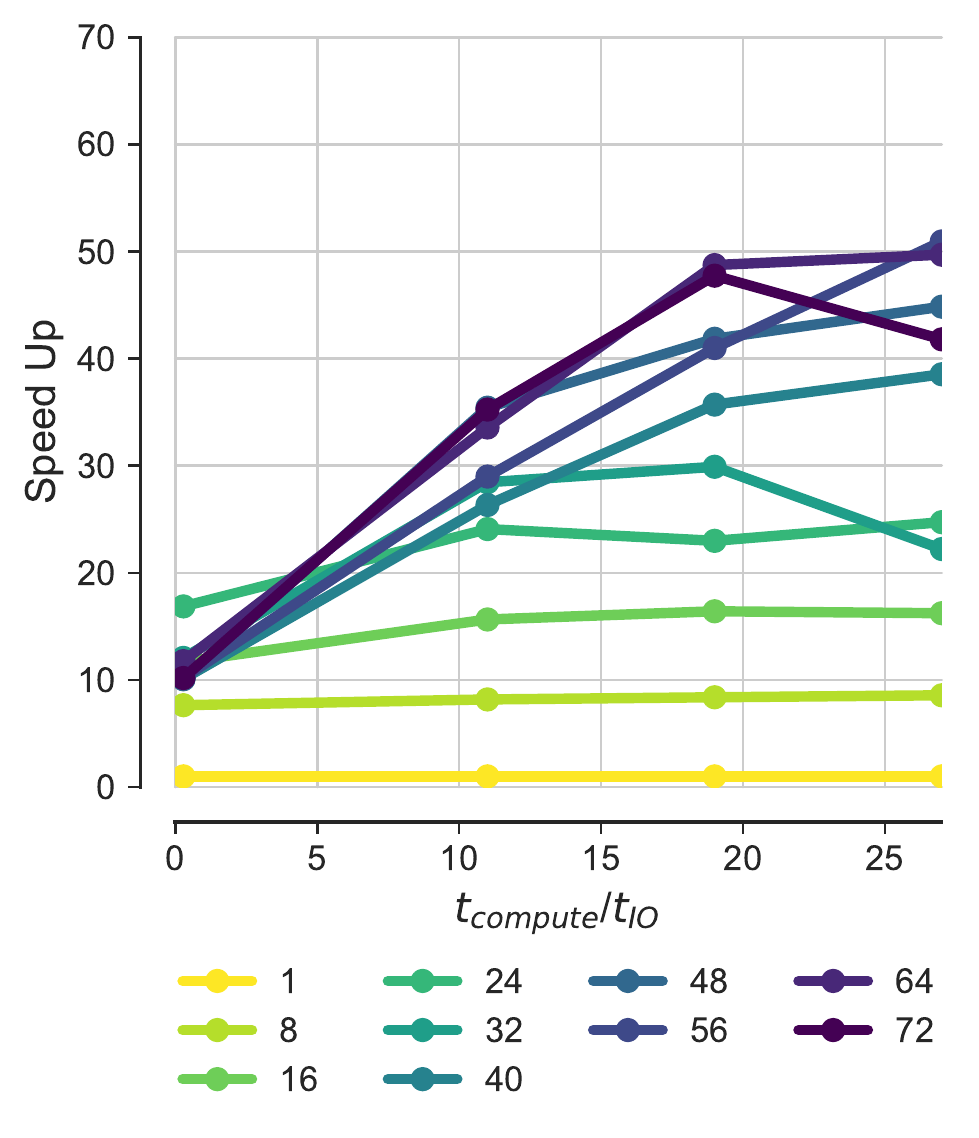}
    \caption{Speed-Up}
    \label{fig:S2_tcomp_tIO_effect}
  \end{subfigure}
  \hfill
  \begin{subfigure}{.3\textwidth}
    \includegraphics[width=\linewidth]{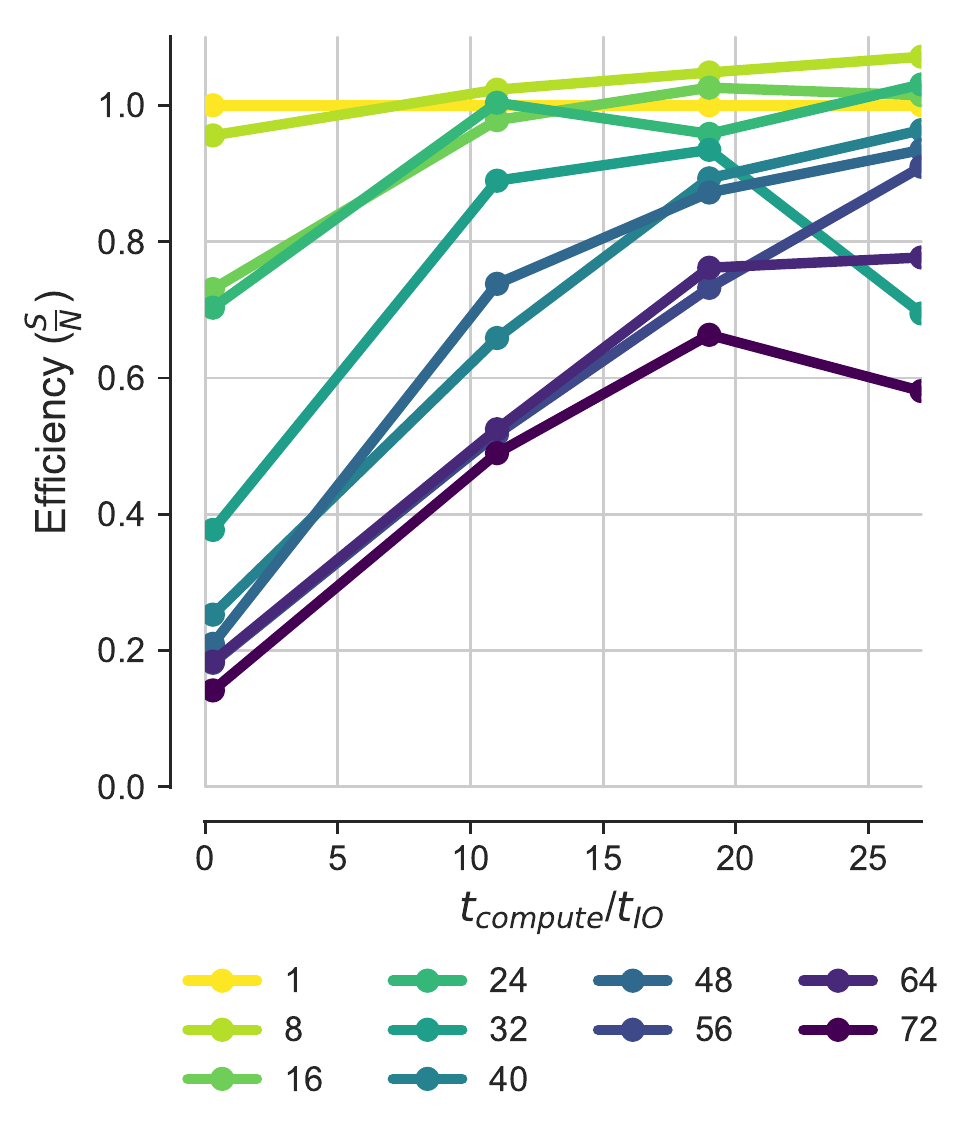}
    \caption{Efficiency}
    \label{fig:E_tcomp_tIO_effect}
  \end{subfigure}
  \caption{Effect of $\RcompIO$ ratio on performance of the RMSD task on \emph{SDSC Comet}. We tested performance for $\RcompIO$ ratios of 0.3, 11, 19, 27, which correspond to $1\times$ RMSD, $40\times$ RMSD, $70\times$ RMSD, and $100\times$ RMSD respectively.
    (a) Effect of $\RcompIO$ on the speed-up.
    (b) Change in speed-up with respect to $\RcompIO$ for different processor counts.
    (c) Change in the efficiency with respect to $\RcompIO$ for different processor counts.}
  \label{fig:tcomp_tIO_effect}
\end{figure}

We performed the experiments with increased workload to measure the effect of the $\RcompIO$ ratio (Eq.~\ref{eq:Compute-IO}) on performance (Figure~\ref{fig:tcomp_tIO_effect}).
The strong scaling performance as measured by the speed-up $S(N)$ improved with increasing $\RcompIO$ ratio (Figure \ref{fig:S1_tcomp_tIO_effect}).
The calculations consistently showed better scaling performance to larger numbers of cores for higher $\RcompIO$ ratios, e.g., $N=56$ cores for the $70\times$ RMSD task. 
The speed-up and efficiency approached their ideal value for each processor count with increasing $\RcompIO$ ratio (Figures \ref{fig:S2_tcomp_tIO_effect} and \ref{fig:E_tcomp_tIO_effect}).
Even for moderately compute-bound workloads, such as the $40\times$ and $70\times$ RMSD tasks, increasing the computational workload over I/O reduced the impact of stragglers even though they still contributed to large variations in timing across different ranks and thus irregular scaling.



\subsection{Effect of \Rcompcomm on Scaling Performance}
\label{sec:tcomm}

In Section \ref{sec:I/O}, we improved scaling limitations due to read I/O by splitting the trajectory, but scaling remained far from ideal because of increased communication costs.
These results motivated an analysis in terms of the communication costs.
In addition to the compute to I/O ratio \RcompIO discussed in Section \ref{sec:increasedworkload} we defined another performance parameter called the compute to communication ratio $\Rcompcomm$ (Eq.~\ref{eq:Compute-comm}).

\begin{figure}[!htb]
  \centering
  \begin{subfigure}{.3\textwidth}
    \includegraphics[width=\linewidth]{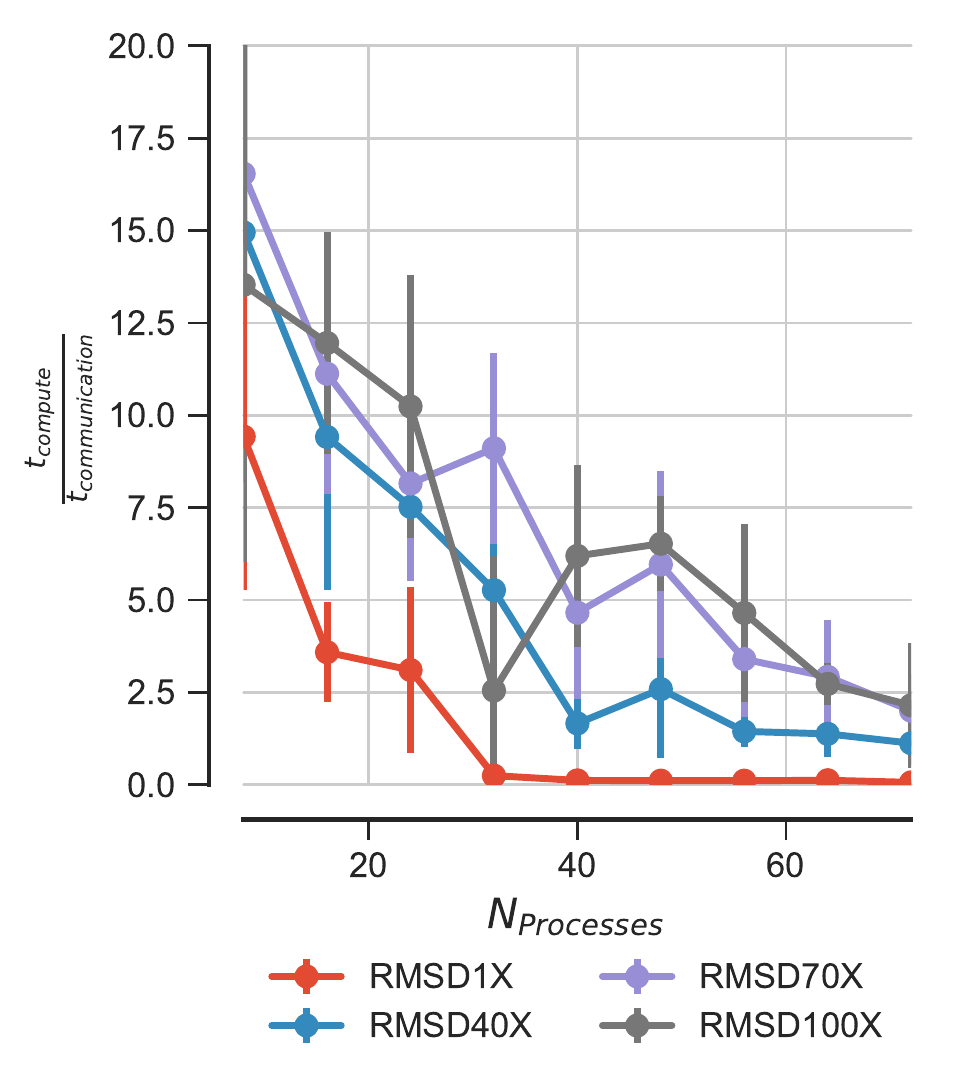}
    \captionsetup{format=hang}
    \caption{Compute to communication ratio \Rcompcomm}
    \label{fig:tcomp_tcomm_ratio}
  \end{subfigure}
  \hspace{1em}
  \begin{subfigure}{.33\textwidth}
    \includegraphics[width=\linewidth]{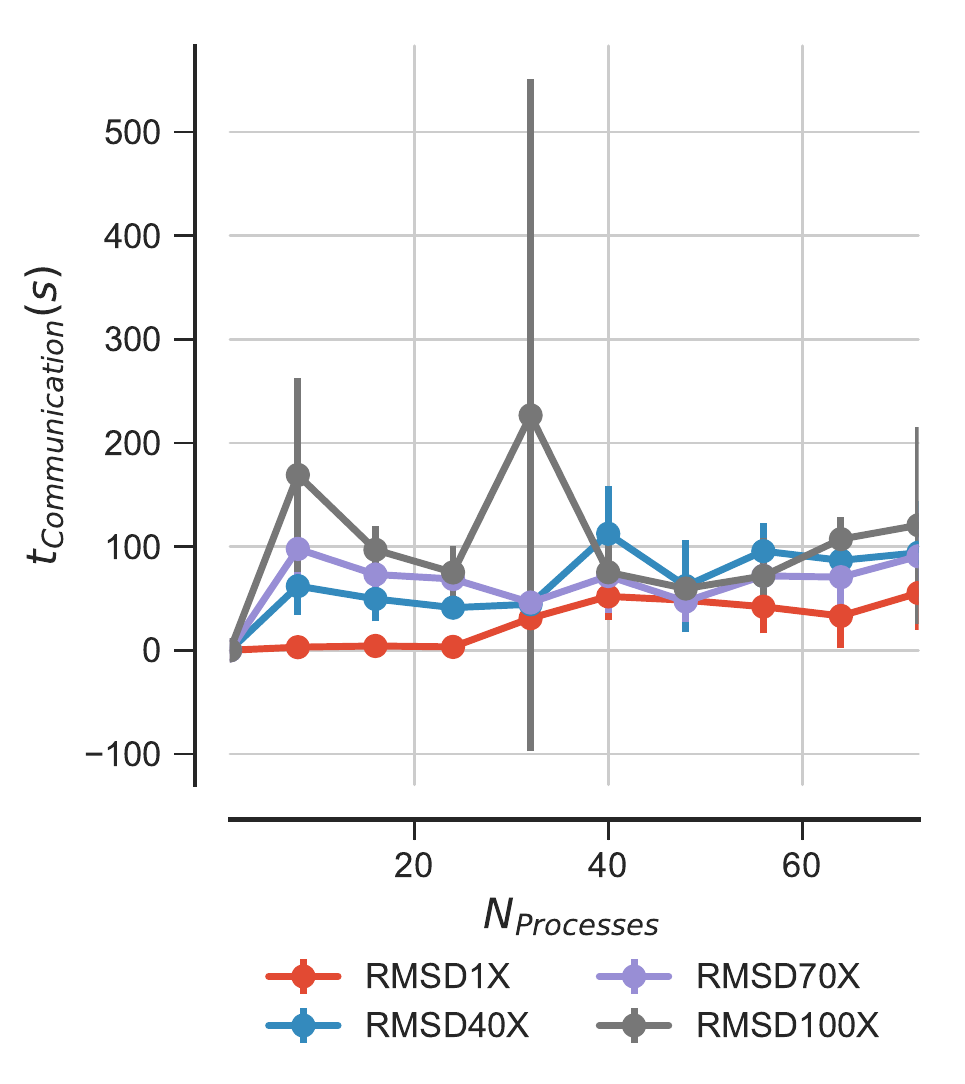}
    \caption{Communication time \tcomm}
    \label{fig:Comm_time_tcomp_tcomm_effect}
  \end{subfigure}
  \caption{Effect of the ratio of compute to communication time \Rcompcomm on scaling performance on \emph{SDSC Comet}.
    These are the same data as shown in Figure~\protect\ref{fig:S1_tcomp_tIO_effect} but analyzed with respect to the communication load.
    (a) Change in \Rcompcomm with the number of processes $N$ for different workloads. 
    (b) Comparison of communication time for different RMSD workloads.
    Five repeats were performed to collect statistics and error bars show standard deviation with respect to mean.}
  \label{fig:tcom_tcomm_effect}
\end{figure}

We analyzed the data for variable workloads (see Section \ref{sec:increasedworkload}) in terms of the $\Rcompcomm$ ratio.
The performance clearly depended on the $\Rcompcomm$ ratio (Figure \ref{fig:tcom_tcomm_effect}).
Performance improved with increasing $\Rcompcomm$ ratios (Figures \ref{fig:tcomp_tcomm_ratio} and \ref{fig:S1_tcomp_tIO_effect}) even if the communication time was larger (Figure \ref{fig:Comm_time_tcomp_tcomm_effect}).
Although we still observed stragglers due to communication at larger $\Rcompcomm$ ratios ($70\times$ RMSD and $100\times$ RMSD), their effect on performance remained modest because the overall performance was dominated by the compute load. 
Evidently, as long as overall performance is dominated by a component such as compute that scales well, then performance problems with components such as communication will be masked and overall acceptable performance can still be achieved (Figures \ref{fig:S1_tcomp_tIO_effect} and \ref{fig:tcomp_tcomm_ratio}).

Communication was usually not problematic within one node because of the shared memory environment.
For less than 24 processes, i.e., a single compute node on \emph{SDSC Comet}, the scaling was good and $\Rcompcomm \gg 1$ for all RMSD loads (Figures \ref{fig:S1_tcomp_tIO_effect} and \ref{fig:tcomp_tcomm_ratio}).
However, beyond a single compute node ($>$ 24 cores), scaling appeared to improve with increasing $\Rcompcomm$ ratio while the communication overhead decreased in importance (Figures \ref{fig:S1_tcomp_tIO_effect} and \ref{fig:tcomp_tcomm_ratio}).

\section{Performance of the ChainReader for Split Trajectories}
\label{sec:chainreader}

In section \ref{sec:splitting-traj} we showed how subfiling (splitting the trajectories) could help overcome I/O limitations and improve scaling. 
However, the number of trajectories may not necessarily be equal to the number of processes.
For example, trajectories from MD simulations on supercomputers are often kept in small segments in individual files that need to be concatenated later to form a trajectory that can be analyzed with common tools.
Such segments might be useful for subfiling but making sure that the number of processes is equal to the number of trajectory files will not always be feasible. 
\package{MDAnalysis} can transparently represent multiple trajectories as one virtual trajectory using the \emph{ChainReader}.
This feature is convenient for serial analysis when trajectories are maintained as segments.
In the current implementation of ChainReader, each process opens all the trajectory segment files but I/O will only happen from a specific block of the trajectory specific to that process only.

We wanted to test if the ChainReader would benefit from the gains measured for the subfiling approach.
Specifically, we measured if the MPI-parallelized RMSD task (with $N_{\text{p}}$ ranks) would benefit if the trajectory was split into $N_{\text{seg}} = N_{\text{p}}$ trajectory segments, corresponding to an ideal scenario.
 
\begin{figure}[!htb]
\centering
\begin{subfigure}{.4\textwidth}
  \includegraphics[width=\linewidth]{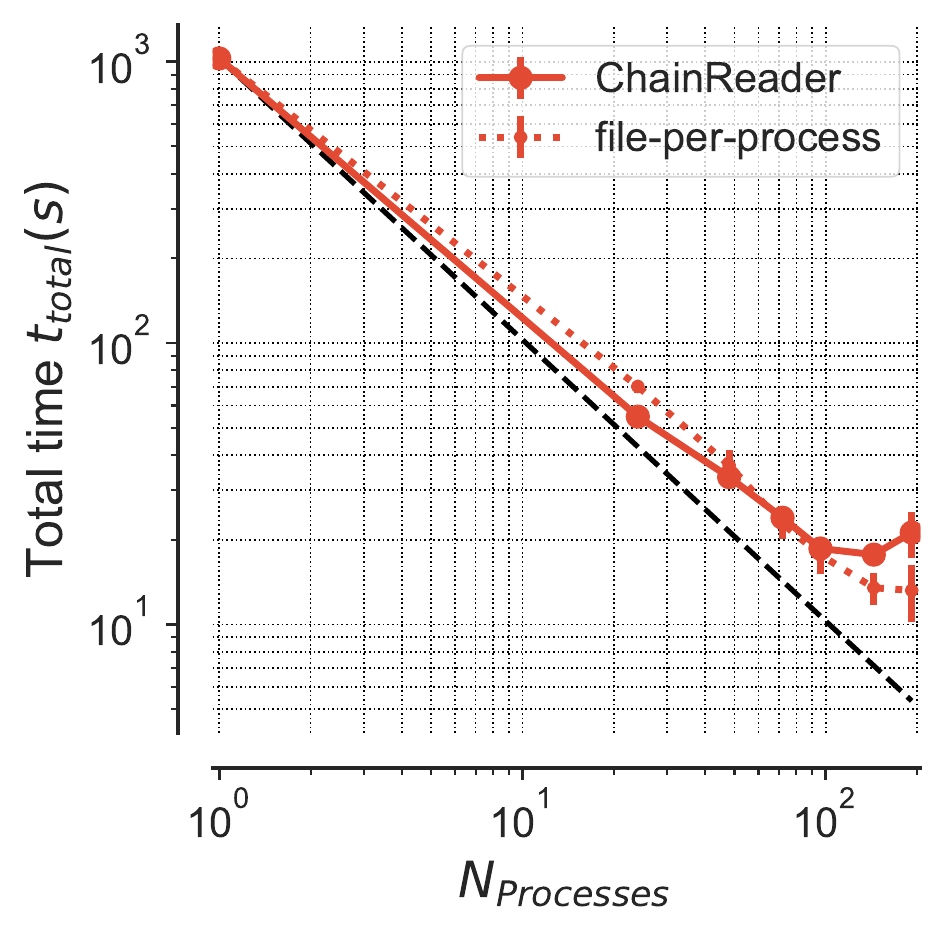}
  \caption{Scaling total}
  \label{fig:MPItottime-chain-reader}
\end{subfigure}
\hfill
\begin{subfigure}{.4\textwidth}
  \includegraphics[width=\linewidth]{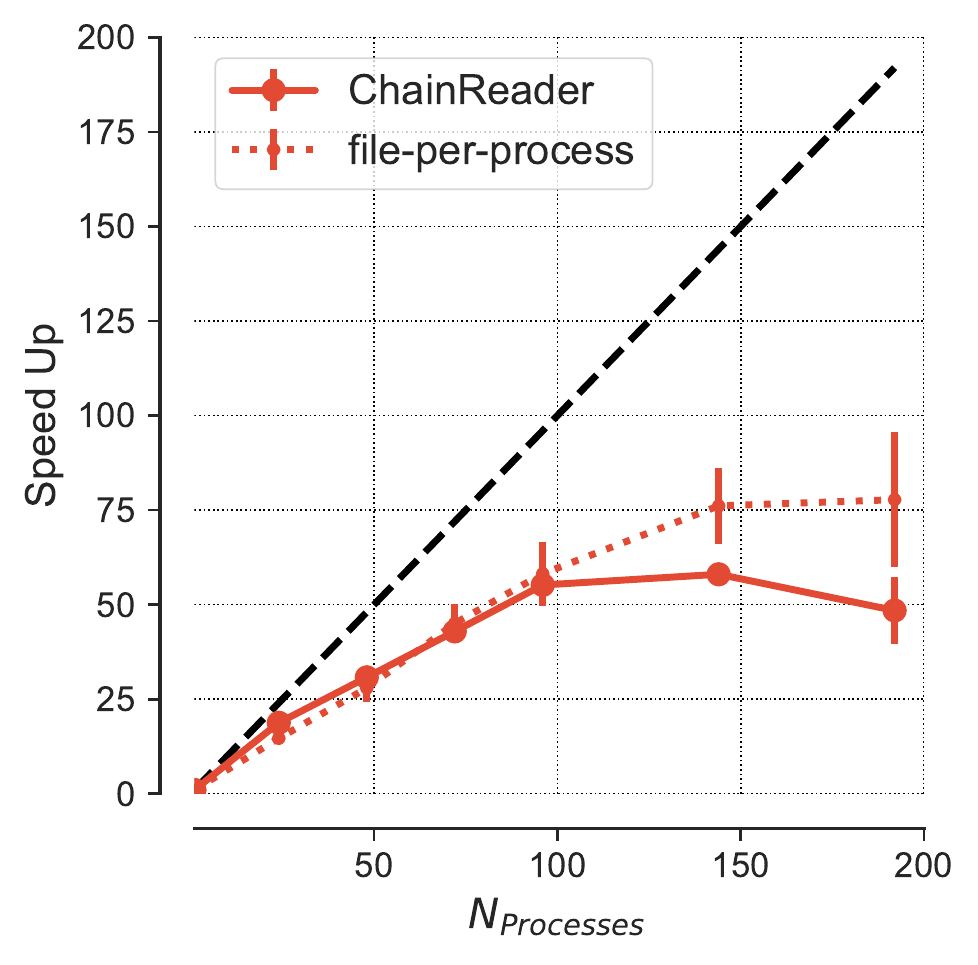}
  \caption{Speed-up}
  \label{fig:MPIspeedup-chain-reader}
\end{subfigure}
\bigskip
\begin{subfigure}{.45\textwidth}
  \includegraphics[width=\linewidth]{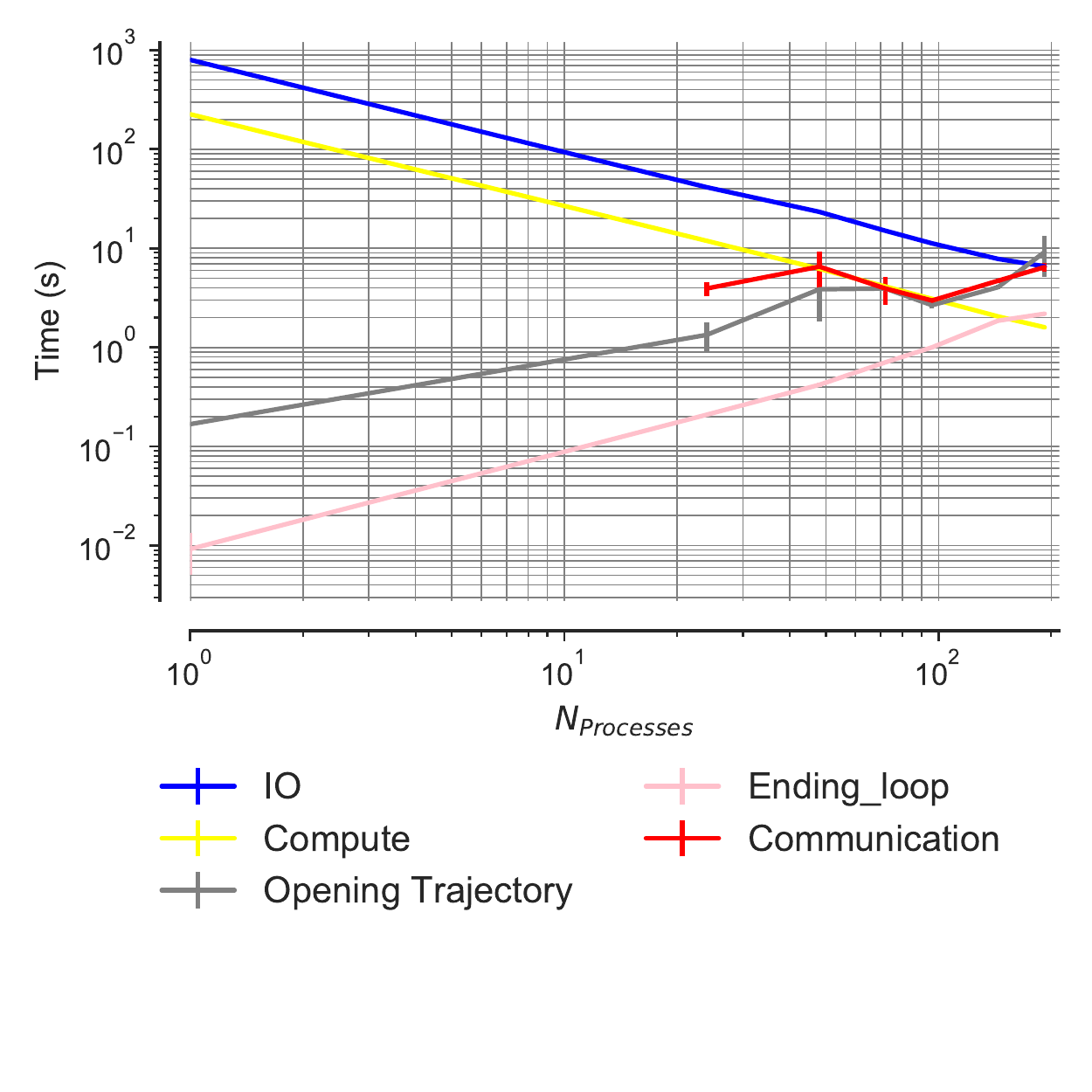}
  \captionsetup{format=hang}
  \caption{Scaling for different components}
  \label{fig:MPIscaling-chain-reader}
\end{subfigure}
\hfill
\begin{subfigure}{.5\textwidth}
  \includegraphics[width=\linewidth]{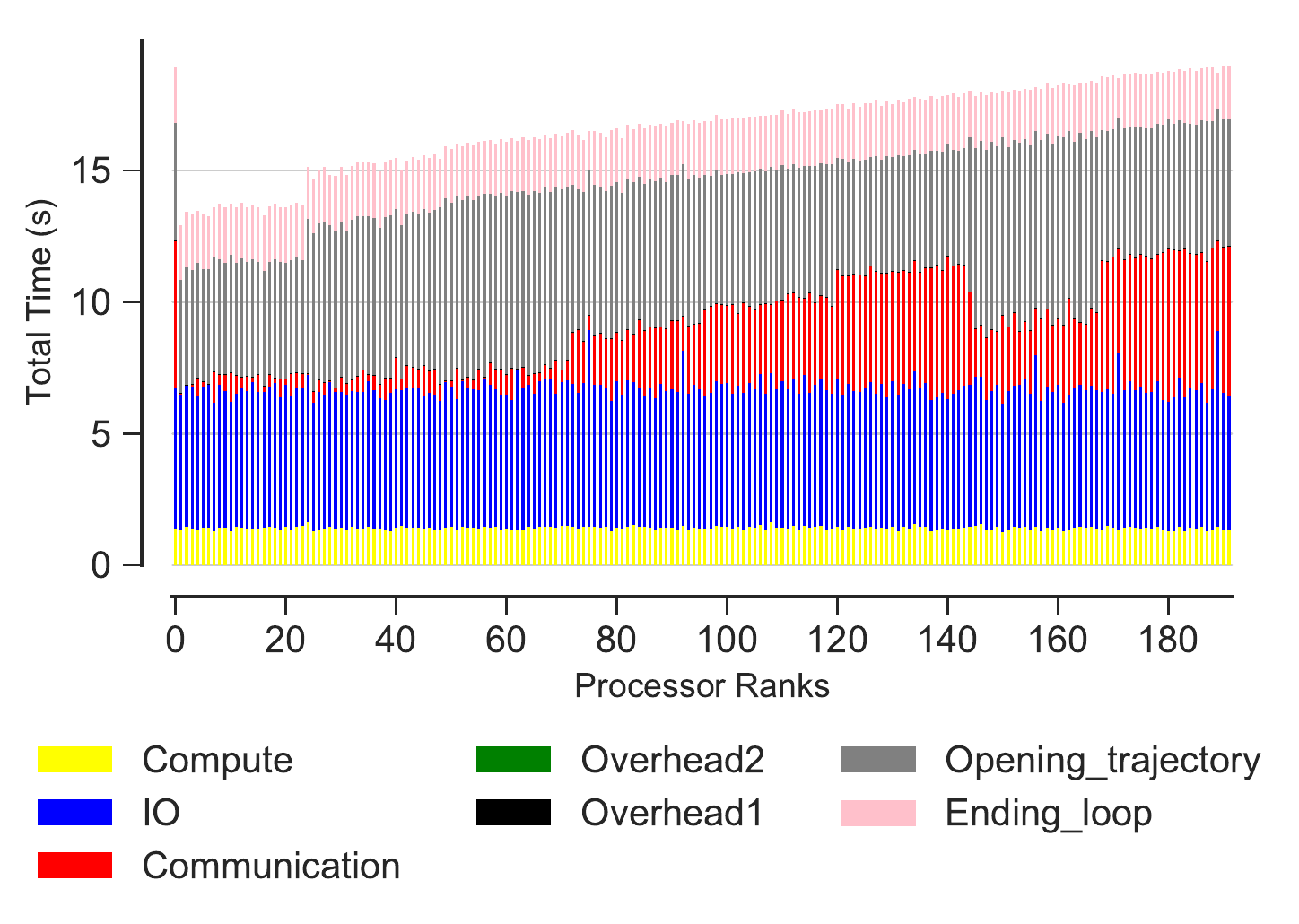}
  \captionsetup{format=hang}
  \caption{Time comparison of different parts of the calculations per MPI rank}
  \label{fig:MPIranks-split-chain-reader}
\end{subfigure}

\caption{Subfiling with the MDAnalysis \emph{ChainReader} for the RMSD task on \emph{SDSC Comet}.
Five repeats were performed to collect statistics.
The data for subfiling with one file per process (from Figure~\protect{\ref{fig:MPIwithIO-split}}) are shown for comparison as dotted lines in (a) and (b).  
(a-c) The error bars show standard deviation with respect to the mean.
(d) Compute \tcomp, read I/O \tIO, communication \tcomm, opening the trajectory $t_{\text{opening\_trajectory}}$, ending the for loop $t_{\text{end\_loop}}$ (includes closing the trajectory file),  and overheads $t_{\text{overhead1}}$, $t_{\text{overhead2}}$ per MPI rank. (See Table \ref{tab:notation} for the definitions.)}
\label{fig:MPIwithIO-split-chain-reader}
\end{figure}

In order to perform our experiments we had to work around an issue with the XTC format reader in \package{MDAnalysis} that was related to the XTC random-access functionality that the \texttt{MDAnalysis.coordinates.XTC.XTCReader} class provides:
The Gromacs XTC format \cite{Lindahl01, Spangberg:2011zr} is a lossy-compression, XDR-based file format that was never designed for random access and the compressed format itself does not support fast random seeking.
The \texttt{XTCReader} stores persistent offsets for trajectory frames to disk \cite{Gowers:2016aa} in order to enable efficient access to random frames.
These offsets will be generated automatically the first time a trajectory is opened and the offsets are stored in hidden \texttt{*.xtc\_offsets.npz} files. 
The advantage of these persistent offset files is that after opening the trajectory for the first time, opening the same file will be very fast, and random access is immediately available. 
However, stored offsets can get out of sync with the trajectory they refer to. 
To prevent the use of stale offset data, trajectory file data (number of atoms, size of the file and last modification time) are also stored for validation.
If any of these parameters change the offsets are recalculated. 
If the XTC changes but the offset file is not updated then the offset file can be detected as invalid.
With ChainReader in parallel, each process opens all the trajectories because each process builds its own \texttt{MDAnalysis.Universe} data structure.
If an invalid offset file is detected (perhaps because of wrong file modification timestamps across nodes), several processes might want to recalculate these parameters and rebuild the offset file, which can lead to a race condition.
In order to avoid the race condition, we removed this check from MDAnalysis for the purpose of the measurements presented here, but this comes at the price of not checking the validity of the offset files at all; future versions of MDAnalysis may lift this limitation.  
We obtained the results for the ChainReader with this modified version of \package{MDAnalysis} that eliminates the race condition by assuming that XTC index files are always valid.

Figure \ref{fig:MPIwithIO-split-chain-reader} shows the results for performance of the ChainReader for the RMSD task.
Strong scaling plateaued for more than 92 cores and performance was worse than what was achieved when each MPI process was assigned its own trajectory segment as described in Section \ref{sec:splitting-traj} and shown for comparison as dotted lines in Figures~\ref{fig:MPItottime-chain-reader} and \ref{fig:MPIspeedup-chain-reader}.
The strong scaling performance did not suffer because of the computation and the read I/O because both \tcomp and \tIO showed excellent strong scaling up to 196 cores (Figure \ref{fig:MPIscaling-chain-reader}).
Instead the time for ending the \texttt{for} loop $t_{\text{end\_loop}}$, which includes the time for closing the trajectory file, and opening the trajectory $t_{\text{opening\_trajectory}}$ appeared to be the scaling bottleneck.
These results differed from the subfiling results (section \ref{sec:splitting-traj}) where neither $t_{\text{end\_loop}}$ nor $t_{\text{opening\_trajectory}}$ limited scaling (Figure \ref{fig:MPIranks-split}). 

Although we did not further investigate the deeper cause for the reduced scaling performance of the ChainReader, we speculate that the primary problem is related to each MPI rank having to open all trajectory files in their ChainReader instance even though they will only read from a small subset.
For $N_{\text{p}}$ ranks and $N_{\text{seg}}$ file segments, in total, $N_{p } N_{\text{seg}}$ file opening/closing operations have to be performed. 
Each server that is part of a Lustre file system can only handle a limited number of I/O requests (read, write, stat, open, close, etc.) per second.
A large number of such requests, from one or more users and one or more jobs, can lead to contention for storage resources. 
For $N_{\text{p}} = N_{\text{seg}} = 100$, the Lustre file system has to perform 10,000 of these operations almost simultaneously, which might degrade performance.

These considerations indicate that the ChainReader in its current implementation limits scaling performance to less than 100 processes due to the large number of file opening operations.
For better performance, the ChainReader would need to be rewritten for MPI such that the layout of the trajectory files is initially obtained by MPI rank 0 (which has to sequentially open all trajectory segments once) and then communicated to all other ranks; each rank then only opens the trajectories it needs to access, thus reducing file access to a minimum.


\end{document}